\documentclass[acmlarge]{acmart}
\usepackage{amsmath}
\usepackage{caption}
\usepackage{subcaption}
\usepackage{hyperref}
\usepackage{url}
\usepackage{lipsum}
\usepackage{graphicx}
\usepackage{tabularx}
\usepackage{array}
\usepackage{multirow}
\usepackage{tikz}
\usepackage{tikz-qtree}
\usetikzlibrary{trees}

\AtBeginDocument{%
  \providecommand\BibTeX{{%
    \normalfont B\kern-0.5em{\scshape i\kern-0.25em b}\kern-0.8em\TeX}}}

\setcopyright{acmcopyright}
\copyrightyear{2023}
\acmYear{2023}

\begin{document}
\title{Investigating Online Financial Misinformation and Its Consequences: A Computational Perspective}

\author{Aman Rangapur}
\email{arangapur@hawk.iit.edu}
\orcid{0000-0003-1362-0498}
\affiliation{%
  \institution{Illinois Institute of Technology}
  \city{Chicago}
  \state{Illinois}
  \country{USA}
  \postcode{60616}
}

\author{Haoran Wang}
\email{hwang219@hawk.iit.edu}
\affiliation{%
  \institution{Illinois Institute of Technology}
  \city{Chicago}
  \country{USA}
}


\author{Kai Shu}
\email{kshu@hawk.iit.edu}
\affiliation{%
  \institution{Illinois Institute of Technology}
  \city{Chicago}
  \country{USA}
}

\renewcommand{\shortauthors}{Aman and Haoran, et al.}
\begin{abstract}
The rapid dissemination of information through digital platforms has revolutionized the way we access and consume news and information, particularly in the realm of finance. However, this digital age has also given rise to an alarming proliferation of financial misinformation, which can have detrimental effects on individuals, markets, and the overall economy. This research paper aims to provide a comprehensive survey of online financial misinformation, including its types, sources, and impacts. We first discuss the characteristics and manifestations of financial misinformation, encompassing false claims and misleading content. We explore various case studies that illustrate the detrimental consequences of financial misinformation on the economy. These real-world examples demonstrate the extent to which misinformation can disrupt markets, harm investors, and create economic instability. Examples range from fraudulent investment schemes to misleading news articles or social media posts aimed at manipulating stock prices or influencing market sentiment. Such misinformation poses serious risks, including market volatility, investor losses, and erosion of trust in financial institutions. Finally, we highlight the potential impact and implications of detecting financial misinformation. Early detection and mitigation strategies can help protect investors, enhance market transparency, and preserve financial stability. We emphasize the importance of greater awareness, education, and regulation to address the issue of online financial misinformation and safeguard individuals and businesses from its harmful effects. In conclusion, this research paper sheds light on the pervasive issue of online financial misinformation and its wide-ranging consequences. By understanding the types, sources, and impacts of misinformation, stakeholders can work towards implementing effective detection and prevention measures to foster a more informed and resilient financial ecosystem.

\end{abstract}


\maketitle
\section{Introduction}
\label{introduction}
The internet has revolutionized the way people access and disseminate information \cite{Rosmani2020,carpenter_2023}. It has transformed various aspects of daily life, including the way individuals conduct financial transactions. The advent of online banking, investment, and trading platforms has made it easier for individuals to access financial information and conduct financial transactions from the comfort of their homes \cite{Broby2021}. However, the widespread availability of financial information online has also led to an increase in the spread of financial misinformation \cite{Muhammed_T2022}. It is crucial to differentiate between misinformation and financial misinformation, as they have distinct implications and consequences. Misinformation refers to the spread of false or inaccurate information, often unintentionally, leading to a misinterpretation of facts or events \cite{allcott2017social}. It can arise from various sources, such as human error, rumors, or incomplete data. On the other hand, financial misinformation specifically pertains to the dissemination of false or misleading information concerning financial matters, with the intention of manipulating markets, investors, or financial systems for personal gain \cite{mohankumar2023financial01}. Financial misinformation can include false investment advice, deceptive financial statements, or fraudulent schemes aimed at defrauding individuals or institutions \cite{chung2022theory1}. Differentiating between these misinformation types is crucial to combat their harmful effects on public understanding, decision-making, and financial stability. With increasing online transactions, the impact of financial misinformation has become significant, warranting a deeper understanding.

In recent years, the recognition of studying online financial misinformation has grown among researchers and practitioners \cite{zhang2022theory009,liu2022role02,mohankumar2023financial01,di2021fake12,chung2022theory1,carpenter_2023}. Studies have been conducted to explore this phenomenon and its impact on consumers. Notably, individuals with higher levels of financial literacy have exhibited greater resistance to the influence of fake financial news, suggesting that financial education could be an effective strategy to combat online financial misinformation \cite{10.1093/rof/rfac058}. However, regulatory bodies face the challenge of striking a balance between protecting consumers and preserving the advantages of innovation and competition in the financial sector \cite{kirilenko2019regulating}.

Financial education has shown promise in reducing individual's vulnerability to financial scams and misinformation \cite{liu2020financial}. Furthermore, the role of cognitive reflection in the dissemination of fake news, including financial misinformation, has been investigated \cite{pennycook2019falls}. Individuals who exhibit higher levels of cognitive reflection demonstrate a lower likelihood of sharing false news stories, emphasizing the significance of critical thinking and skepticism in the online environment. Examining the field of e-commerce, researchers have developed frameworks and conducted studies on various forms of deception. \cite{xiao_benbasat_2011asdd} proposed a theoretical framework explaining how sellers manipulate information, presentation, and generation of products to deceive consumers. Additionally, the impact of fake reviews on the visibility of hotels in the market has been demonstrated, showing the substantial influence even a small number of fabricated reviews can have \cite{lappas_sabnis_valkanas_2016as}. Moreover, the issue of identity deception in e-commerce has become significant, particularly due to physical distancing measures \cite{vishwanath_2014_social}. In the context of the supply chain, the economic impact of agents falsifying attributes to gain favorable decisions has been explored \cite{Cezar2020AdversarialCI}.

The characteristics of fake news, such as originality, recursive and periodic nature, and the role of social bots, contribute to its rapid and extensive dissemination \cite{vosoughi_roy_aral_2018adf,shin_jian_driscoll_bar_2018df,shao_ciampaglia_varol_yang_flammini_menczer_2018}. Efforts to combat fake news have involved investigating credibility labeling and reputation ratings of news sources \cite{kim_moravec_dennis_2019ass}. Manipulated financial reports, often generated through misconduct by inside managers and outside auditing firms, represent the most prevalent form of financial disinformation \cite{beneish1999detection,huang_lin_chiu_yenxc_2016}. Inaccurate and biased information resulting from the media's incentive to publish sensational news can lead to abnormal market reactions \cite{ahern_sosyuraa1_2015}. Research has explored the relationship between a firm's unusual news flow and its stock return, shedding light on the impact of financial misinformation \cite{bali_bodnaruk_scherbinaaq_tang_2018}. Moreover, social media has facilitated low-cost manipulation of the financial market by firms \cite{tardelli_avvenutii_tesconii_cresci_2022}.

While the literature on financial misinformation continues to grow, encompassing the various types of misinformation, the motivations behind dissemination, and its impact on individuals and society \cite{atske_2021,carpenter_2023}, there is a need for a comprehensive understanding of the different datasets, methods, and challenges related to this problem. Figure \ref{fig:hierarchy} illustrates the hierarchical structure of online financial misinformation, showcasing the interconnected layers of dissemination, amplification, and impact in a research paper.

\usetikzlibrary{arrows,shapes,positioning,shadows,trees}
\tikzset{
  basic/.style  = {draw, text width=7cm,  text height=0.25cm, drop shadow, font=\sffamily, rectangle},
  root/.style   = {basic, rounded corners=2pt, thin, align=left,
                   fill=green!60},
  level 2/.style = {basic, rounded corners=4pt, thin,align=left, fill=green!30,
                   text width=20em},
  level 3/.style = {basic, thin, align=left, fill=pink!30, text width=10em}
}
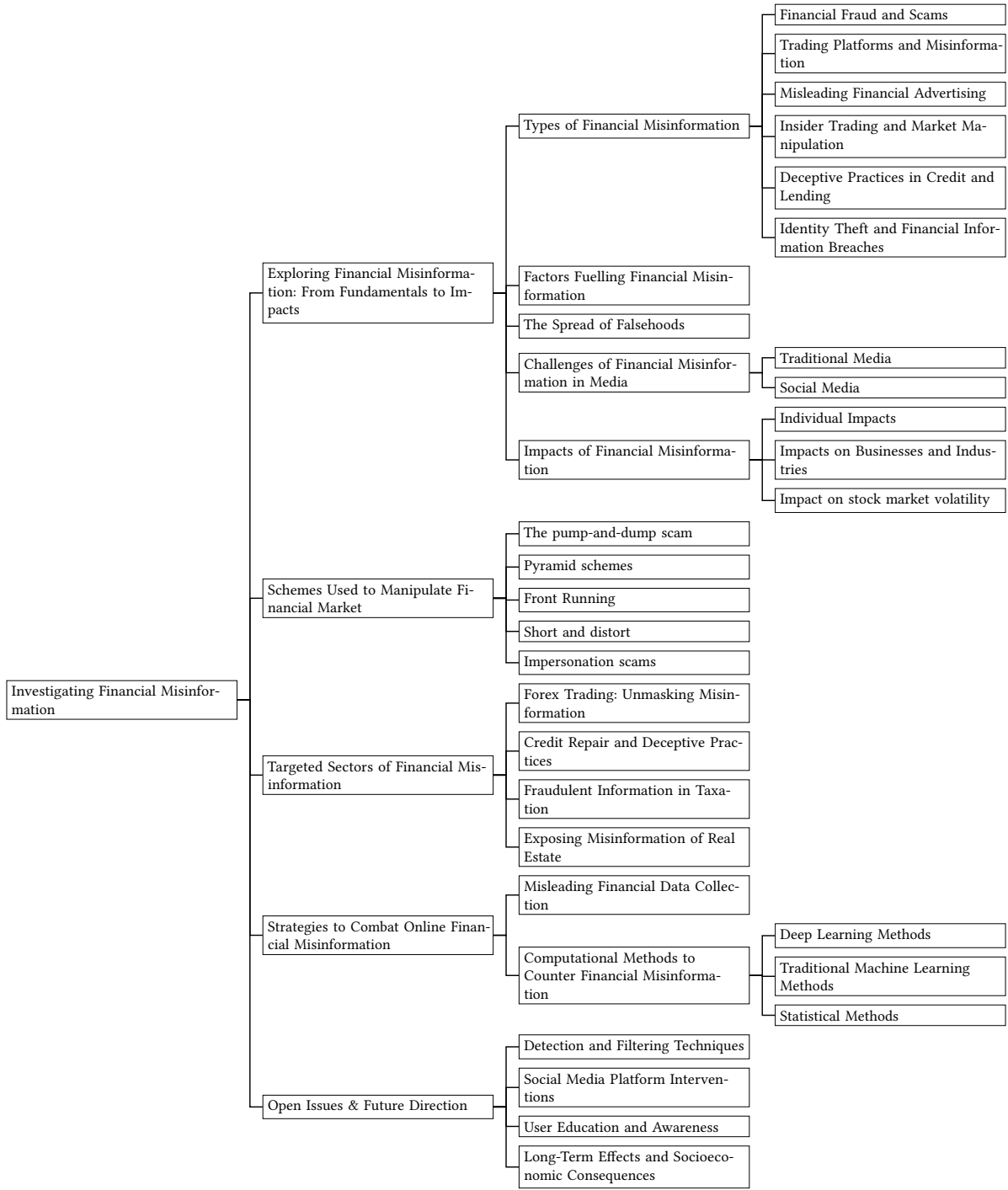
\begin{figure}
    \caption{Hierarchy of the paper.}
    \label{fig:hierarchy}
    \resizebox{1.0\textwidth}{!}{
\begin{tikzpicture}[
  grow'=right,
  level distance=2.35in,
  sibling distance=.08in,
]
\tikzset{
  edge from parent/.style={
    thick,
    draw,
    edge from parent fork right,
  },
  every tree node/.style={
    draw,
    minimum width=1.0in,
    text width=2.02in,
    align=left,
  },
}
\Tree
[
  .{Investigating Financial Misinformation}
  [
    .{\hyperref[exploring_financial_misinformation]{Exploring Financial Misinformation: From Fundamentals to Impacts}}
    [
      .{Types of Financial Misinformation}
      [.{Financial Fraud and Scams} ]
      [.{Trading Platforms and Misinformation} ]
      [.{Misleading Financial Advertising} ]
      [.{Insider Trading and Market Manipulation} ]
      [.{Deceptive Practices in Credit and Lending} ]
      [.{Identity Theft and Financial Information Breaches} ]
    ]
    [.{Factors Fuelling Financial Misinformation } ]
    [.{The Spread of Falsehoods } ]
    [.{Challenges of Financial Misinformation in Media } 
    [.{Traditional Media} ]
    [.{Social Media} ]
    ]
    [.{Impacts of Financial Misinformation } 
    [.{Individual Impacts} ]
    [.{Impacts on Businesses and Industries} ]
    [.{Impact on stock market volatility} ]
    ]
  ]
  [
    .{Schemes Used to Manipulate Financial Market}
    [.{The pump-and-dump scam} ]
    [.{Pyramid schemes} ]
    [.{Front Running} ]
    [.{Short and distort} ]
    [.{Impersonation scams} ]   
  ]
  [
  .{Targeted Sectors of Financial Misinformation}
    [.{Forex Trading: Unmasking Misinformation} ]
    [.{Credit Repair and Deceptive Practices} ]
    [.{Fraudulent Information in Taxation} ]
    [.{Exposing Misinformation of Real Estate} ]
  ] 
  [
    .{Strategies to Combat Online Financial Misinformation}
    [.{Misleading Financial Data Collection} ]
    [.{Computational Methods to Counter Financial Misinformation}
    [.{Deep Learning Methods} ]
    [.{Traditional Machine Learning Methods} ]
    [.{Statistical Methods} ]
    ]
  ] 
  [
    .{Open Issues \& Future Direction} 
    [.{Detection and Filtering Techniques } ]
    [.{Social Media Platform Interventions } ]
    [.{User Education and Awareness } ]
    [.{Long-Term Effects and Socioeconomic Consequences } ]
  ]
]
\end{tikzpicture}
}
\end{figure}

This paper aims to offer a comprehensive understanding of online financial misinformation, encompassing various aspects such as types, impacts, factors, detection, and strategies to combat it. The paper intends to provide an extensive analysis of existing datasets, methods, and challenges related to this subject. Furthermore, the paper examines the various methods used to identify, classify, and mitigate online financial misinformation. Finally, the paper highlights the challenges and future scope, and research directions in this field.

To achieve these goals, the paper is structured as follows. Section \ref{Financial_Dynamics_and_Misinformation} gives a study of financial dynamics and misinformation. Section \ref{exploring_financial_misinformation} provides an overview of the literature on financial misinformation. Section \ref{Schemes_Used_to_Manipulate_Financial_Market} discusses the different schemes used to manipulate the financial market. Section \ref{targeted_sectors} showcases targeted sectors of financial misinformation. Section \ref{Strategies_to_Combat_Online_Financial_Misinformation} explores the strategies to combat financial misinformation which includes different datasets, and computational methods used to identify, classify, and mitigate online financial misinformation. Section \ref{open_issues} highlights the challenges and future research directions in this field. Finally, section \ref{conclusion} concludes the paper with a summary of the main findings and a discussion of the implications of this research.

\section{Financial Dynamics and Misinformation}
\label{Financial_Dynamics_and_Misinformation}
In today's interconnected world, finance plays a crucial role in shaping economies, markets, and investment decisions. However, the proliferation of misinformation has emerged as a significant challenge in the realm of finance. This section explores the intricate relationship between financial dynamics and misinformation, examining how false or misleading information can impact the basics of finance, stock markets, cryptocurrencies, and economies, and ultimately influence investment decisions. By understanding the implications of financial misinformation, investors can navigate the complexities of the modern financial landscape more effectively. In addition, we will present a series of compelling case studies that illuminate the profound impact of financial misinformation, offering tangible examples that underscore its detrimental effects. These case studies will provide real-world scenarios from diverse financial contexts, enabling us to grasp the extent and complexity of the issue at hand. By examining these cases, we can gain invaluable insights into the consequences of financial misinformation and the urgent need for effective strategies to combat its spread.

\begin{figure}
    \centering
    \includegraphics[width=0.45\linewidth]{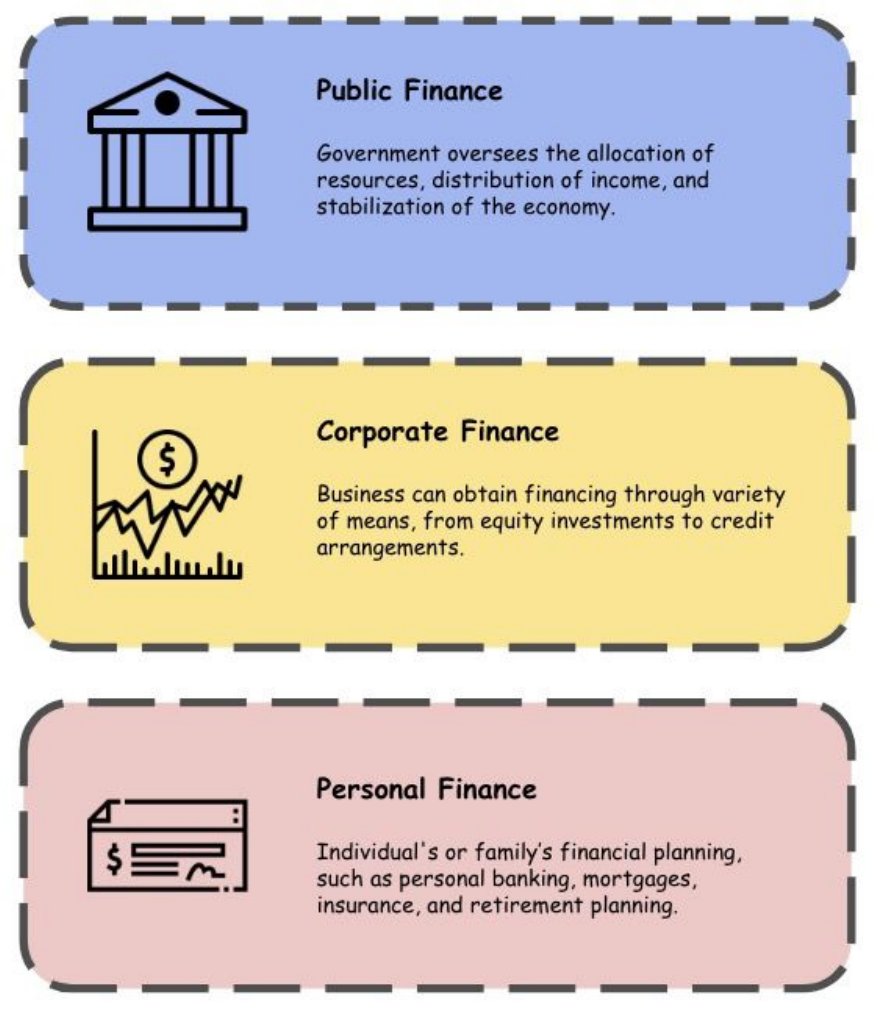}
    \vspace{0.25cm}
    \caption{Finance Categories}
    \label{fig:category}
    \vspace{-0.6cm}
\end{figure}

\textbf{Finance} is a broad field that encompasses the management, allocation, and study of money, investments, and financial systems \cite{shoup2017public}. The concept is so broad that it could cover anything from the Federal Reserve raising interest rates to 5-year-old managing quarters in his Piggy Bank. Finance can be categorized into three main areas: public finance, corporate finance, and personal finance. Public finance involves managing government funds, budgeting, and taxation. Corporate finance focuses on financial decisions within organizations, such as investment analysis and capital structure management. Personal finance involves managing individual financial resources, including budgeting, saving, investing, and retirement planning. Understanding these categories provides insights into how finance operates at the government, corporate, and individual levels \cite{garman2014personal}. Figure \ref{fig:category} illustrates the different categories in finance.

\textbf{Public Finance}: Public finance encompasses the financial operations and policies of governments and public institutions, encompassing budgeting, taxation, public spending, and debt management. Financial misinformation within the domain of public finance can emerge from political biases, misleading narratives concerning public spending priorities, or misconceptions regarding the economic impact of government policies \cite{garman2014personal}. The dissemination of inaccurate information possesses the potential to influence public opinion, shape policy decisions, and yield extensive consequences. To counteract financial misinformation in the realm of public finance, it is crucial to educate the public about fundamental economic concepts and foster critical thinking skills.

The COVID-19 pandemic has engendered an environment conducive to online financial scams. Exploiting individual's fears and uncertainties surrounding the pandemic, scammers have been promoting fraudulent investment opportunities, engaging in phishing activities through emails, and establishing counterfeit charitable organizations \cite{vedova_technology_2023}. These scams frequently exploit people's inclination to assist others or seek quick profits during times of economic instability. Scammers may endorse spurious remedies like colloidal silver or essential oils, or they may endorse unverified drugs or vaccines. Other forms of scams involve phishing emails masquerading as government agencies or health organizations, or counterfeit charities purporting to aid those affected by the pandemic \cite{oecdcovidd199123}. According to the Federal Trade Commission (FTC), Americans suffered losses exceeding \$5.8 billion due to COVID-19-related scams in 2020 \cite{vedova_technology_2023}.

\textbf{Corporate Finance}: Corporate finance encompasses the financial activities of corporations, including capital investment, financial planning, and cash flow management. This field also involves fund procurement, investment analysis, and financial risk evaluation. Financial misinformation within the domain of corporate finance can have severe ramifications, such as stock market manipulation, insider trading, or false financial reporting \cite{10.1093/rof/rfac058}. Ensuring accurate and transparent financial reporting, regulatory oversight, and investor education is crucial for combating financial misinformation in corporate finance.

The Enron Corporation scandal of 2001 exposed a shocking case of corporate fraud and financial misinformation. Enron, formerly a prominent energy company, engaged in deceptive accounting practices, deliberately misleading investors about its true financial state \cite{lii2010case}. The company employed intricate financial structures and manipulated financial statements to conceal its mounting debt and inflate reported profits. Through off-balance-sheet transactions and the utilization of special purpose entities, Enron created a facade of financial strength and profitability \cite{markham2015financial}. However, once the truth behind Enron's deceptive practices came to light, the company experienced a rapid collapse and eventually filed for bankruptcy. The Enron scandal served as a wake-up call, emphasizing the necessity to enhance corporate governance practices and detect financial misinformation \cite{petrick2003enron}. It shed light on the potential consequences of such practices, including the erosion of investor trust and the devastating impact on employees, shareholders, and the broader financial market.

The Theranos scandal of 2015 involved the downfall of a healthcare technology company that claimed to have developed a revolutionary blood-testing technology. Led by its founder and CEO, Elizabeth Holmes, Theranos attracted significant investments by promoting false claims and disseminating misleading information about the effectiveness and capabilities of their technology \cite{rogal2020ssecrets}. Investors were enticed by the promise of a groundbreaking diagnostic tool capable of performing a wide range of tests using only a small amount of blood. However, subsequent investigations conducted by regulatory agencies and journalists revealed that the Theranos technology did not function as advertised \cite{griffin02022promises}. The company's claims were built on a foundation of falsehoods, and the technology's reliability and accuracy were severely compromised. The scandal led to the eventual collapse of Theranos and the initiation of criminal charges against Holmes, underscoring the extensive consequences of financial misinformation in the healthcare industry \cite{fialaa2019theranos}.

\textbf{Personal Finance}: Personal finance encompasses the financial decisions and strategies made by individuals or households to manage their money, savings, and investments. It involves budgeting, tax planning, retirement planning, and other aspects of personal financial management. Financial misinformation in the context of personal finance often arises from a lack of understanding or accessibility to accurate information. Misleading advice, scams, and false promises can lead individuals astray and result in poor financial choices \cite{liu2020financial}. Therefore, it is crucial for individuals to have access to reliable educational resources that enable them to make informed decisions and protect themselves from falling victim to financial misinformation.

Ponzi schemes and financial misinformation are inherently interconnected due to the deceptive practices employed within these schemes \cite{moore20121postmodern}. Ponzi schemes thrive on disseminating false or misleading information about investment opportunities and promised returns. Scheme operators intentionally provide inaccurate or exaggerated information to entice unsuspecting investors into participating \cite{frankel20121ponzi}. One notorious example of a Ponzi scheme is the case of Bernie Madoff, which exemplifies the deceptive nature of such investment schemes. Madoff's scheme operated for decades, defrauding investors of billions of dollars \cite{azim20160bernard}. By making false claims of consistently high returns, Madoff successfully lured individuals, financial institutions, and charitable organizations into investing substantial sums \cite{azim20160bernard,quisenberry2017pponzi}. The severe consequences of this Ponzi scheme shed light on the devastating impact of financial misinformation, emphasizing the critical importance of due diligence in investment decisions. It urges investors to exercise skepticism and diligently verify the legitimacy of investment opportunities \cite{sarna2010hhistory}.

Finance serves as the foundation for understanding the functioning of economic systems and the management of money. It encompasses various concepts such as budgeting, lending, borrowing, investing, and risk management. Accurate and reliable information plays a crucial role in making informed financial decisions, as individuals, businesses, and governments heavily rely on it to assess opportunities and effectively manage their resources. However, when misinformation permeates the financial realm, it can distort perceptions and lead to misguided decisions. Economic growth, which entails increases in gross domestic product (GDP), job creation, and improved living standards, relies on accurate information for effective decision-making. The impact of misinformation on economic perceptions cannot be underestimated \cite{NBERw30994}. Misinformation regarding economic indicators, government policies, or market expectations can mislead investors, businesses, and policymakers, hindering their ability to make sound investment decisions and impeding overall economic progress \cite{ahsan2013effects01}. Financial misinformation can hamper economic growth by distorting market perceptions, diverting resources to less productive avenues, and eroding confidence in financial systems.

\textbf{Stock market} serves as a vital indicator of economic health, reflecting investor sentiment and influencing economic activity. The stock market is a prime area where financial dynamics intersect with misinformation \cite{aggarwal2006stock}. Investors rely on accurate and timely information to assess the value and potential of stocks. Misinformation, whether in the form of false rumors, manipulated financial statements, or exaggerated news reports, can lead to market volatility and irrational trading behavior \cite{bali_bodnaruk_scherbinaaq_tang_2018}. This can cause stock prices to deviate from their intrinsic values, undermining the efficiency of the market and impacting investment decisions. It highlights the importance of verifying information from credible sources before making investment choices. In January 2021, a group of retail investors on Reddit organized a buying frenzy for the stock of video game retailer GameStop \cite{hasso2022participated, university89937337}. The goal was to drive up the price of the stock and make a profit, but the move was also seen as a challenge to Wall Street and the financial establishment. The buying frenzy was fueled by a wave of online hype and misinformation, with many investors believing that the stock was undervalued and poised for a significant increase in price. However, the buying frenzy eventually caused the stock price to skyrocket, reaching a high of \$347.51 on January 27, 2021 \cite{hasso2022participated}. This price increase was driven in part by short sellers who had bet against the stock and were forced to buy shares to cover their positions. The buying frenzy was not sustainable, and the stock eventually crashed back down, resulting in significant losses for many investors \cite{lucangeli2021gamestop}.

The emergence of \textbf{cryptocurrencies} has introduced a new dimension to the challenges posed by financial misinformation. Cryptocurrency markets are highly volatile and susceptible to speculation and market manipulation \cite{vedova_technology_2021,RePEcjbsaltfin201704_gcbs}. Dissemination of false information about cryptocurrency projects, regulatory developments, or market trends can lead to dramatic price fluctuations and misinformed investment decisions \cite{Trozze2022}. The lack of regulation and oversight in the cryptocurrency space amplifies the risks associated with financial misinformation, necessitating a closer examination of its impact on investors and the stability of digital asset markets \cite{royalsociety_2022}. Cryptocurrency scams are increasingly prevalent as the popularity of cryptocurrencies like Bitcoin and Ethereum continues to grow. These scams typically involve fraudulent investment opportunities that promise high returns with little to no risk \cite{vedova_technology_2021}. Scammers often utilize social media platforms like Twitter and Facebook to promote these schemes, and they may employ fake testimonials or endorsements from celebrities to lend credibility to their claims \cite{RePEcjbsaltfin201704_gcbs}. One notable example is the Bitconnect scheme, which promised investors returns of up to 40\% per month. However, the scheme was ultimately exposed as a 'Ponzi' scheme, with new investor's money being used to pay off earlier investors. The scheme eventually collapsed in 2018, leading to the U.S. Securities and Exchange Commission (SEC) filing charges against its organizers \cite{Trozze2022}.

\textbf{Investment decisions} rely heavily on accurate and reliable information \cite{MohdPadil2022}. However, the proliferation of financial misinformation can lead to misguided investment choices and substantial financial losses. Investors who fall victim to misleading information may make ill-informed decisions, such as investing in fraudulent schemes or failing to recognize market risks \cite{finraorg_2021}. The consequences of financial misinformation extend beyond individual investors to broader market stability, investor confidence, and the overall health of the economy \cite{vedova_technology_2021}.

Given the interconnectedness between these financial concepts and the pervasive impact of financial misinformation, this paper aims to delve into the prevalence, sources, dissemination methods, and consequences of financial misinformation. By conducting a comprehensive survey, this study seeks to shed light on the challenges posed by financial misinformation and provide valuable insights for developing effective strategies to combat its detrimental effects on individuals, financial markets, and the overall economy. Understanding the dynamics of financial misinformation is crucial for fostering informed decision-making, promoting market integrity, and safeguarding the well-being of investors and the financial system as a whole.

\section{Exploring Financial Misinformation: From Fundamentals to Market Impacts}
\label{exploring_financial_misinformation}
\subsection{Types of Online Financial Misinformation}
\label{Types_of_Online_Financial_Misinformation}
The internet has made it easier for people to access and share information, but it has also created new opportunities for financial misinformation and deception \cite{blanton2012rise,williams6383930}. Online financial misinformation can take various forms, including fake news and rumors, misleading advertisements, fraudulent investment schemes, phishing and identity theft, and impersonation scams \cite{article1}.

Fake news and rumors are often spread on social media and other online platforms \cite{williams6383930}. They can be about anything from stock market trends to government policies. These stories are usually sensational and designed to attract clicks and shares. Unfortunately, they can also be completely untrue or based on partial information \cite{carpenter_2023}. Fake news and rumors can be damaging to investors who rely on accurate information to make informed decisions. Misleading ads often promise high returns with little or no risk, which is a red flag for potential fraud. These ads can be targeted at vulnerable people, such as those with little financial knowledge or those who are struggling financially \cite{harvard_business_review_2019001283}. Fraudulent investment schemes are one of the most common types of online financial misinformation \cite{williams6383930}. These schemes promise high returns on investments but are usually nothing more than a 'Ponzi' scheme, where early investors are paid off with money from later investors. These scams can take various forms, such as fake cryptocurrency investments or phony stock tips \cite{NBERw30994}. Let's undertake a brief yet comprehensive examination of the various categories of financial misinformation.
\subsubsection{Financial Fraud and Scams}
\label{Financial_Fraud_and_Scams}
Financial fraud and scams represent a persistent and ever-evolving threat that poses significant risks to individuals, businesses, and the overall integrity of the financial system \cite{reurink2019financial}. These fraudulent activities exploit the trust and vulnerability of victims, leading to devastating financial consequences \cite{burnes2017prevalence,gamble2014ccauses}. Financial fraud can have far-reaching effects, not only on individual victims but also on financial institutions and the broader economy. The loss of financial assets and resources can lead to severe financial hardships, bankruptcy, and even the collapse of businesses \cite{rathinarajjj2010financial}. Raising awareness and promoting financial literacy among individuals is essential in mitigating the risks associated with financial fraud \cite{reurink2019financial}. This can include providing guidance on verifying the legitimacy of investment opportunities, being cautious with sharing personal information online, and regularly monitoring financial accounts for any suspicious activities \cite{gamble2014ccauses,rathinarajjj2010financial}. Fraudsters often leverage technology to carry out sophisticated scams, but advancements can be employed by financial institutions and regulators to detect patterns, identify potential fraud cases, and respond swiftly \cite{perols2011ffinancial}.

\subsubsection{Online Trading Platforms and Misinformation}
\label{Online_Trading_Platforms_and_Misinformation}
Online trading platforms have revolutionized the financial landscape, providing individuals with easy access to a wide range of investment opportunities \cite{tay2022ccomparison}. However, the rise of these platforms has also introduced challenges associated with misinformation. Misinformation on online trading platforms can manifest in various ways, including the spread of false rumors, market manipulation schemes, and misleading investment advice \cite{tay2022ccomparison}. These forms of misinformation can have significant repercussions on financial markets and individual investors. Social media platforms play a pivotal role in the dissemination of misinformation related to online trading \cite{braun2019fake}. The viral nature of social media amplifies the reach and impact of misleading information, potentially leading to panic, irrational trading decisions, and market volatility \cite{demartini2020hhuman}.

Addressing misinformation on online trading platforms requires a multi-faceted approach. Stronger regulatory frameworks and increased enforcement can help curb deceptive practices and promote market integrity \cite{bizzi2019doublee}. Online trading platforms should establish guidelines and ethical standards for information sharing, including measures to address conflicts of interest \cite{tay2022ccomparison}. Striking the right balance between freedom of expression and the need for accurate information is crucial to maintain trust and credibility.

\subsubsection{Misleading Financial Advertising and Promotions}
\label{Misleading_Financial_Advertising_and_Promotions}
Misleading financial advertising and promotions have emerged as a significant challenge within the financial industry, necessitating further examination of its implications and the implementation of effective regulatory approaches \cite{rubin2010financiall}. These misleading practices can take various forms, such as exaggerated claims, incomplete information, or deceptive tactics to entice consumers into purchasing financial products or services \cite{aven2016ambushh}. The consequences of misleading financial advertising extend beyond individual investors, as they erode trust in the financial industry as a whole. Regulators have responded to these concerns by establishing guidelines and enforcing advertising standards to protect consumers \cite{adams2017advertt}. However, the evolving landscape of advertising, including the widespread use of social media and online platforms, poses ongoing challenges in effectively monitoring and regulating financial promotions \cite{lawrence2021mmisleading}. Collaborative efforts among regulators, financial institutions, and advertising platforms are crucial to enhance transparency, strengthen enforcement mechanisms, and promote financial literacy. By fostering a regulatory environment that emphasizes consumer protection and responsible advertising practices, the risks associated with misleading financial advertising can be mitigated.

\subsubsection{Insider Trading and Market Manipulation}
\label{Insider_Trading_and_Market_Manipulation}
Insider trading and market manipulation are two interrelated issues that pose significant threats to the fairness, transparency, and integrity of financial markets worldwide \cite{chitimira2016unpacking}. Insider trading involves the exploitation of non-public, material information by individuals with privileged access, such as corporate insiders or professionals, to trade securities for personal gain \cite{john1997kmarket,chitimira2016unpacking}. By trading based on confidential information, insiders can generate substantial profits, but at the expense of ordinary investors who lack access to such privileged insights \cite{austin2017insiider}.

Market manipulation, on the other hand, involves deliberate actions aimed at distorting market conditions or prices to create artificial outcomes that benefit the manipulators. Various tactics are employed in market manipulation, including spreading false information, engaging in pump-and-dump schemes, engaging in wash trades, or colluding with other market participants \cite{bromberg2017fffinancial}. These activities can deceive investors, drive artificial price movements, and create a false perception of market demand or supply. Market manipulation not only compromises market efficiency and fairness but also erodes investor confidence and trust in the financial system \cite{engelen2007ethicss}. Securities commissions and financial regulatory authorities are responsible for combating insider trading and market manipulation \cite{putnicnvs20122mmarket}. Insider trading regulations typically require insiders to disclose their trades and prohibit them from trading based on non-public information. Market manipulation is addressed through regulations that prohibit fraudulent practices, enhance transparency, and foster fair competition.

\subsubsection{Deceptive Practices in Credit and Lending}
\label{Deceptive_Practices_in_Credit_and_Lending}
Deceptive practices in credit and lending have become a growing concern in the financial industry, as they can exploit consumers and undermine their financial well-being \cite{aberggf2016case}. Common deceptive practices include misrepresenting interest rates, fees, or repayment terms, concealing crucial information, and engaging in predatory lending practices that target vulnerable individuals. Such deceptive practices not only harm borrowers by trapping them in cycles of debt and financial distress but also erode trust in the financial system as a whole \cite{barrettt1970truth}. Regulatory bodies play a vital role in combating deceptive practices by establishing and enforcing consumer protection laws, enhancing disclosure requirements, and imposing penalties on institutions that engage in deceptive behavior \cite{staddller2011predatory}. Ensuring transparency, fairness, and ethical lending practices is crucial to promote consumer trust, fostering financial inclusion, and maintaining a healthy and sustainable credit and lending ecosystem.

\subsubsection{Identity Theft and Financial Information Breaches}
\label{Identity_Theft_and_Financial_Information_Breaches}
Identity theft and financial information breaches pose significant risks to individuals, businesses, and financial institutions in today's digital age. Identity theft involves the unauthorized acquisition and use of personal information, such as social security numbers, credit card details, or bank account information, for fraudulent purposes \cite{perretti2008data}. Financial information breaches occur when sensitive financial data, either stored electronically or transmitted online, is compromised by hackers or cybercriminals. These breaches can result in substantial financial losses, reputational damage, and emotional distress for the victims \cite{anderson2008identity}. Personal and financial information obtained through identity theft and breaches can be used for various illicit activities, including fraudulent financial transactions, opening unauthorized accounts, or even selling the information on the dark web \cite{gargg2003quantifying}.

Financial institutions also bear the burden of ensuring the security and confidentiality of customer data \cite{anderson2008identity}. The breach of customer information can erode trust, damage the institution's reputation, and lead to legal and regulatory consequences \cite{finkllea2009identity}. To mitigate the risks associated with identity theft and financial information breaches, both individuals and organizations must prioritize cybersecurity and implement robust security measures \cite{romanoskyy2010data}. Individuals should remain vigilant and exercise caution when sharing personal information online, being wary of phishing attempts and suspicious emails or websites.

\subsection{Factors Fuelling Online Financial Misinformation}
Online financial misinformation is a growing concern in today's digital age, and its spread can be attributed to various factors. The widespread use of social media and the internet has created an environment where online financial misinformation can easily spread \cite{publisher_nasdaq}. Social media platforms are designed to share information quickly and easily, and this can facilitate the spread of misinformation. A study by the Pew Research Center found that around two-thirds of American adults get their news from social media, and this includes financial news and information \cite{atske_2021}. The internet serves as a platform for the dissemination of false information and fraudulent schemes, enabling malicious actors to perpetrate scams. Particularly, social media emerges as the predominant channel utilized by scammers to target potential victims \cite{atske_202112344589}. Individuals are susceptible to psychological biases and cognitive limitations that can make them vulnerable to online financial misinformation \cite{kahneman2011thinking}. Confirmation bias can cause individuals to seek out and believe information that confirms their existing beliefs, even if it is false. Other biases, such as the availability heuristic and the anchoring effect, can also influence an individual's perception of financial information. Cognitive limitations, such as limited attention and information overload, can make it difficult for individuals to discern the veracity of financial information \cite{simon1971designing}.

Economic and political factors also contribute to the spread of online financial misinformation \cite{vedova_technology_2023}. Economic instability and uncertainty can create an environment where individuals are more susceptible to financial scams and false information. Political events, such as elections and policy changes, create opportunities for misinformation to be disseminated \cite{clarke2018fake}. The desire for quick financial gains can cloud individual's judgment and critical thinking abilities, making them more likely to accept and share information without thoroughly verifying its accuracy. Misinformation spreaders take advantage of this mindset by crafting persuasive narratives that align with people's aspirations, offering false promises of easy wealth or secret investment strategies. During the COVID-19 pandemic, scammers took advantage of economic uncertainty and government aid programs to target individuals with false investment opportunities and phishing scams \cite{vedova_technology_2023}. A lack of regulatory oversight and enforcement contributes to the spread of online financial misinformation. The internet and social media have created a global platform for financial transactions and information sharing, and this has created challenges for regulators \cite{doi:10.1080/23311983.2022.2037229}. In some cases, regulatory frameworks have not kept pace with the rapidly evolving digital landscape. This has created gaps in oversight and enforcement, which can be exploited by malicious actors \cite{royalsociety_2022}.

\subsection{The Spread of Falsehoods and Misinformation on the Internet}
\label{The_Spread_of_Falsehoods_and_Misinformation_on_the_Internet}
The phenomenon of false information and deception on the internet has gained significant attention from researchers across various disciplines. Early investigations into online misinformation explored the effects of fabricated news stories on reader's attitudes and beliefs \cite{metzger2007making}. This study found that individuals were more likely to believe false information when it aligned with their preexisting attitudes, highlighting the influential role of confirmation bias in the dissemination of misinformation. During the 2016 US Presidential election, the role of social media platforms in the spread of fake news came under scrutiny \cite{allcott2017social}. The researchers discovered that false stories were more widely shared on Facebook compared to true stories, reaching a larger audience than legitimate news articles.

In recent years, researchers have focused on developing automated tools to detect and combat online deception. \cite{shu2017fake} proposed a deep learning approach for fake news detection, achieving high accuracy in identifying fabricated stories. \cite{wang2020grover} developed a model capable of generating realistic fake news articles to assist researchers and policymakers in understanding the potential impact of false information. Psychological factors influencing the spread of fake news have also been investigated. Studies have explored the influence of emotions and social identity \cite{lewandowsky2012misinformation,del2016spreading}. Additionally, interventions designed to counteract misinformation, such as fact-checking and debunking strategies, have been examined \cite{nyhan2010corrections,van2015debunking}.
\subsection{Challenges of Financial Misinformation in Traditional Media and Social Media}
\label{Challenges_of_Financial_Misinformation_in_Traditional_Media_and_Social_Media}
\subsubsection{Traditional media}
Traditional media, such as newspapers, television, and radio, typically follows a controlled and curated process, subjecting information to editorial review and fact-checking before dissemination \cite{marta2019information09}. Journalists and news organizations are expected to adhere to ethical guidelines and maintain credibility \cite{elvestad2018can112}. In contrast, social media platforms lack similar gatekeeping measures, allowing anyone to post or share information without rigorous fact-checking or editorial review \cite{gainous20199traditional}. The speed and reach of information dissemination differ greatly between traditional media and social media \cite{marta2019information09}. In traditional media, content is typically produced by professional journalists and subject matter experts who undergo editorial scrutiny before publication \cite{williams20191preferences}.

\subsubsection{Social Media}
Social media platforms have revolutionized the speed of information sharing, enabling the rapid spread of false or misleading financial information to a wide audience within minutes \cite{stromback2020news1}. In contrast, traditional media follows a more controlled and time-consuming process before the information reaches the public \cite{marta2019information09}. User-generated content is a prominent feature of social media platforms. Individuals can freely express their opinions and share information. While this promotes freedom of speech, it also increases the likelihood of financial misinformation being disseminated by well-intentioned individuals or deliberate actors. Another notable difference is the algorithmic amplification on social media platforms \cite{turcotte20151news}. Algorithms personalize content based on user preferences and engagement, prioritizing content that generates strong reactions or shares. This can lead to the amplification of sensational or false financial information, contributing to the rapid spread of misinformation. Traditional media outlets generally do not employ similar algorithms for content distribution.

In terms of source credibility, traditional media outlets often possess established reputations and a history of producing reliable financial news and analysis \cite{elvestad2018can112}. Readers and viewers may trust traditional media sources based on their track record and adherence to journalistic standards. Social media, however, lacks a standardized credibility assessment process, making it more challenging for users to determine the reliability and accuracy of information shared on these platforms \cite{marta2019information09}. These differences between financial misinformation in traditional media and social media have implications for the prevalence, speed, and impact of misinformation within each medium. Recognizing these disparities is essential for devising effective strategies to combat financial misinformation and promote financial literacy among individuals.

\subsection{Impacts of Online Financial Misinformation}
\label{Impacts_of_Online_Financial_Misinformation}
The widespread dissemination of online financial misinformation can exert substantial detrimental effects on individuals, businesses, and the overall economy. One of the most immediate impacts of online financial misinformation is the financial losses that can result for individuals and businesses \cite{finraorg_2021}. Fraudulent investment schemes, phishing scams, and other forms of financial deception can leave victims with significant losses \cite{hebert_hernandez_perkins_puig_tressler_ftc_2023,fscs_2022}. For individuals, this can mean lost savings or investments, while businesses may suffer losses that impact their bottom line and ability to operate. When people are exposed to false or misleading information, they may begin to question the credibility of the sources that provide financial information \cite{doi:10.1509/jmkr.45.6.633}. This can make it more difficult for legitimate financial institutions to gain the trust of their customers, and for markets to function efficiently.

\subsubsection{Individual Impacts}
The proliferation of online financial misinformation has significant ramifications for individuals, as highlighted by various studies. One of the key risks faced by individuals relates to falling victim to financial scams and frauds that are disseminated through false information \cite{fscs_2022}. Misleading investment advice or fraudulent schemes can lead individuals to make ill-informed financial decisions, resulting in substantial financial losses and potential depletion of personal savings or retirement funds \cite{doi:10.1509/jmkr.45.6.633}.

In addition to financial implications, the emotional and psychological toll on individuals should not be underestimated. Continuous exposure to misinformation can generate feelings of stress, anxiety, and confusion as individuals grapple with the challenge of differentiating between reliable sources and deceptive information \cite{beatty2016uneven}. Consequently, this can undermine individuals' confidence in making informed financial decisions and erode their trust in financial institutions \cite{williams6383930}.

To alleviate the adverse impacts on individuals, it is crucial for them to develop robust media literacy skills, enhance critical thinking abilities, and cultivate the capacity to independently verify information \cite{bruhn2015personal}. By acquiring these skills, individuals can become more discerning consumers of financial information, thereby reducing their vulnerability to misinformation and fraudulent activities. Furthermore, the collective efforts of financial institutions, educational institutions, and government entities are pivotal in promoting financial literacy and providing resources that assist individuals in safely navigating the digital landscape.

\subsubsection{Impacts on Businesses and Industries}
Online financial misinformation poses significant challenges and risks to businesses and industries. False information circulating online can have detrimental effects on companies, leading to reputational damage and potential financial losses \cite{clarke2018fake,finraorg_2021}. Companies may face a barrage of false rumors or misleading information that can quickly spread across social media platforms, affecting investor sentiment and stock prices \cite{Polak2012THEMEA}. Moreover, the dissemination of inaccurate information can disrupt business operations and decision-making processes. Executives and managers may find themselves navigating through a sea of misinformation, making it challenging to develop sound strategies and allocate resources effectively \cite{ahsan2013effects01}. To mitigate these impacts, companies need to closely monitor online narratives, engage in proactive communication, and establish robust mechanisms to combat misinformation and protect their brand reputation. Additionally, collaboration between industry stakeholders and regulators is crucial to develop guidelines, standards, and best practices to address the issue effectively and maintain trust in the marketplace.

Online financial misinformation also has negative effects on economic growth and stability \cite{clarke2018fake}. When people make decisions based on inaccurate or misleading information, this can lead to market inefficiencies and misallocations of resources \cite{williams6383930}. This can, in turn, impact economic growth and stability. If investors are misled into making bad investment decisions, this can impact the performance of the companies they invest in, which can ultimately harm the broader economy. Online financial misinformation damages the reputation of individuals and businesses \cite{publisher_nasdaq}. If individuals are targeted by phishing scams or other forms of financial fraud, this can harm their reputation and make it more difficult for them to access financial services in the future.

\subsubsection{Impact on stock market volatility}
Misinformation thrives within the realm of stock market data, primarily due to the inherent uncertainty associated with making predictions based on current events. This uncertainty factor complicates the task of discerning between accurate and incorrect information, amplifying the vulnerability of investors and the financial system at large \cite{Polak2012THEMEA}. The dynamic nature of the stock market, influenced by a multitude of factors such as economic indicators, geopolitical events, and company-specific developments, makes it a fertile ground for misinformation to propagate \cite{Jureviien2013BehaviouralFTI}.

The challenge lies in accurately interpreting the complex interplay of various variables that shape the market's trajectory. Analysts and experts often rely on historical data, technical analysis, and fundamental indicators to form their predictions, but even with these tools at their disposal, the outcome remains uncertain \cite{Polak2012THEMEA,Jureviien2013BehaviouralFTI}. The inherent unpredictability of the market, coupled with the rapid dissemination of information in the digital age, creates an environment where misinformation can easily take root and spread \cite{ahsan2013effects01}. Misinformation in the stock market can manifest in different forms. It may involve spreading false rumors about companies, exaggerating the impact of certain events, or manipulating data to fit a particular narrative \cite{how_news_mar110982}. The motivations behind disseminating such misinformation can vary, ranging from individual attempts to manipulate stock prices for personal gain to orchestrated efforts aimed at creating panic or instability within the market \cite{Kar2022HowDMII}.

The connection between trading volume, stock price, and media coverage was explored in relation to EntreMed, a company focused on developing therapeutic treatments for cancer and other diseases. On a specific day, when The Times published an article about a new cancer drug from EntreMed, the company's stock price saw a significant surge of 28.4 percent. These findings indicate that media coverage has the potential to impact both the trading volume and pricing of security by drawing attention to it \cite{HUBERMAN_REGEV2004}. In Asian markets, an intriguing trend emerges when it comes to shareholder's reactions to news originating from the United States. Notably, negative news tends to elicit a more significant response compared to positive news \cite{Doong2005ResponseAI}. Financial media and political events play a crucial role in addressing the issue of information asymmetry within the marketplace \cite{Tetlock_2008,Kim_Mei_1999}. The relationship between news announcements from Dow Jones and their impact on stock returns was analyzed, revealing a direct correlation with trading volume and market returns \cite{Mitchell_Mulherin_impact}. Positive cover page stories in media publications are correlated with positive stock performance. Similarly, their findings also supported the argument that negative news is associated with negative stock performance \cite{Arnold_Earl_North_2007}. In other words, there is a direct correlation between the sentiment of news coverage (positive or negative) and the subsequent performance of stocks.

Investors, especially those lacking deep financial expertise, face the daunting task of navigating through a sea of information, often conflicting and unreliable, to make informed investment decisions \cite{how_news_mar110982}. The challenge becomes even more pronounced when individuals rely on social media platforms or unverified sources for market insights. In these cases, misinformation can quickly spread, leading to irrational investment choices and potentially significant financial losses \cite{10.1093/rof/rfac058}.

To mitigate the impact of misinformation, investors must exercise caution and employ critical thinking skills when consuming and evaluating market information \cite{abbasi_albrechtt_vance_hansenn_2012}. Verifying the credibility of sources, seeking out multiple perspectives, and conducting thorough research are crucial steps in distinguishing between reliable information and misleading claims. Regulators and financial institutions also play a vital role in combating misinformation by enforcing stricter guidelines for information dissemination, promoting transparency, and implementing mechanisms to detect and prevent market manipulation \cite{kirilenko2019regulating}.

\section{Schemes Used to Manipulate Financial Market}
\label{Schemes_Used_to_Manipulate_Financial_Market}
\subsection{The pump-and-dump scam}
The pump and dump scam is a fraudulent scheme that targets stock markets with the intention of artificially inflating the price of a particular stock \cite{dunham2007pump}. This deceptive tactic involves spreading false or misleading information about a company or its stock, enticing unsuspecting investors to buy the stock at inflated prices. The orchestrators of the scam then sell their shares once the price reaches its peak, causing the stock price to plummet and resulting in substantial losses for investors who bought during the pump phase \cite{renault2014pump}. The pump and dump scheme typically involves several stages:

\noindent \textbf{Accumulation Phase}: The orchestrators quietly accumulate a large number of shares of a low-priced or thinly traded stock. \\
\noindent \textbf{Promotion Phase}: False or exaggerated information is disseminated through various channels to create a buzz and generate interest in the targeted stock. \\
\noindent \textbf{Buying Frenzy}: Unsuspecting investors, attracted by promises of quick profits, start buying the stock, driving up the price. \\
\noindent \textbf{Dumping Phase}: The orchestrators sell their shares in large volumes, causing the stock price to rapidly decline. \\
The consequences of falling victim to a pump-and-dump scam can be severe, resulting in significant financial losses for investors. The anonymous nature of the orchestrators often makes it challenging for investors to seek recourse or hold them accountable. It is important for investors to exercise caution, conduct thorough research, and consult reputable sources before making investment decisions to avoid becoming victims of such fraudulent schemes.

\subsection{Pyramid schemes}
Pyramid schemes are deceptive investment models that rely on recruitment and promise unrealistic returns, luring individuals with the allure of quick wealth and financial stability \cite{nat2002marketingg}. These schemes operate by continuously enrolling new participants who must pay an upfront fee to join. The commissions earned by recruiters come primarily from the investments of new members rather than legitimate product sales or investments \cite{keep02014multilevel,gastwirth1984two}. Pyramid schemes exploit individual's trust, often using financial misinformation to deceive and manipulate them. The dissemination of false or misleading information plays a crucial role in perpetuating these schemes by creating an illusion of legitimacy and enticing more people to participate \cite{jarvis200000rise,hidajat2020people}.

Furthermore, pyramid schemes often employ various tactics to propagate financial misinformation \cite{gastwirth1977probability,hidajat2020people}. They may use persuasive marketing strategies that exaggerate potential earnings, downplay risks, and create a sense of urgency to entice individuals into joining. These schemes often rely on testimonials and success stories that are either fabricated or based on a small fraction of participants who managed to profit at the expense of the majority \cite{bosley2019decision,nat2002marketingg}. The widespread use of social media platforms and online communities has further amplified the reach and impact of financial misinformation, making it easier for pyramid schemes to target and recruit unsuspecting victims \cite{muncy2004ethicall}.

\subsection{Front Running}
Front running is an unethical practice wherein a broker or trader exploits advanced knowledge of pending orders from their clients for personal gain \cite{bernhardt2008front,scopino2014questionable}. Instead of executing the client's orders immediately, the front runner uses this insider information to execute trades for their own benefit before executing the client's orders \cite{bernhardt2008front}. This allows the front runner to profit from the expected price movement resulting from the execution of the pending order, often to the detriment of the client \cite{sannikov2016dynamic}.

The mechanics of front running involve acquiring inside information about pending orders that are not publicly available. The front runner then executes trades on their own behalf, taking advantage of the anticipated price movement caused by the pending client orders. After profiting from these personal trades, the front runner executes the clients' orders, potentially at less favorable prices due to their prior market influence \cite{baum2021sok}.

Front running is widely regarded as unethical because it prioritizes the personal interests of the front runner over the best interests of their clients \cite{sannikov2016dynamic}. This manipulative practice undermines trust and fairness in the financial markets, as the front-runner profits at the expense of clients who seek optimal execution of their trades.

\subsection{Short and distort}
Short and distort is a deceptive trading strategy that involves two key elements: short-selling and the dissemination of false or negative information about a particular stock \cite{mitts2020short,weiner2017growing}. The ultimate goal of this strategy is to profit from the decline in the stock's price by creating a negative perception among other market participants \cite{weiner2017growing}.

The first step in the short and distort strategy is for the trader to borrow shares of the targeted stock from a broker. These borrowed shares are then sold in the market, creating a short position. By selling borrowed shares, the trader is effectively betting that the stock's price will decrease. Simultaneously, the trader engages in the dissemination of false or negative information about the targeted stock. This can involve spreading rumors, publishing negative news articles, or making exaggerated claims that may damage the company's reputation or undermine investor confidence \cite{elll2016time}. The purpose of spreading this misinformation is to create panic and induce selling pressure among other investors.

As the negative sentiment spreads and more investors begin to sell the stock, its price starts to decline. At this point, the trader executes the second part of the strategy. They buy back the shares they initially borrowed at the lower price to cover their short position \cite{weiner2019sec}. By returning the borrowed shares to the broker, the trader realizes a profit from the difference between the higher initial selling price and the lower buying price. The consequences of the short and distorted strategy can be significant for both individual investors and the overall market. For individual investors, relying on false or negative information can lead to substantial financial losses if they make decisions based on the manipulated perception of the stock. Moreover, the strategy undermines the integrity of the financial system by manipulating stock prices based on fabricated or exaggerated information \cite{li2023doeeoes}. This can erode investor confidence and trust in the market, negatively impacting the overall efficiency and fairness of the financial ecosystem.

\subsection{Impersonation scams}
Impersonation scams are another form of online financial misinformation \cite{MohdPadil2022}. They involve someone pretending to be someone else, such as a representative from a legitimate financial institution. These scammers often use social engineering techniques to gain the trust of their victims, such as pretending to be a friend or family member. Once they have gained the victim's trust, they will ask for money or sensitive information. Online financial misinformation can take various forms, and it is essential to be aware of the potential risks. Investors and consumers must be vigilant when receiving financial information online and should take steps to verify the accuracy of any information before making decisions \cite{ebem20177iinternet}. Financial institutions and regulators also have a role to play in combatting online financial misinformation by increasing awareness, providing education, and enforcing regulations to protect consumers \cite{reznik2012identity}.

\section{Targeted Sectors of Financial Misinformation}
\label{targeted_sectors}
\subsection{Forex Trading: Unmasking Misinformation}
\label{Forex_Trading}
Forex trading, also referred to as foreign exchange trading, is a decentralized global market where participants engage in the buying, selling, and exchanging of currencies \cite{evans2018fforex}. With daily trading volumes reaching trillions of dollars, it stands as one of the largest financial markets worldwide. However, this market is not immune to the challenges posed by financial misinformation \cite{carapucco2018reinforcement}. While legitimate Forex trading involves diligent market analysis, effective strategies, and risk management, the prevalence of misinformation within this domain presents significant obstacles \cite{mendes2012fforex,bennn2018uninformative}.

Misinformation in Forex trading can stem from various sources, including unverified trading signals, fraudulent trading platforms, and individuals or entities promoting quick and unrealistic wealth accumulation schemes without adequately disclosing the associated risks \cite{mendes2012fforex,evans2018fforex}. The influence of financial misinformation in Forex trading can have profound consequences for traders and the overall market. Traders who fall victim to misleading information may make ill-informed decisions, leading to financial losses and detrimental effects on their financial well-being \cite{dymova2016fforex,lopezz2018betting}. Additionally, financial misinformation can undermine market integrity by eroding trust in the Forex trading industry through false claims and deceptive practices. Promoting financial literacy and education among traders is paramount, enabling them to distinguish accurate information from misinformation and make informed decisions \cite{neelyy2013lessons}.

\subsection{Credit Repair and Deceptive Practices}
\label{Credit_Repair}
Credit repair is a service designed to enhance an individual's creditworthiness by addressing negative items on their credit reports \cite{leonarddf2019credit}. While legitimate credit repair services can be beneficial for individuals aiming to improve their credit, the industry is plagued by financial misinformation \cite{american2009americannn}. Misinformation in credit repair can take various forms, including false assurances of immediate credit score improvement, deceptive methods to remove accurate negative information from credit reports, and misleading advertisements that exploit individual's financial vulnerabilities \cite{dippenaaar2021wary,headdyy2000finance}.

This financial misinformation can have significant consequences for individuals and the credit repair industry as a whole. Those who fall prey to misleading information may make uninformed decisions, leading to financial losses and adverse effects on their credit standing \cite{american2009americannn}. To address this issue, it is crucial to promote transparency, education, and responsible practices within the credit repair industry. Providing accurate information, debunking common myths, and establishing regulatory measures can help individuals make informed choices and protect themselves from the impact of financial misinformation in credit repair \cite{nehfl1991legislative}.

\subsection{Fraudulent Information in Taxation}
\label{Taxation}
Taxation plays a crucial role in modern society by providing governments with the necessary resources to support public goods and services \cite{salanie2011economics,burgess1993taxation,mcbarnet1991whiter}. However, discussions about taxation often become mired in financial misinformation, leading to widespread misconceptions and confusion. One common form of misinformation involves misrepresenting tax rates, where individuals mistakenly believe that a single tax rate applies to their entire income, disregarding the progressive nature of most tax systems \cite{torglerr2008we,ramsey1927contribution}. Additionally, misconceptions about tax deductions, credits, and loopholes can create a distorted perception of tax evasion or preferential treatment, when individuals or corporations are actually utilizing legitimate provisions within the tax code.

Addressing this issue requires promoting financial literacy, transparency, and accurate information dissemination to foster a better understanding of taxation's purpose, benefits, and fairness \cite{burgess1993taxation,knightth1992criminal}. Governments should prioritize education initiatives to provide clear explanations of tax policies, rates, and the role of taxation in funding public services \cite{azrina2014integrative,widiyati2021role}. Media outlets, fact-checking organizations, and financial experts also play a crucial role in combating misinformation by providing accurate information and debunking misleading claims \cite{lennoxxxon2013tax}. By fostering an environment of informed discussions and evidence-based decision-making, we can ensure that taxation is perceived and implemented in a manner that aligns with the principles of equity and societal progress, benefiting both individuals and society as a whole \cite{dewwede2020tax}.

\subsection{Exposing Misinformation of Real Estate}
\label{Real_Estate_Fraud}
The real estate sector is a significant and intricate industry that is prone to financial misinformation. Misconceptions and inaccuracies surrounding real estate can lead to distorted perceptions and misguided decisions. Financial misinformation in real estate can manifest in various ways, including inflated property valuations, misleading claims about investment returns, and unrealistic expectations regarding housing market trends.

One prevalent form of financial misinformation in real estate is the exaggeration of property values \cite{bucknall2008real}. Inflated valuations can create false impressions of wealth and prompt individuals to make ill-informed buying or selling choices. Additionally, misleading assertions about investment returns in real estate can mislead potential investors, portraying it as a surefire path to rapid and substantial profits \cite{eltweri2021applications,cahyani2021fraudulent}. In reality, real estate investments entail inherent risks and necessitate meticulous analysis and research \cite{rahmatika2019detection}. Another area susceptible to financial misinformation is housing market trends. Forecasts or claims about future market conditions can often be speculative or influenced by personal biases. Misinformation regarding market trends can misguide buyers or sellers, potentially resulting in overpriced purchases or undervalued sales \cite{rahmatika2019detection,eltweri2021applications}.

Addressing financial misinformation in real estate requires a combination of education, research, and critical thinking \cite{wicaksari2023diamond}. Individuals should seek reliable and unbiased information from trusted sources, such as real estate professionals, market analysts, and government agencies \cite{cahyani2021fraudulent}. Furthermore, promoting financial literacy and fostering an understanding of the factors that drive real estate values can empower individuals to make informed decisions and avoid falling victim to misinformation.

\section{AI in misinformation}
The rise of AI has introduced new challenges, particularly in combating misinformation in various domains. In the financial sector, AI's potential to enhance decision-making processes and streamline operations is accompanied by the risk of its manipulation to generate and disseminate false information, leading to significant consequences \cite{zhou2023synthetic}.

AI-generated misinformation in finance raises concerns about market manipulation. Through advanced algorithms, AI can process vast amounts of data and uncover patterns that may elude human observers. This capability can be exploited by malicious actors who employ AI to spread deceptive information about financial assets such as stocks or cryptocurrencies. By creating convincing narratives and leveraging the speed of social media and online platforms, AI-generated misinformation can cause substantial fluctuations in market prices, resulting in financial losses for unsuspecting investors \cite{najee2021towards}.

Addressing AI-generated misinformation in finance necessitates a comprehensive approach. Firstly, there is a need to increase awareness and provide education to investors, financial professionals, and the general public regarding the potential risks associated with AI-generated misinformation \cite{jahanbakhsh2023exploring}. Promoting digital literacy and critical thinking skills can empower individuals to evaluate information effectively and make informed decisions.

Moreover, AI algorithms can manipulate social media platforms and online forums to propagate false rumors or artificially inflate the popularity of specific financial assets. This can create an illusory market demand and attract unwary investors to invest in overhyped or fraudulent opportunities \cite{rubin2022misinformationnk}. Additionally, AI can generate manipulated media assets, such as deepfake videos or audio recordings, featuring prominent financial experts or industry insiders making false statements or predictions \cite{lu2022eheffects}. These deceptive media assets can significantly impact market sentiment and investor behavior, potentially resulting in financial repercussions.

An example of AI-generated misinformation in finance is the production of fake news articles or market analysis reports by language models like Bloomberg-GPT \cite{wu2023bloomberggpt}. The veracity of the content provided by such models is not substantiated. For instance, an AI-generated article might falsely assert that a specific company is on the brink of bankruptcy, inducing panic among investors and causing a sharp decline in its stock price. It is crucial to note that while AI-generated misinformation in finance is a concern, it is not confined to a specific AI model like Bloomberg-GPT. Various AI models, including GPT-3 and others, can be employed to create and disseminate misinformation in the financial domain. Therefore, addressing the broader issue of AI-generated misinformation is imperative, rather than singling out any particular model.
\section{Strategies to Combat Online Financial Misinformation}
\label{Strategies_to_Combat_Online_Financial_Misinformation}
The section highlights the importance of education, awareness initiatives, regulations, enforcement measures, detection techniques, and collaborative efforts in combating financial scams and misinformation.

Education and awareness initiatives play a crucial role in combating financial scams and protecting individuals from fraudulent activities. These initiatives can be implemented through various mediums, such as public service announcements, online tutorials, and workshops \cite{secemblem_2017}. The primary objective is to educate individuals on how to recognize and avoid financial scams while increasing their knowledge about the potential dangers of investing in unregulated or deceitful schemes \cite{scskksnssw_10239}. One effective approach to disseminating information and raising awareness is through public service announcements. These announcements can be broadcasted on television, radio, or social media platforms, reaching a wide audience \cite{secemblem_2017}. By highlighting real-life examples and providing practical tips, these messages can empower individuals with the knowledge to identify red flags and protect themselves from scams.

Online tutorials and workshops are also valuable tools for educating individuals about financial scams. These mediums offer interactive and engaging platforms where experts can share insights and strategies to avoid fraudulent activities \cite{scskksnssw_10239}. Through step-by-step guidance, individuals can learn how to conduct due diligence, verify the legitimacy of investment opportunities, and make informed financial decisions. To complement these educational efforts, the implementation of regulations and enforcement measures is essential. Regulatory bodies play a critical role in deterring fraudulent activities and protecting consumers from financial misinformation \cite{venkataramakrishnan_2023,scskksnssw_10239}. They can establish policies and guidelines that mandate financial institutions to maintain elevated levels of transparency and accountability.

One key aspect of regulation is the requirement for financial institutions to conduct regular audits. These audits help identify and prevent fraudulent activities by ensuring that proper controls and procedures are in place \cite{fintec_gov12309}. By regularly reviewing financial records and transactions, institutions can detect irregularities or suspicious patterns that may indicate fraudulent behavior.

Enforcement agencies also play a vital role in combating financial misinformation. They can impose penalties and fines on individuals and corporations found guilty of engaging in fraudulent activities. Additionally, collaboration with international counterparts allows for the pursuit of perpetrators of financial crimes across borders, making it harder for them to escape justice \cite{fintec_gov12309}. Collaborative efforts among stakeholders are crucial in developing effective strategies to combat financial misinformation. Financial institutions, tech companies, governments, and regulatory bodies should work together to address the challenges posed by fraudulent activities \cite{harvard_business_review_2021aaa}. By sharing information and exchanging best practices, these groups can enhance their collective knowledge and understanding of emerging threats. 

Moreover, the advancement of technology has played a significant role in identifying and preventing financial fraud. Sophisticated algorithms and machine learning techniques have been developed to analyze large datasets and detect patterns of financial misinformation. By leveraging these tools, authorities can identify suspicious activities and alert consumers and regulatory bodies to potential risks.

\subsection{Misleading Financial Data Collection}
\label{Datasets}
\begin{table}[!htbp]

\caption{Representative Datasets on Financial Misinformation.}
\resizebox{\textwidth}{!}{%
\begin{tabular}{@{}llllll@{}}
\toprule
Dataset & Topic & Instances & Source & Language & Year \\ \midrule
\texttt{Clarke et al. \cite{clarke2018fake}}  & Investor Attention & 157K & Seeking Alpha & English & 2018 \\ \midrule
\texttt{BanFakeNews-Financial \cite{hossain2020banfakenews}} & Multi-Domain & 1226 & Jaachai, BDFactCheck        & Bangla & 2020 \\ \midrule
\texttt{Kogan et al. \cite{kogan2020fake}}  & Trading & 350K  & \begin{tabular}[c]{@{}l@{}}Seeking Alpha \&\\ Motley Fool\end{tabular} & English   & 2020 \\ \midrule
\texttt{Bryan Fong \cite{fong2021analysing1}}    & Multi-sector  & - & \begin{tabular}[c]{@{}l@{}}2019 Chinese ADR\\ Delisting Threat\end{tabular} & English & 2021 \\ \midrule
\texttt{WEIBO21-Financial \cite{nan2021mdfend}}  & Multi-sector       & 1321  & Sina Weibo  & Chinese  & 2021 \\ \midrule
\texttt{Zhi et al. \cite{zhi2021financial01}}   & Multi-sector & 8K  & \begin{tabular}[c]{@{}l@{}}East Money Information,\\ Sina, Headline subjects, etc\\\end{tabular} & English  & 2021 \\ \midrule
\texttt{WELFake \cite{verma2021wwelfake,mohankumar2023financial01}}    & -  & 72K  & \begin{tabular}[c]{@{}l@{}}Kaggle, McIntire,\\ Reuters, BuzzFeed \end{tabular}  & English  & 2021 \\ \midrule
\texttt{Liu et al. \cite{liu2022role02}}   & Multi sector  & 125k  & Seeking Alpha     & English & 2022 \\ \midrule
\texttt{Chung et al. \cite{chung2022theory1}} & Stock Market   & 10.4K  & Twitter, StockTwits  & English  & 2022 \\ \midrule
\texttt{Zhang et al. \cite{zhang2022theory009}}  & Stock Market  & 7247  & Seeking Alpha & English  & 2022 \\ \midrule
\texttt{Liu et al. \cite{liu2023tracking118}}      &  Multi-sector   & 87K    & \begin{tabular}[c]{@{}l@{}}Accounting and \\ Auditing Enforcement \\ Releases (AAERs)\end{tabular}  & English & 2023 \\ \midrule
\texttt{Boehm et al. \cite{NBERw30994}}    & \begin{tabular}[c]{@{}l@{}}GDP, Capacity ,\\ Utilization, Retail\\ Sales, etc\end{tabular}  & 17k       & Bloomberg          & English & 2023 \\
\midrule
\texttt{Fin-Fact}    & Multi-sector  & 3.6k       & PolitiFact, FactCheck & English & 2023 \\
\bottomrule

\end{tabular}
}
\label{financial_misinformation_data}
\end{table}

General-content misinformation datasets are widely available, but there are only a handful of datasets related to online financial misinformation. Table \ref{financial_misinformation_data} presents a collection of datasets related to financial topics such as stock-market, investor attention, trading, retail sales, etc.
\begin{itemize}

    \item \texttt{Clarke et al. \cite{clarke2018fake}}: The dataset comprised 383 fake news articles, alongside information on the 157,253 legitimate news articles published on Seeking Alpha between August 1, 2012, and March 31, 2013, encompassing articles related to investor attention.

    \item \texttt{BanFakeNews-Financial \cite{hossain2020banfakenews}:} The BanFakeNews dataset, a collection of Bangla multi-sector fake news, was compiled by sourcing both fake and real news articles from the Jaachai and BDFactCheck websites. Within this dataset, there are 1226 instances of financial fake news.

    \item \texttt{Kogan et al. \cite{kogan2020fake}}: A comprehensive dataset was compiled by collecting data from Seeking Alpha and a competitor platform, Motley Fool. Seeking Alpha provided 203,545 articles spanning the period from 2005 to 2015, while Motley Fool contributed 147,916 articles from 2009 to 2014. The dataset encompasses information on 7,700 publicly traded firms.

    \item \texttt{Bryan Fong \cite{fong2021analysing1}}: The dataset is collected from the 2019 Chinese ADR Delisting Threat. The market model indices used are the NYSE and NASDAQ Composite indices, respective to each firm’s listing during the estimation and event windows from April 08 to October 18, 2019.
    
    \item \texttt{WEIBO21-Financial \cite{nan2021mdfend}:} Weibo21 a Chinese multi-sector fake news dataset, collected fake and real news from Sina Weibo, spanning between December 2014 and March 2021. This dataset contains financial fake news with 1321 instances.

    \item \texttt{Zhi et al. \cite{zhi2021financial01}}: The dataset was gathered from various financial websites, such as East Money Information, Sina, Headline, and others. Market data was analyzed by comparing the price movement with the previous day. News articles published on non-trading days were excluded from the dataset. Ultimately, a total of 8000 labeled samples were obtained for analysis.

    \item \texttt{WELFake \cite{verma2021wwelfake,mohankumar2023financial01}}: A larger and more generic Word Embedding over Linguistic Features for Fake News Detection (WELFake) dataset was developed, comprising 72,134 news articles, including 35,028 instances of real news and 37,106 instances of fake news. This dataset was created by merging four widely used news datasets, namely Kaggle, McIntire, Reuters, and BuzzFeed Political.

    \item \texttt{Liu et al. \cite{liu2022role02}}: Data was obtained from Seeking Alpha, encompassing all articles written from 2006 through 2018. The final sample consisted of 125,475 articles spanning 37,864 firm-quarters.

    \item \texttt{Chung et al. \cite{chung2022theory1}}: This dataset includes data from social media messages and stock prices of four Dow Jones Industrial Average (DJIA) technology companies: Apple, Microsoft, Intel, and Cisco. The data was collected at five-minute intervals on U.S. trading days between 11 July 2017 and 15 August 2018.

    \item \texttt{Zhang et al. \cite{zhang2022theory009}}: Seeking Alpha was selected as the source of financial news articles. The collected articles were published between August 2011 and March 2014. Among them, we identified 381 instances of financial disinformation and 6866 articles that were legitimate financial news.

    \item \texttt{Liu et al. \cite{liu2023tracking118}}: A comprehensive final dataset was acquired by manually compiling Accounting and Auditing Enforcement Releases (AAERs). Among the 4,012 SEC AAERs issued between 1994 and 2016, it was verified that the dataset encompasses a comprehensive collection. This final dataset consists of 87,765 firm-year observations, encompassing 11,303 firms, with 720 confirmed instances of fraud.

    \item \texttt{Boehm et al. \cite{NBERw30994}}: The data encompassing macroeconomic financial news releases were collected from Bloomberg's US Economic Calendar. A total of 17,253 financial news instances were gathered for analysis.\\

The lack of clear labeling and justifications in the aforementioned financial misinformation datasets is a cause for concern. Although these datasets contribute to the advancement of algorithms and machine learning, the absence of transparency in labeling raises questions about the reliability of the results they produce. To address this issue, we have created a benchmark dataset in the financial domain known as \textsc{Fin-Fact}\footnote{\url{https://github.com/IIT-DM/Fin-Fact}}. What sets it apart is its authentic data and multimodal structure, encompassing both text and image data to represent a wide range of financial information. Additionally, Fin-Fact includes expert fact-checker comments, empowering models to provide comprehensive explanations.

\end{itemize}

\subsection{Computational Methods to Counter Financial Misinformation}
\label{Methodologies}
In recent years, there has been a growing interest in understanding the mechanisms through which fake news exerts statistically significant influences on financial markets and prices, despite contradicting the efficient-market hypothesis. Numerous studies have investigated the impact of financial misinformation on market dynamics, investor behavior, and price volatility \cite{zhouu_zhangg_2008,liu2022role02,clarke2018fake,fong2021analysing1,chung2022theory1,zhi2021financial01,liu2023tracking118}. Additionally, several studies have explored the role of accounting information in the production of fake financial news. \cite{liu2022role02} provided insights into the nature of contemporary fake financial news by examining descriptive statistics of trends in the content and volume of fake news articles. The study employed bunching analyses to uncover two key observations regarding the interaction between accounting information and incentives for producing fake news. Firstly, it was found that fake news authors tend to publish more fake articles in close proximity to earnings announcements, as these events attract widespread market attention. Secondly, there is a strong preference for publishing fake articles before earnings announcements, particularly when the accounting information environment is relatively weaker compared to the post-announcement period. To complement the bunching results, regression analyses were conducted, revealing evidence that supports a robust accounting information environment disincentivizes the production of fake news and mitigates the market reaction to such news. These findings contribute to our understanding of the complex dynamics between accounting information, fake financial news, and market behavior. 

By utilizing data from \href{https://www.seekingalpha.com/}{Seeking Alpha\textsuperscript{*}}, researchers found that fake news stories generated significantly more attention compared to a control sample of legitimate articles \cite{clarke2018fake}. However, there was no evidence to suggest that commenters on the articles could detect fake news, and the ability of the Seeking Alpha editors to identify fake news was found to be modest. Nonetheless, machine learning algorithms demonstrated success in identifying fake news based on linguistic features. Furthermore, the stock market was observed to price fake news accurately. Although abnormal trading volume increased upon the release of fake news, it was lower compared to that associated with legitimate news. Moreover, the stock price reaction to fake news was discounted in comparison to legitimate news articles.

In the realm of understanding the impacts of fake news on financial markets and prices, an innovative model has emerged to shed light on the underlying dynamics \cite{fong2021analysing1}. It provides valuable insights into how fake news can significantly influence markets, even challenging the widely accepted efficient-market hypothesis. It not only explores the empirical price impacts of fake news but also addresses the intriguing patterns of market reactions. By reconciling the observed overreactions to fake news and the underreactions to real news, the model offers a fresh perspective on the intricate dynamics at play. Moreover, it unveils a previously unexplored aspect of fake news, namely the potential amplification of underreactions to subsequent real news related to the specific security in question. Through comprehensive evaluation against real-world events, such as the debunking of a significant fake news event, this research further validates the model's dynamics and predictions.

The spreading of disinformation in social media posed a threat to cybersecurity and undermined market efficiency. Detecting disinformation was particularly challenging due to the large volumes of social media content and the rapidly changing nature of the environment. \cite{chung2022theory1} aimed to address this issue by developing and validating a novel deep-learning approach called TRNN (Temporal and Contextual Recurrent Neural Network) for disinformation detection. Grounded in social and psychological theories, TRNN utilized deep learning techniques and data-centric augmentation to enhance disinformation detection in the realm of financial social media. TRNN outperformed traditional machine learning techniques, achieving superior precision, recall, F-score, and accuracy. The model encoded temporal and contextual information using LSTM recurrent neurons to capture hidden patterns. A case study on Apple Inc.'s stock price movement showcased TRNN's potential for secure knowledge management. 

Another noteworthy contribution is a novel approach that combines the power of convolutional neural networks (CNN) and long short-term memory (LSTM) models to enhance the detection accuracy of fake news in the financial domain \cite{zhi2021financial01}. While previous research has demonstrated the effectiveness of CNN and LSTM individually for fake news detection \cite{bahad2019fake,balwant2019bidirectional,vyas2021fake,li2020multii}, this study proposes a multi-fact CNN-LSTM model that integrates both models to capture local and global dependencies in textual data. By leveraging the strengths of CNN and LSTM, this approach shows promising potential in improving the identification of financial fake news. The development of more reliable and robust systems for detecting misinformation in financial markets is a crucial step forward. \cite{mohankumar2023financial01} introduced a novel approach for financial fake news detection. Financial fake news datasets were prepared using topic modeling and Transformer-based techniques. The approach utilized two cross-joint networks, CAEN (Context-Aware Embedding Network) and CSRN (Contextual Sequential Representation Network), to capture linguistic and financial context and generate contextual sequential representations. Evaluation of benchmark datasets demonstrated impressive results, outperforming state-of-the-art and neural network-based baselines.

\begin{table}[!htbp]
\centering
\caption{Prior Financial Misinformation Detection Studies Using Financial Data}
\resizebox{1.05\linewidth}{!}{
\begin{tabular}{m{3.7em} m{16em} m{9em} m{9em} m{7em} m{5em}}
\hline
Category & Description & Dataset & Technique & Task & Work\\
\hline 
Statistical Methods & Aims to elucidate the mechanisms through which fake news exerts statistically significant influences on financial markets and prices, despite contradicting the efficient-market hypothesis. & \texttt{Bryan Fong} & Unified representative agent(URA). & Classification & URA \cite{fong2021analysing1}\\
\\
& Discovered the number of fake articles exhibits a bimodal pattern over our sample period, revealing various patterns. & \texttt{Liu et al.} & Regression Analysis & Regression & RA \cite{liu2022role02}\\
\\
Machine Learning & ML methods, like SVM, decision trees, and logistic regression, use labeled instances to train classifiers that learn disinformation patterns. & Various firm-years documents & LR, SVM, ID3, etc.. & Classification & ML \cite{persons1995using,green1997assessing,fanning1998neural,summers1998fraudulently,beneish1999detection,spathis2002detecting,spathis2002detecting012,lin2003fuzzy,kaminski2004can01,kirkos2007data01,gaganis2009classification01,cecchini2010detecting01,dikmen2010detection01,dechow2011predicting01}\\
\\
& & \texttt{Clarke et al.} & Gradient Boosting, RF, and SVM based on the LIWC linguistic features. & Classification & SVM-LIWC \cite{clarke2018fake}\\
\\
Deep Learning & TRNN approach that incorporated contextual and temporal information from social media and synchronized financial market data. & \texttt{Chung et al.} & Temporal Recurrent Neural Network(TRNN) & Classification & TRNN \cite{chung2022theory1} \\
\\
& The proposed model is a novel multi-fact CNN-LSTM architecture that combines news body, comments, sources, and market data. & \texttt{Zhi et al.} & CNN-LSTM & Classification & CNN-LSTM \cite{zhi2021financial01}\\
\\
& The research paper combines TDT features with LSTM and BERT textual features to train a random forest (RF) model for analyzing financial news. & \texttt{Zhang et al.} &TDT with RNNs & Classification & TDT-RNN \cite{zhang2022theory009}\\
\\
& A new approach for financial fake news detection is proposed, utilizing two cross-joint networks with context-aware embedding and sequential representation. & \texttt{WELFake} & Context-aware cross joint-based embedding network (CAEN) & Classification & CAEN \cite{mohankumar2023financial01}\\
\\
& A sophisticated approach was proposed to detect fraud by tracking detailed changes in disclosures over time. An optimized method was used to align paragraphs in consecutive MD\&As, facilitating the identification of specific changes. & \texttt{Liu et al.} & LSTM, Hierarchical Vote Collective of Transformation-based Ensembles(HIVE) & Classification & HIVE \cite{liu2023tracking118}\\
\\
\hline
\end{tabular}
}
\label{table:FinancialFraudDetection}
\end{table}

The influence of maliciously false information, or disinformation, on people's beliefs and behaviors, has significant social and economic implications. \cite{zhang2022theory009} focuses on examining news articles related to financial markets on crowd-sourced digital platforms. By assembling a unique dataset comprising financial news articles that were investigated and prosecuted by the Securities and Exchange Commission, along with propagation data on digital platforms and the financial performance data of the focal firm, the researchers develop a well-justified machine learning system for detecting financial disinformation on social media platforms. The design of the system is rooted in the truth-default theory, which emphasizes the importance of communication context and motive, coherence, information correspondence, propagation, and sender demeanor as key constructs for assessing deceptive communication. Through extensive analyses, the performance and efficacy of the proposed system are evaluated. The study not only contributes to advancing theoretical understanding but also holds practical value in terms of its implications for detecting and combatting financial disinformation. By leveraging machine learning techniques and considering crucial factors in assessing deceptive communication, this research provides insights that can inform the development of more effective strategies for addressing disinformation in the financial domain.

In the domain of detecting financial fraud, \cite{abbasi_albrechtt_vance_hansenn_2012} proposed a meta-learning framework that utilized firm's financial ratios to detect financial fraud. Their approach aimed to identify patterns and anomalies in financial data that could signal fraudulent activities. By leveraging machine learning techniques, the framework analyzed various financial ratios such as profitability, liquidity, and solvency indicators, and applied classification algorithms to differentiate between fraudulent and non-fraudulent cases. This research contributed to the development of early detection systems for financial fraud, enabling companies and regulators to mitigate risks and take proactive measures.

Building upon the work of \cite{abbasi_albrechtt_vance_hansenn_2012}, \cite{dongg_liaoo_zhangg_2018} improved the algorithm by incorporating financial social media data to assess corporate fraud. Recognizing the growing impact of social media on financial markets and information dissemination, their research aimed to leverage the vast amount of user-generated content to enhance fraud detection capabilities. The authors proposed an analytic framework guided by systemic functional linguistics (SFL) theory \cite{schleppegrell2013systemic}, which leveraged unstructured data from financial social media platforms to assess the risk of corporate fraud. They assembled a unique dataset consisting of 64 fraudulent firms and a matched sample of 64 non-fraudulent firms, along with social media data prior to the firm's alleged fraud violations in Accounting and Auditing Enforcement Releases (AAERs). The framework automatically extracted various signals, including sentiment features, emotion features, topic features, lexical features, and social network features from the financial social media data. These signals were then utilized as inputs for machine learning classifiers to detect fraud. To evaluate the performance of their algorithm, the authors compared it against baseline approaches using only financial ratios and language-based features separately. They further validated the robustness of their algorithm by detecting leaked information and rumors, testing it on a new dataset, and conducting an applicability check.

\cite{liu2023tracking118} proposed a nuanced method for detecting frauds by tracking granular changes in disclosures over time. Paragraph alignment between consecutive disclosures was achieved by optimizing their similarities. Three types of changed contents were identified: recurrent, newly added, and deleted, and their changes were measured using fraud-relevant linguistic features. A deep learning model was employed to predict frauds based on the firm's Management Discussion and Analysis change trajectory, a multivariate time series of granular metrics. Extensive experiments demonstrated the superior performance of the model compared to benchmark models, with performance increasing as the length of the change trajectory grew. Specific types of changes, such as weak modal or reward words in newly added or deleted contents, were strongly associated with fraud. This work contributed to the fraud detection literature by providing an optimization-based method to define change trajectories and trace information mutation in narratives, along with an effective deep learning architecture. Table. \ref{table:FinancialFraudDetection} shows the existing works in detecting Financial Misinformation.

In conclusion, the existing works in the detection of financial misinformation have employed a range of techniques, including deep learning models such as LSTM and CNN, as well as advanced language models like BERT. Additionally, approaches like TRNN have leveraged financial social media data for disinformation detection. Statistical and machine learning-based methods have also been utilized to analyze linguistic cues and other features associated with financial misinformation. These studies collectively contribute to our understanding of the challenges and potential solutions in identifying and combating financial misinformation. They highlight the importance of leveraging advanced technologies and data-driven approaches to effectively detect and address the spread of deceptive information in the financial domain. Moving forward, further research is necessary to refine and improve these techniques, considering the dynamic nature of financial misinformation in online platforms, and to explore the potential of new approaches for more accurate and efficient detection.

\section{Open Issues and Future Direction}
\label{open_issues}
\subsection{Detection and Filtering Techniques}
\label{Detection_and_Filtering_Techniques}
One of the critical challenges in combating financial misinformation lies in the development of robust and accurate detection and filtering techniques. While existing approaches have shown promise \cite{chung2022theory1,dongg_liaoo_zhangg_2018,10.1093/rof/rfac058}, there is a need for continuous improvement. To address this challenge, it is crucial to emphasize the importance of enhancing algorithms that can effectively detect and filter financial misinformation. This involves staying ahead of the ever-evolving tactics used by malicious actors to spread false or misleading information. By actively improving detection and filtering techniques, we can minimize the impact of financial misinformation on individuals, markets, and society as a whole. Furthermore, exploring advanced natural language processing (NLP) techniques and machine learning models could significantly improve the capabilities in this area. NLP has made significant strides in recent years, enabling machines to better understand and process human language. By leveraging the power of NLP, algorithms can better comprehend the contextual nuances, linguistic patterns, and sentiment associated with financial content. This deeper understanding can contribute to the accuracy of detecting misinformation.

Machine learning models, especially those based on deep learning techniques, have demonstrated effectiveness in various domains of misinformation detection. Their application to financial information could enhance detection accuracy and enable faster response times to emerging misinformation. These models can be trained on large datasets to learn patterns and indicators of financial misinformation, allowing them to identify and flag potentially misleading content more efficiently. To ensure continuous improvement in combating financial misinformation, it is crucial to foster collaboration between researchers, industry experts, and policymakers. This collaborative effort can help in the exchange of knowledge, sharing of best practices, and development of standardized approaches to misinformation detection and filtering.

\subsection{Social Media Platform Interventions}
\label{SocialMediaPlatform_Interventions}
Social media platforms have emerged as prominent channels for the dissemination of financial misinformation due to their extensive reach and influence \cite{doi:10.1080/07421222.2021.1990612,atske_202112344589}. This widespread distribution of false or misleading financial information poses a significant challenge that requires effective interventions to mitigate its impact. To address this issue, it is crucial for stakeholders to pool their expertise and resources, working together to enhance the effectiveness of interventions. Collaboration between social media platforms, financial institutions, regulatory bodies, and researchers can lead to the development of comprehensive strategies for combating the spread of financial misinformation. By sharing knowledge and best practices, these stakeholders can create a collective defense against the dissemination of false information.

Improving reporting mechanisms is an important aspect of tackling financial misinformation on social media platforms. Users need accessible and efficient channels to report false or misleading content, ensuring that platforms can take appropriate action in a timely manner. By streamlining and enhancing the reporting process, platforms can effectively identify and address instances of financial misinformation, minimizing its impact on users. In addition, evaluating the impact of platform policies is vital in combating the spread of financial misinformation. Policies related to the visibility and prominence of misinformation, user engagement algorithms, and content recommendation systems play a significant role in shaping the dissemination dynamics on social media platforms. Understanding the influence of these policies can inform the development of more targeted and effective intervention strategies. By analyzing the impact of different policies, platforms can identify areas for improvement and implement changes that limit the reach and influence of financial misinformation.

Furthermore, it is essential to leverage innovative strategies in the fight against financial misinformation on social media. This may involve the use of artificial intelligence and machine learning algorithms to automatically detect and flag potentially misleading content. Developing advanced algorithms that can analyze patterns, linguistic cues, and user engagement metrics can contribute to more efficient and accurate identification of financial misinformation. By adopting these innovative approaches, platforms can stay one step ahead of malicious actors who seek to spread false or misleading information.

\subsection{User Education and Awareness}
\label{User_Education_and_Awareness}
Enhancing user education and awareness is a fundamental aspect of combating the risks associated with financial misinformation. Research studies, such as those conducted by \cite{nguyen2022mindsponge00} and  \cite{pennycook2019falls}, highlight the importance of investigating the effectiveness of financial literacy programs in equipping users with the necessary skills to identify and mitigate the risks associated with financial misinformation.

Financial literacy programs can play a significant role in empowering individuals with the knowledge and tools to navigate the complex landscape of financial information. These programs can educate users on key concepts, such as investment strategies, financial markets, and common tactics used in spreading financial misinformation. By fostering an understanding of these topics, users can develop critical thinking skills and become more discerning when encountering potentially misleading information. In addition to financial literacy programs, promoting media literacy education is crucial in addressing the challenges posed by financial misinformation. Media literacy focuses on equipping individuals with the ability to critically evaluate and analyze information from various sources. By promoting media literacy, users can develop a better understanding of the techniques used to manipulate and distort financial information \cite{dell2019fake001}. They can also learn how to verify sources, fact-check claims, and assess the credibility of information before making financial decisions based on it.

Strategies to increase user awareness of the prevalence and potential consequences of financial misinformation are essential. This can involve targeted campaigns that highlight real-life examples of financial misinformation and its impact on individuals, markets, and society. By illustrating the potential consequences of falling victim to false or misleading financial information, users can become more cautious and proactive in verifying the information they encounter. Collaboration between educators, financial institutions, regulatory bodies, and social media platforms is instrumental in promoting user education and awareness. By working together, these stakeholders can develop educational resources, disseminate information through various channels, and integrate financial and media literacy into existing educational frameworks. These efforts can ensure that users, regardless of their background or level of financial knowledge, have access to the necessary tools and information to navigate the digital landscape safely.

\subsection{Long-Term Effects and Socioeconomic Consequences}
\label{Long-Term_Effects_and_Socioeconomic_Consequences}
The long-term impact of financial misinformation extends beyond immediate consequences and can have far-reaching effects on individuals, financial markets, and broader socioeconomic systems \cite{bali_bodnaruk_scherbinaaq_tang_2018,ahsan2013effects01}. \cite{Mitchell_Mulherin_impact} highlight the importance of examining the long-term consequences on individual investor's trust in financial institutions, their willingness to participate in financial markets, and the potential erosion of investor confidence. The widespread dissemination of false or misleading financial information can contribute to the formation of financial bubbles or market instability \cite{pennycook2019falls,shin_jian_driscoll_bar_2018df,bali_bodnaruk_scherbinaaq_tang_2018,atske_202112344589}. Understanding the mechanisms through which financial misinformation can distort market dynamics, exacerbate volatility, and affect asset prices is crucial for maintaining market integrity and stability \cite{renault2014pump,bali_bodnaruk_scherbinaaq_tang_2018,Mitchell_Mulherin_impact}. Additionally, exploring the psychological and behavioral implications of exposure to financial misinformation on investor decision-making can provide valuable insights \cite{bastick2021would112}. Factors such as cognitive biases, emotional responses, and susceptibility to manipulation can influence an individual's choices, further underscoring the need to address the consequences of financial misinformation from a behavioral perspective.

\section{Conclusion}
\label{conclusion}
This research paper addresses the critical issue of online financial misinformation and its wide-ranging implications. With the rapid expansion of digital platforms, the accessibility and consumption of financial news and information have transformed significantly. However, this progress has also led to a concerning proliferation of financial misinformation, posing substantial risks to individuals, markets, and the overall economy. Through an extensive survey, the paper explores the various forms and manifestations of online financial misinformation, including false claims, deceptive practices, and misleading content. The paper also highlights the significance of early detection of financial misinformation and the implementation of effective mitigation strategies. By proactively identifying and addressing misinformation, stakeholders can protect investors, promote market transparency, and preserve financial stability. The detrimental effects of such misinformation are evident, resulting in market volatility, investor losses, and a decline in trust towards financial institutions.

\bibliographystyle{ACM-Reference-Format}
\bibliography{output}


\begin{thebibliography}{211}


\ifx \showCODEN    \undefined \def \showCODEN     #1{\unskip}     \fi
\ifx \showDOI      \undefined \def \showDOI       #1{#1}\fi
\ifx \showISBNx    \undefined \def \showISBNx     #1{\unskip}     \fi
\ifx \showISBNxiii \undefined \def \showISBNxiii  #1{\unskip}     \fi
\ifx \showISSN     \undefined \def \showISSN      #1{\unskip}     \fi
\ifx \showLCCN     \undefined \def \showLCCN      #1{\unskip}     \fi
\ifx \shownote     \undefined \def \shownote      #1{#1}          \fi
\ifx \showarticletitle \undefined \def \showarticletitle #1{#1}   \fi
\ifx \showURL      \undefined \def \showURL       {\relax}        \fi
\providecommand\bibfield[2]{#2}
\providecommand\bibinfo[2]{#2}
\providecommand\natexlab[1]{#1}
\providecommand\showeprint[2][]{arXiv:#2}

\bibitem[Abbasi et~al\mbox{.}(2012)]%
        {abbasi_albrechtt_vance_hansenn_2012}
\bibfield{author}{\bibinfo{person}{Ahmed Abbasi}, \bibinfo{person}{Conan
  Albrecht}, \bibinfo{person}{Anthony Vance}, {and} \bibinfo{person}{James
  Hansen}.} \bibinfo{year}{2012}\natexlab{}.
\newblock \showarticletitle{Metafraud: a meta-learning framework for detecting
  financial fraud}.
\newblock \bibinfo{journal}{\emph{Mis Quarterly}} (\bibinfo{year}{2012}),
  \bibinfo{pages}{1293--1327}.
\newblock


\bibitem[Aberg(2016)]%
        {aberggf2016case}
\bibfield{author}{\bibinfo{person}{Eric~M Aberg}.}
  \bibinfo{year}{2016}\natexlab{}.
\newblock \showarticletitle{The Case for UDAAP-Based Credit Card Lending
  Regulations: Providing Greater Financial Security for America and American
  Consumers}.
\newblock \bibinfo{journal}{\emph{Geo. Wash. L. Rev.}}  \bibinfo{volume}{84}
  (\bibinfo{year}{2016}), \bibinfo{pages}{1029}.
\newblock


\bibitem[Adams and Smart(2017)]%
        {adams2017advertt}
\bibfield{author}{\bibinfo{person}{Paul Adams} {and} \bibinfo{person}{Laura
  Smart}.} \bibinfo{year}{2017}\natexlab{}.
\newblock \showarticletitle{From advert to action: behavioural insights into
  the advertising of financial products}.
\newblock \bibinfo{journal}{\emph{FCA Occasional Paper}} \bibinfo{number}{26}
  (\bibinfo{year}{2017}).
\newblock


\bibitem[Adjin-Tettey(2022)]%
        {doi:10.1080/23311983.2022.2037229}
\bibfield{author}{\bibinfo{person}{Theodora~Dame Adjin-Tettey}.}
  \bibinfo{year}{2022}\natexlab{}.
\newblock \showarticletitle{Combating fake news, disinformation, and
  misinformation: Experimental evidence for media literacy education}.
\newblock \bibinfo{journal}{\emph{Cogent Arts \& Humanities}}
  \bibinfo{volume}{9}, \bibinfo{number}{1} (\bibinfo{year}{2022}),
  \bibinfo{pages}{2037229}.
\newblock
\urldef\tempurl%
\url{https://doi.org/10.1080/23311983.2022.2037229}
\showDOI{\tempurl}
\showeprint{https://doi.org/10.1080/23311983.2022.2037229}


\bibitem[Aggarwal and Wu(2006)]%
        {aggarwal2006stock}
\bibfield{author}{\bibinfo{person}{Rajesh~K Aggarwal} {and}
  \bibinfo{person}{Guojun Wu}.} \bibinfo{year}{2006}\natexlab{}.
\newblock \showarticletitle{Stock market manipulations}.
\newblock \bibinfo{journal}{\emph{The Journal of Business}}
  \bibinfo{volume}{79}, \bibinfo{number}{4} (\bibinfo{year}{2006}),
  \bibinfo{pages}{1915--1953}.
\newblock


\bibitem[Ahern and Sosyura(2015)]%
        {ahern_sosyuraa1_2015}
\bibfield{author}{\bibinfo{person}{Kenneth~R Ahern} {and}
  \bibinfo{person}{Denis Sosyura}.} \bibinfo{year}{2015}\natexlab{}.
\newblock \showarticletitle{Rumor has it: Sensationalism in financial media}.
\newblock \bibinfo{journal}{\emph{The Review of Financial Studies}}
  \bibinfo{volume}{28}, \bibinfo{number}{7} (\bibinfo{year}{2015}),
  \bibinfo{pages}{2050--2093}.
\newblock


\bibitem[Ahsan et~al\mbox{.}(2013)]%
        {ahsan2013effects01}
\bibfield{author}{\bibinfo{person}{AFM~Mainul Ahsan},
  \bibinfo{person}{Mohammad~Osman Gani}, {and} \bibinfo{person}{Md~Bokhtiar
  Hasan}.} \bibinfo{year}{2013}\natexlab{}.
\newblock \showarticletitle{Effects of Misinformation on the Stock Return: A
  Case Study}.
\newblock \bibinfo{journal}{\emph{Advances in Economics and Business}}
  \bibinfo{volume}{1} (\bibinfo{year}{2013}), \bibinfo{pages}{282--289}.
\newblock


\bibitem[Allcott and Gentzkow(2017)]%
        {allcott2017social}
\bibfield{author}{\bibinfo{person}{Hunt Allcott} {and} \bibinfo{person}{Matthew
  Gentzkow}.} \bibinfo{year}{2017}\natexlab{}.
\newblock \showarticletitle{Social media and fake news in the 2016 election}.
\newblock \bibinfo{journal}{\emph{Journal of economic perspectives}}
  \bibinfo{volume}{31}, \bibinfo{number}{2} (\bibinfo{year}{2017}),
  \bibinfo{pages}{211--236}.
\newblock


\bibitem[Anderson et~al\mbox{.}(2008)]%
        {anderson2008identity}
\bibfield{author}{\bibinfo{person}{Keith~B Anderson}, \bibinfo{person}{Erik
  Durbin}, {and} \bibinfo{person}{Michael~A Salinger}.}
  \bibinfo{year}{2008}\natexlab{}.
\newblock \showarticletitle{Identity theft}.
\newblock \bibinfo{journal}{\emph{Journal of Economic Perspectives}}
  \bibinfo{volume}{22}, \bibinfo{number}{2} (\bibinfo{year}{2008}),
  \bibinfo{pages}{171--192}.
\newblock


\bibitem[Arnold et~al\mbox{.}(2007)]%
        {Arnold_Earl_North_2007}
\bibfield{author}{\bibinfo{person}{Tom Arnold}, \bibinfo{person}{John~H
  Earl~Jr}, {and} \bibinfo{person}{David~S North}.}
  \bibinfo{year}{2007}\natexlab{}.
\newblock \showarticletitle{Are cover stories effective contrarian indicators?}
\newblock \bibinfo{journal}{\emph{Financial Analysts Journal}}
  \bibinfo{volume}{63}, \bibinfo{number}{2} (\bibinfo{year}{2007}),
  \bibinfo{pages}{70--75}.
\newblock


\bibitem[Association(2009)]%
        {american2009americannn}
\bibfield{author}{\bibinfo{person}{American~Bar Association}.}
  \bibinfo{year}{2009}\natexlab{}.
\newblock \bibinfo{booktitle}{\emph{The American Bar Association Guide to
  Credit and Bankruptcy: Everything You Need to Know about Credit Repair,
  Staying Or Getting Out of Debt, and Personal Bankruptcy}}.
\newblock \bibinfo{publisher}{Random House Reference \&}.
\newblock


\bibitem[Atske(2021a)]%
        {atske_2021}
\bibfield{author}{\bibinfo{person}{Sara Atske}.}
  \bibinfo{year}{2021}\natexlab{a}.
\newblock \showarticletitle{Misinformation and competing views of reality
  abounded throughout 2020}.
\newblock \bibinfo{journal}{\emph{Pew Research Center's Journalism Project}}
  (\bibinfo{date}{Feb} \bibinfo{year}{2021}).
\newblock
\urldef\tempurl%
\url{https://www.pewresearch.org/journalism/2021/02/22/misinformation-and-competing-views-of-reality-abounded-throughout-2020/}
\showURL{%
\tempurl}


\bibitem[Atske(2021b)]%
        {atske_202112344589}
\bibfield{author}{\bibinfo{person}{Sara Atske}.}
  \bibinfo{year}{2021}\natexlab{b}.
\newblock \showarticletitle{News consumption across social media in 2021}.
\newblock \bibinfo{journal}{\emph{Pew Research Center's Journalism Project}}
  (\bibinfo{date}{Sep} \bibinfo{year}{2021}).
\newblock
\urldef\tempurl%
\url{https://www.pewresearch.org/journalism/2021/09/20/news-consumption-across-social-media-in-2021/}
\showURL{%
\tempurl}


\bibitem[Austin(2017)]%
        {austin2017insiider}
\bibfield{author}{\bibinfo{person}{Janet Austin}.}
  \bibinfo{year}{2017}\natexlab{}.
\newblock \bibinfo{booktitle}{\emph{Insider trading and market manipulation:
  investigating and prosecuting across borders}}.
\newblock \bibinfo{publisher}{Edward Elgar Publishing}.
\newblock


\bibitem[Authority(2015)]%
        {scskksnssw_10239}
\bibfield{author}{\bibinfo{person}{Financial~Conduct Authority}.}
  \bibinfo{year}{2015}\natexlab{}.
\newblock \showarticletitle{Regulatory sandbox}.
\newblock \bibinfo{journal}{\emph{Financial Conduct Authority}}
  \bibinfo{volume}{11} (\bibinfo{year}{2015}), \bibinfo{pages}{30}.
\newblock


\bibitem[Av{\'e}n and Hed{\'e}n(2016)]%
        {aven2016ambushh}
\bibfield{author}{\bibinfo{person}{Malin Av{\'e}n} {and} \bibinfo{person}{Carl
  Hed{\'e}n}.} \bibinfo{year}{2016}\natexlab{}.
\newblock \bibinfo{title}{Ambush Marketing Strategies’ Misleading Influence
  on Consumers}.
\newblock
\newblock


\bibitem[Azim and Azam(2016)]%
        {azim20160bernard}
\bibfield{author}{\bibinfo{person}{Mohammad Azim} {and}
  \bibinfo{person}{MD~Saiful Azam}.} \bibinfo{year}{2016}\natexlab{}.
\newblock \showarticletitle{Bernard Madoff's' Ponzi scheme': Fraudulent
  behaviour and the role of auditors}.
\newblock \bibinfo{journal}{\emph{Accountancy Business and the Public
  interest}}  \bibinfo{volume}{15} (\bibinfo{year}{2016}),
  \bibinfo{pages}{122--137}.
\newblock


\bibitem[Azrina Mohd~Yusof and Ling~Lai(2014)]%
        {azrina2014integrative}
\bibfield{author}{\bibinfo{person}{Nor Azrina Mohd~Yusof} {and}
  \bibinfo{person}{Ming Ling~Lai}.} \bibinfo{year}{2014}\natexlab{}.
\newblock \showarticletitle{An integrative model in predicting corporate tax
  fraud}.
\newblock \bibinfo{journal}{\emph{Journal of Financial Crime}}
  \bibinfo{volume}{21}, \bibinfo{number}{4} (\bibinfo{year}{2014}),
  \bibinfo{pages}{424--432}.
\newblock


\bibitem[Bahad et~al\mbox{.}(2019)]%
        {bahad2019fake}
\bibfield{author}{\bibinfo{person}{Pritika Bahad}, \bibinfo{person}{Preeti
  Saxena}, {and} \bibinfo{person}{Raj Kamal}.} \bibinfo{year}{2019}\natexlab{}.
\newblock \showarticletitle{Fake news detection using bi-directional
  LSTM-recurrent neural network}.
\newblock \bibinfo{journal}{\emph{Procedia Computer Science}}
  \bibinfo{volume}{165} (\bibinfo{year}{2019}), \bibinfo{pages}{74--82}.
\newblock


\bibitem[Bali et~al\mbox{.}(2018)]%
        {bali_bodnaruk_scherbinaaq_tang_2018}
\bibfield{author}{\bibinfo{person}{Turan~G Bali}, \bibinfo{person}{Andriy
  Bodnaruk}, \bibinfo{person}{Anna Scherbina}, {and} \bibinfo{person}{Yi
  Tang}.} \bibinfo{year}{2018}\natexlab{}.
\newblock \showarticletitle{Unusual news flow and the cross section of stock
  returns}.
\newblock \bibinfo{journal}{\emph{Management Science}} \bibinfo{volume}{64},
  \bibinfo{number}{9} (\bibinfo{year}{2018}), \bibinfo{pages}{4137--4155}.
\newblock


\bibitem[Balwant(2019)]%
        {balwant2019bidirectional}
\bibfield{author}{\bibinfo{person}{Manoj~Kumar Balwant}.}
  \bibinfo{year}{2019}\natexlab{}.
\newblock \showarticletitle{Bidirectional LSTM based on POS tags and CNN
  architecture for fake news detection}. In \bibinfo{booktitle}{\emph{2019 10th
  International conference on computing, communication and networking
  technologies (ICCCNT)}}. IEEE, \bibinfo{pages}{1--6}.
\newblock


\bibitem[Barrett(1970)]%
        {barrettt1970truth}
\bibfield{author}{\bibinfo{person}{Roger~S Barrett}.}
  \bibinfo{year}{1970}\natexlab{}.
\newblock \showarticletitle{Truth in Lending-Advertising}.
\newblock \bibinfo{journal}{\emph{Bus. LAw.}}  \bibinfo{volume}{26}
  (\bibinfo{year}{1970}), \bibinfo{pages}{829}.
\newblock


\bibitem[Bastick(2021)]%
        {bastick2021would112}
\bibfield{author}{\bibinfo{person}{Zach Bastick}.}
  \bibinfo{year}{2021}\natexlab{}.
\newblock \showarticletitle{Would you notice if fake news changed your
  behavior? An experiment on the unconscious effects of disinformation}.
\newblock \bibinfo{journal}{\emph{Computers in human behavior}}
  \bibinfo{volume}{116} (\bibinfo{year}{2021}), \bibinfo{pages}{106633}.
\newblock


\bibitem[Baum et~al\mbox{.}(2021)]%
        {baum2021sok}
\bibfield{author}{\bibinfo{person}{Carsten Baum}, \bibinfo{person}{James
  Hsin-yu Chiang}, \bibinfo{person}{Bernardo David},
  \bibinfo{person}{Tore~Kasper Frederiksen}, {and} \bibinfo{person}{Lorenzo
  Gentile}.} \bibinfo{year}{2021}\natexlab{}.
\newblock \showarticletitle{Sok: Mitigation of front-running in decentralized
  finance}.
\newblock \bibinfo{journal}{\emph{Cryptology ePrint Archive}}
  (\bibinfo{year}{2021}).
\newblock


\bibitem[Beatty and Fothergill(2016)]%
        {beatty2016uneven}
\bibfield{author}{\bibinfo{person}{Christina Beatty} {and}
  \bibinfo{person}{Stephen Fothergill}.} \bibinfo{year}{2016}\natexlab{}.
\newblock \bibinfo{booktitle}{\emph{The uneven impact of welfare reform: The
  financial losses to places and people}}.
\newblock \bibinfo{publisher}{Sheffield Hallam University}.
\newblock


\bibitem[Ben-David et~al\mbox{.}(2018)]%
        {bennn2018uninformative}
\bibfield{author}{\bibinfo{person}{Itzhak Ben-David}, \bibinfo{person}{Justin
  Birru}, {and} \bibinfo{person}{Viktor Prokopenya}.}
  \bibinfo{year}{2018}\natexlab{}.
\newblock \showarticletitle{Uninformative feedback and risk taking: Evidence
  from retail forex trading}.
\newblock \bibinfo{journal}{\emph{Review of Finance}} \bibinfo{volume}{22},
  \bibinfo{number}{6} (\bibinfo{year}{2018}), \bibinfo{pages}{2009--2036}.
\newblock


\bibitem[Beneish(1999)]%
        {beneish1999detection}
\bibfield{author}{\bibinfo{person}{Messod~D Beneish}.}
  \bibinfo{year}{1999}\natexlab{}.
\newblock \showarticletitle{The detection of earnings manipulation}.
\newblock \bibinfo{journal}{\emph{Financial Analysts Journal}}
  \bibinfo{volume}{55}, \bibinfo{number}{5} (\bibinfo{year}{1999}),
  \bibinfo{pages}{24--36}.
\newblock


\bibitem[Bernhardt and Taub(2008)]%
        {bernhardt2008front}
\bibfield{author}{\bibinfo{person}{Dan Bernhardt} {and} \bibinfo{person}{Bart
  Taub}.} \bibinfo{year}{2008}\natexlab{}.
\newblock \showarticletitle{Front-running dynamics}.
\newblock \bibinfo{journal}{\emph{Journal of Economic Theory}}
  \bibinfo{volume}{138}, \bibinfo{number}{1} (\bibinfo{year}{2008}),
  \bibinfo{pages}{288--296}.
\newblock


\bibitem[Bizzi and Labban(2019)]%
        {bizzi2019doublee}
\bibfield{author}{\bibinfo{person}{Lorenzo Bizzi} {and} \bibinfo{person}{Alice
  Labban}.} \bibinfo{year}{2019}\natexlab{}.
\newblock \showarticletitle{The double-edged impact of social media on online
  trading: Opportunities, threats, and recommendations for organizations}.
\newblock \bibinfo{journal}{\emph{Business Horizons}} \bibinfo{volume}{62},
  \bibinfo{number}{4} (\bibinfo{year}{2019}), \bibinfo{pages}{509--519}.
\newblock


\bibitem[Blanton et~al\mbox{.}(2012)]%
        {blanton2012rise}
\bibfield{author}{\bibinfo{person}{Kimberly Blanton} {et~al\mbox{.}}}
  \bibinfo{year}{2012}\natexlab{}.
\newblock \showarticletitle{The rise of financial fraud}.
\newblock \bibinfo{journal}{\emph{Center for Retirement Research Brief}}
  \bibinfo{number}{12-5} (\bibinfo{year}{2012}).
\newblock


\bibitem[Boehm and Kroner(2023)]%
        {NBERw30994}
\bibfield{author}{\bibinfo{person}{Christoph~E Boehm} {and}
  \bibinfo{person}{T.~Niklas Kroner}.} \bibinfo{year}{2023}\natexlab{}.
\newblock \bibinfo{booktitle}{\emph{The US, Economic News, and the Global
  Financial Cycle}}.
\newblock \bibinfo{type}{Working Paper} 30994. \bibinfo{institution}{National
  Bureau of Economic Research}.
\newblock
\urldef\tempurl%
\url{https://doi.org/10.3386/w30994}
\showDOI{\tempurl}


\bibitem[Bosley et~al\mbox{.}(2019)]%
        {bosley2019decision}
\bibfield{author}{\bibinfo{person}{Stacie~A Bosley}, \bibinfo{person}{Marc~F
  Bellemare}, \bibinfo{person}{Linda Umwali}, {and} \bibinfo{person}{Joshua
  York}.} \bibinfo{year}{2019}\natexlab{}.
\newblock \showarticletitle{Decision-making and vulnerability in a pyramid
  scheme fraud}.
\newblock \bibinfo{journal}{\emph{Journal of Behavioral and Experimental
  Economics}}  \bibinfo{volume}{80} (\bibinfo{year}{2019}),
  \bibinfo{pages}{1--13}.
\newblock


\bibitem[Braun and Eklund(2019)]%
        {braun2019fake}
\bibfield{author}{\bibinfo{person}{Joshua~A Braun} {and}
  \bibinfo{person}{Jessica~L Eklund}.} \bibinfo{year}{2019}\natexlab{}.
\newblock \showarticletitle{Fake news, real money: Ad tech platforms,
  profit-driven hoaxes, and the business of journalism}.
\newblock \bibinfo{journal}{\emph{Digital Journalism}} \bibinfo{volume}{7},
  \bibinfo{number}{1} (\bibinfo{year}{2019}), \bibinfo{pages}{1--21}.
\newblock


\bibitem[Broby(2021)]%
        {Broby2021}
\bibfield{author}{\bibinfo{person}{Daniel Broby}.}
  \bibinfo{year}{2021}\natexlab{}.
\newblock \showarticletitle{Financial technology and the future of banking}.
\newblock \bibinfo{journal}{\emph{Financial Innovation}} \bibinfo{volume}{7},
  \bibinfo{number}{1} (\bibinfo{date}{18 Jun} \bibinfo{year}{2021}),
  \bibinfo{pages}{47}.
\newblock
\showISSN{2199-4730}
\urldef\tempurl%
\url{https://doi.org/10.1186/s40854-021-00264-y}
\showDOI{\tempurl}


\bibitem[Bromberg et~al\mbox{.}(2017)]%
        {bromberg2017fffinancial}
\bibfield{author}{\bibinfo{person}{Lev Bromberg}, \bibinfo{person}{George
  Gilligan}, {and} \bibinfo{person}{Ian Ramsay}.}
  \bibinfo{year}{2017}\natexlab{}.
\newblock \showarticletitle{Financial market manipulation and insider trading:
  an international study of enforcement approaches}.
\newblock \bibinfo{journal}{\emph{Journal of Business Law, Issue}}
  \bibinfo{number}{8} (\bibinfo{year}{2017}), \bibinfo{pages}{652--679}.
\newblock


\bibitem[Bruhn(2015)]%
        {bruhn2015personal}
\bibfield{author}{\bibinfo{person}{Aaron~G Bruhn}.}
  \bibinfo{year}{2015}\natexlab{}.
\newblock \showarticletitle{Personal and social impacts of significant
  financial loss}.
\newblock \bibinfo{journal}{\emph{Australian Journal of Management}}
  \bibinfo{volume}{40}, \bibinfo{number}{3} (\bibinfo{year}{2015}),
  \bibinfo{pages}{459--477}.
\newblock


\bibitem[Bucknall(2008)]%
        {bucknall2008real}
\bibfield{author}{\bibinfo{person}{Brian Bucknall}.}
  \bibinfo{year}{2008}\natexlab{}.
\newblock \showarticletitle{Real estate fraud and systems of title
  registration: The paradox of certainty}.
\newblock \bibinfo{journal}{\emph{Can. Bus. LJ}}  \bibinfo{volume}{47}
  (\bibinfo{year}{2008}), \bibinfo{pages}{1}.
\newblock


\bibitem[Burgess and Stern(1993)]%
        {burgess1993taxation}
\bibfield{author}{\bibinfo{person}{Robin Burgess} {and}
  \bibinfo{person}{Nicholas Stern}.} \bibinfo{year}{1993}\natexlab{}.
\newblock \showarticletitle{Taxation and development}.
\newblock \bibinfo{journal}{\emph{Journal of economic literature}}
  \bibinfo{volume}{31}, \bibinfo{number}{2} (\bibinfo{year}{1993}),
  \bibinfo{pages}{762--830}.
\newblock


\bibitem[Burnes et~al\mbox{.}(2017)]%
        {burnes2017prevalence}
\bibfield{author}{\bibinfo{person}{David Burnes}, \bibinfo{person}{Charles~R
  Henderson~Jr}, \bibinfo{person}{Christine Sheppard}, \bibinfo{person}{Rebecca
  Zhao}, \bibinfo{person}{Karl Pillemer}, {and} \bibinfo{person}{Mark~S
  Lachs}.} \bibinfo{year}{2017}\natexlab{}.
\newblock \showarticletitle{Prevalence of financial fraud and scams among older
  adults in the United States: A systematic review and meta-analysis}.
\newblock \bibinfo{journal}{\emph{American journal of public health}}
  \bibinfo{volume}{107}, \bibinfo{number}{8} (\bibinfo{year}{2017}),
  \bibinfo{pages}{e13--e21}.
\newblock


\bibitem[Cahyani et~al\mbox{.}(2021)]%
        {cahyani2021fraudulent}
\bibfield{author}{\bibinfo{person}{Anik~Mega Cahyani}, \bibinfo{person}{Elva
  Nuraina}, {and} \bibinfo{person}{Farida Styaningrum}.}
  \bibinfo{year}{2021}\natexlab{}.
\newblock \showarticletitle{Fraudulent financial reporting on property, real
  estate, and building construction companies}.
\newblock \bibinfo{journal}{\emph{Assets: Jurnal Akuntansi Dan Pendidikan}}
  \bibinfo{volume}{10}, \bibinfo{number}{2} (\bibinfo{year}{2021}),
  \bibinfo{pages}{132--146}.
\newblock


\bibitem[Carapu{\c{c}}o et~al\mbox{.}(2018)]%
        {carapucco2018reinforcement}
\bibfield{author}{\bibinfo{person}{Jo{\~a}o Carapu{\c{c}}o},
  \bibinfo{person}{Rui Neves}, {and} \bibinfo{person}{Nuno Horta}.}
  \bibinfo{year}{2018}\natexlab{}.
\newblock \showarticletitle{Reinforcement learning applied to Forex trading}.
\newblock \bibinfo{journal}{\emph{Applied Soft Computing}}
  \bibinfo{volume}{73} (\bibinfo{year}{2018}), \bibinfo{pages}{783--794}.
\newblock


\bibitem[Carpenter(2023)]%
        {carpenter_2023}
\bibfield{author}{\bibinfo{person}{Perry Carpenter}.}
  \bibinfo{year}{2023}\natexlab{}.
\newblock \showarticletitle{Council post: Get the 411 on misinformation,
  disinformation and malinformation}.
\newblock \bibinfo{journal}{\emph{Forbes}} (\bibinfo{date}{Jan}
  \bibinfo{year}{2023}).
\newblock
\urldef\tempurl%
\url{https://www.forbes.com/sites/forbesbusinesscouncil/2023/01/13/get-the-411-on-misinformation-disinformation-and-malinformation/?sh=41f7d685256a}
\showURL{%
\tempurl}


\bibitem[Cecchini et~al\mbox{.}(2010)]%
        {cecchini2010detecting01}
\bibfield{author}{\bibinfo{person}{Mark Cecchini}, \bibinfo{person}{Haldun
  Aytug}, \bibinfo{person}{Gary~J Koehler}, {and} \bibinfo{person}{Praveen
  Pathak}.} \bibinfo{year}{2010}\natexlab{}.
\newblock \showarticletitle{Detecting management fraud in public companies}.
\newblock \bibinfo{journal}{\emph{Management Science}} \bibinfo{volume}{56},
  \bibinfo{number}{7} (\bibinfo{year}{2010}), \bibinfo{pages}{1146--1160}.
\newblock


\bibitem[Cezar et~al\mbox{.}(2020)]%
        {Cezar2020AdversarialCI}
\bibfield{author}{\bibinfo{person}{Asunur Cezar}, \bibinfo{person}{Srinivasan
  Raghunathan}, {and} \bibinfo{person}{Sumit Sarkar}.}
  \bibinfo{year}{2020}\natexlab{}.
\newblock \showarticletitle{Adversarial classification: Impact of agents’
  faking cost on firms and agents}.
\newblock \bibinfo{journal}{\emph{Production and Operations Management}}
  \bibinfo{volume}{29}, \bibinfo{number}{12} (\bibinfo{year}{2020}),
  \bibinfo{pages}{2789--2807}.
\newblock


\bibitem[Chitimira(2016)]%
        {chitimira2016unpacking}
\bibfield{author}{\bibinfo{person}{Howard Chitimira}.}
  \bibinfo{year}{2016}\natexlab{}.
\newblock \showarticletitle{Unpacking selected key elements of the insider
  trading and market manipulation offences in South Africa}.
\newblock \bibinfo{journal}{\emph{Journal of Corporate and Commercial Law and
  Practice}} \bibinfo{volume}{2}, \bibinfo{number}{2} (\bibinfo{year}{2016}),
  \bibinfo{pages}{24--41}.
\newblock


\bibitem[Chung et~al\mbox{.}(2022)]%
        {chung2022theory1}
\bibfield{author}{\bibinfo{person}{Wingyan Chung}, \bibinfo{person}{Yinqiang
  Zhang}, {and} \bibinfo{person}{Jia Pan}.} \bibinfo{year}{2022}\natexlab{}.
\newblock \showarticletitle{A theory-based deep-learning approach to detecting
  disinformation in financial social media}.
\newblock \bibinfo{journal}{\emph{Information Systems Frontiers}}
  (\bibinfo{year}{2022}), \bibinfo{pages}{1--20}.
\newblock


\bibitem[Clarke et~al\mbox{.}(ming)]%
        {clarke2018fake}
\bibfield{author}{\bibinfo{person}{Jonathan Clarke}, \bibinfo{person}{Hailiang
  Chen}, \bibinfo{person}{Ding Du}, {and} \bibinfo{person}{Yu~Jeffrey Hu}.}
  \bibinfo{year}{Forthcoming}\natexlab{}.
\newblock \showarticletitle{Fake news, investor attention, and market
  reaction}.
\newblock \bibinfo{journal}{\emph{Information Systems Research}}
  (\bibinfo{year}{Forthcoming}).
\newblock
\newblock
\shownote{45 Pages Posted: 1 Aug 2018 Last revised: 18 Dec 2019}.


\bibitem[de~La~Feria(2020)]%
        {dewwede2020tax}
\bibfield{author}{\bibinfo{person}{Rita de La~Feria}.}
  \bibinfo{year}{2020}\natexlab{}.
\newblock \showarticletitle{Tax fraud and selective law enforcement}.
\newblock \bibinfo{journal}{\emph{Journal of law and Society}}
  \bibinfo{volume}{47}, \bibinfo{number}{2} (\bibinfo{year}{2020}),
  \bibinfo{pages}{240--270}.
\newblock


\bibitem[Dechow et~al\mbox{.}(2011)]%
        {dechow2011predicting01}
\bibfield{author}{\bibinfo{person}{Patricia~M Dechow}, \bibinfo{person}{Weili
  Ge}, \bibinfo{person}{Chad~R Larson}, {and} \bibinfo{person}{Richard~G
  Sloan}.} \bibinfo{year}{2011}\natexlab{}.
\newblock \showarticletitle{Predicting material accounting misstatements}.
\newblock \bibinfo{journal}{\emph{Contemporary accounting research}}
  \bibinfo{volume}{28}, \bibinfo{number}{1} (\bibinfo{year}{2011}),
  \bibinfo{pages}{17--82}.
\newblock


\bibitem[Del~Vicario et~al\mbox{.}(2016)]%
        {del2016spreading}
\bibfield{author}{\bibinfo{person}{Michela Del~Vicario},
  \bibinfo{person}{Alessandro Bessi}, \bibinfo{person}{Fabiana Zollo},
  \bibinfo{person}{Fabio Petroni}, \bibinfo{person}{Antonio Scala},
  \bibinfo{person}{Guido Caldarelli}, \bibinfo{person}{H~Eugene Stanley}, {and}
  \bibinfo{person}{Walter Quattrociocchi}.} \bibinfo{year}{2016}\natexlab{}.
\newblock \showarticletitle{Spreading of misinformation online}.
\newblock \bibinfo{journal}{\emph{Proceedings of the National Academy of
  Sciences}} \bibinfo{volume}{113}, \bibinfo{number}{3} (\bibinfo{year}{2016}),
  \bibinfo{pages}{554--559}.
\newblock


\bibitem[Dell(2019)]%
        {dell2019fake001}
\bibfield{author}{\bibinfo{person}{Marin Dell}.}
  \bibinfo{year}{2019}\natexlab{}.
\newblock \showarticletitle{Fake news, alternative facts, and disinformation:
  the importance of teaching media literacy to law students}.
\newblock \bibinfo{journal}{\emph{TOURo L. REv.}}  \bibinfo{volume}{35}
  (\bibinfo{year}{2019}), \bibinfo{pages}{619}.
\newblock


\bibitem[Demartini et~al\mbox{.}(2020)]%
        {demartini2020hhuman}
\bibfield{author}{\bibinfo{person}{Gianluca Demartini},
  \bibinfo{person}{Stefano Mizzaro}, {and} \bibinfo{person}{Damiano Spina}.}
  \bibinfo{year}{2020}\natexlab{}.
\newblock \showarticletitle{Human-in-the-loop Artificial Intelligence for
  Fighting Online Misinformation: Challenges and Opportunities.}
\newblock \bibinfo{journal}{\emph{IEEE Data Eng. Bull.}} \bibinfo{volume}{43},
  \bibinfo{number}{3} (\bibinfo{year}{2020}), \bibinfo{pages}{65--74}.
\newblock


\bibitem[Di~Domenico et~al\mbox{.}(2021)]%
        {di2021fake12}
\bibfield{author}{\bibinfo{person}{Giandomenico Di~Domenico},
  \bibinfo{person}{Jason Sit}, \bibinfo{person}{Alessio Ishizaka}, {and}
  \bibinfo{person}{Daniel Nunan}.} \bibinfo{year}{2021}\natexlab{}.
\newblock \showarticletitle{Fake news, social media and marketing: A systematic
  review}.
\newblock \bibinfo{journal}{\emph{Journal of Business Research}}
  \bibinfo{volume}{124} (\bibinfo{year}{2021}), \bibinfo{pages}{329--341}.
\newblock


\bibitem[Dikmen and K{\"u}{\c{c}}{\"u}kkocao{\u{g}}lu(2010)]%
        {dikmen2010detection01}
\bibfield{author}{\bibinfo{person}{Burcu Dikmen} {and}
  \bibinfo{person}{G{\"u}ray K{\"u}{\c{c}}{\"u}kkocao{\u{g}}lu}.}
  \bibinfo{year}{2010}\natexlab{}.
\newblock \showarticletitle{The detection of earnings manipulation: the
  three-phase cutting plane algorithm using mathematical programming}.
\newblock \bibinfo{journal}{\emph{Journal of Forecasting}}
  \bibinfo{volume}{29}, \bibinfo{number}{5} (\bibinfo{year}{2010}),
  \bibinfo{pages}{442--466}.
\newblock


\bibitem[Dippenaar(2021)]%
        {dippenaaar2021wary}
\bibfield{author}{\bibinfo{person}{Annelene Dippenaar}.}
  \bibinfo{year}{2021}\natexlab{}.
\newblock \showarticletitle{Be wary of credit repair scammers on the prowl}.
\newblock \bibinfo{journal}{\emph{Personal Finance Magazine}}
  \bibinfo{volume}{2021}, \bibinfo{number}{483} (\bibinfo{year}{2021}),
  \bibinfo{pages}{14--14}.
\newblock


\bibitem[Dong et~al\mbox{.}(2018)]%
        {dongg_liaoo_zhangg_2018}
\bibfield{author}{\bibinfo{person}{Wei Dong}, \bibinfo{person}{Shaoyi Liao},
  {and} \bibinfo{person}{Zhongju Zhang}.} \bibinfo{year}{2018}\natexlab{}.
\newblock \showarticletitle{Leveraging financial social media data for
  corporate fraud detection}.
\newblock \bibinfo{journal}{\emph{Journal of Management Information Systems}}
  \bibinfo{volume}{35}, \bibinfo{number}{2} (\bibinfo{year}{2018}),
  \bibinfo{pages}{461--487}.
\newblock


\bibitem[Doong et~al\mbox{.}(2005)]%
        {Doong2005ResponseAI}
\bibfield{author}{\bibinfo{person}{Shuh-Chyi Doong},
  \bibinfo{person}{Sheng-Yung Yang}, {and} \bibinfo{person}{Thomas~C Chiang}.}
  \bibinfo{year}{2005}\natexlab{}.
\newblock \showarticletitle{Response asymmetries in Asian stock markets}.
\newblock \bibinfo{journal}{\emph{Review of Pacific Basin Financial Markets and
  Policies}} \bibinfo{volume}{8}, \bibinfo{number}{04} (\bibinfo{year}{2005}),
  \bibinfo{pages}{637--657}.
\newblock


\bibitem[Dunham(2007)]%
        {dunham2007pump}
\bibfield{author}{\bibinfo{person}{Ken Dunham}.}
  \bibinfo{year}{2007}\natexlab{}.
\newblock \showarticletitle{Pump and dump scams}.
\newblock \bibinfo{journal}{\emph{Information Systems Security}}
  \bibinfo{volume}{16}, \bibinfo{number}{1} (\bibinfo{year}{2007}),
  \bibinfo{pages}{65--71}.
\newblock


\bibitem[Dymova et~al\mbox{.}(2016)]%
        {dymova2016fforex}
\bibfield{author}{\bibinfo{person}{Ludmila Dymova}, \bibinfo{person}{Pavel
  Sevastjanov}, {and} \bibinfo{person}{Krzysztof Kaczmarek}.}
  \bibinfo{year}{2016}\natexlab{}.
\newblock \showarticletitle{A Forex trading expert system based on a new
  approach to the rule-base evidential reasoning}.
\newblock \bibinfo{journal}{\emph{Expert Systems with Applications}}
  \bibinfo{volume}{51} (\bibinfo{year}{2016}), \bibinfo{pages}{1--13}.
\newblock


\bibitem[Ebem et~al\mbox{.}(2017)]%
        {ebem20177iinternet}
\bibfield{author}{\bibinfo{person}{Deborah~Uzoamaka Ebem},
  \bibinfo{person}{Joseph~Chinonye Onyeagba}, {and}
  \bibinfo{person}{Geraldine~Egondu Ugwuonah}.}
  \bibinfo{year}{2017}\natexlab{}.
\newblock \showarticletitle{Internet banking: identity theft and solutions-the
  Nigerian perspective}.
\newblock \bibinfo{journal}{\emph{The Journal of Internet Banking and
  Commerce}} \bibinfo{volume}{22}, \bibinfo{number}{2} (\bibinfo{year}{2017}),
  \bibinfo{pages}{1--15}.
\newblock


\bibitem[El~Haj and Kapogiannis(2016)]%
        {elll2016time}
\bibfield{author}{\bibinfo{person}{Mohamad El~Haj} {and}
  \bibinfo{person}{Dimitrios Kapogiannis}.} \bibinfo{year}{2016}\natexlab{}.
\newblock \showarticletitle{Time distortions in Alzheimer’s disease: a
  systematic review and theoretical integration}.
\newblock \bibinfo{journal}{\emph{NPJ aging and mechanisms of disease}}
  \bibinfo{volume}{2}, \bibinfo{number}{1} (\bibinfo{year}{2016}),
  \bibinfo{pages}{1--5}.
\newblock


\bibitem[Eltweri et~al\mbox{.}(2021)]%
        {eltweri2021applications}
\bibfield{author}{\bibinfo{person}{Ahmed Eltweri}, \bibinfo{person}{Alessio
  Faccia}, {and} \bibinfo{person}{OSAMA KHASSAWNEH}.}
  \bibinfo{year}{2021}\natexlab{}.
\newblock \showarticletitle{Applications of Big Data within Finance: Fraud
  Detection and Risk Management within the Real Estate Industry}. In
  \bibinfo{booktitle}{\emph{2021 3rd International Conference on E-Business and
  E-commerce Engineering}}. \bibinfo{pages}{67--73}.
\newblock


\bibitem[Elvestad et~al\mbox{.}(2018)]%
        {elvestad2018can112}
\bibfield{author}{\bibinfo{person}{Eiri Elvestad}, \bibinfo{person}{Angela
  Phillips}, {and} \bibinfo{person}{Mira Feuerstein}.}
  \bibinfo{year}{2018}\natexlab{}.
\newblock \showarticletitle{Can trust in traditional news media explain
  cross-national differences in news exposure of young people online? A
  comparative study of Israel, Norway and the United Kingdom}.
\newblock \bibinfo{journal}{\emph{Digital journalism}} \bibinfo{volume}{6},
  \bibinfo{number}{2} (\bibinfo{year}{2018}), \bibinfo{pages}{216--235}.
\newblock


\bibitem[Engelen and Van~Liedekerke(2007)]%
        {engelen2007ethicss}
\bibfield{author}{\bibinfo{person}{Peter-Jan Engelen} {and}
  \bibinfo{person}{Luc Van~Liedekerke}.} \bibinfo{year}{2007}\natexlab{}.
\newblock \showarticletitle{The ethics of insider trading revisited}.
\newblock \bibinfo{journal}{\emph{Journal of Business Ethics}}
  \bibinfo{volume}{74} (\bibinfo{year}{2007}), \bibinfo{pages}{497--507}.
\newblock


\bibitem[Evans(2018)]%
        {evans2018fforex}
\bibfield{author}{\bibinfo{person}{Martin~DD Evans}.}
  \bibinfo{year}{2018}\natexlab{}.
\newblock \showarticletitle{Forex trading and the WMR fix}.
\newblock \bibinfo{journal}{\emph{Journal of Banking \& Finance}}
  \bibinfo{volume}{87} (\bibinfo{year}{2018}), \bibinfo{pages}{233--247}.
\newblock


\bibitem[Fanning and Cogger(1998)]%
        {fanning1998neural}
\bibfield{author}{\bibinfo{person}{Kurt~M Fanning} {and}
  \bibinfo{person}{Kenneth~O Cogger}.} \bibinfo{year}{1998}\natexlab{}.
\newblock \showarticletitle{Neural network detection of management fraud using
  published financial data}.
\newblock \bibinfo{journal}{\emph{Intelligent Systems in Accounting, Finance \&
  Management}} \bibinfo{volume}{7}, \bibinfo{number}{1} (\bibinfo{year}{1998}),
  \bibinfo{pages}{21--41}.
\newblock


\bibitem[Fiala and Diamandis(2019)]%
        {fialaa2019theranos}
\bibfield{author}{\bibinfo{person}{Clare Fiala} {and}
  \bibinfo{person}{Eleftherios~P Diamandis}.} \bibinfo{year}{2019}\natexlab{}.
\newblock \bibinfo{title}{Theranos: Almost Complete Absence of Laboratory
  Medicine Input}.
\newblock , \bibinfo{numpages}{749--752}~pages.
\newblock


\bibitem[Finklea(2009)]%
        {finkllea2009identity}
\bibfield{author}{\bibinfo{person}{Kristin~M Finklea}.}
  \bibinfo{year}{2009}\natexlab{}.
\newblock \bibinfo{booktitle}{\emph{Identity theft: Trends and issues}}.
\newblock \bibinfo{publisher}{DIANE Publishing}.
\newblock


\bibitem[Fong(2021)]%
        {fong2021analysing1}
\bibfield{author}{\bibinfo{person}{Bryan Fong}.}
  \bibinfo{year}{2021}\natexlab{}.
\newblock \showarticletitle{Analysing the behavioural finance impact of 'fake
  news' phenomena on financial markets: a representative agent model and
  empirical validation}.
\newblock \bibinfo{journal}{\emph{Financial Innovation}} \bibinfo{volume}{7},
  \bibinfo{number}{1} (\bibinfo{year}{2021}), \bibinfo{pages}{1--30}.
\newblock


\bibitem[Frankel(2012)]%
        {frankel20121ponzi}
\bibfield{author}{\bibinfo{person}{Tamar Frankel}.}
  \bibinfo{year}{2012}\natexlab{}.
\newblock \bibinfo{booktitle}{\emph{The Ponzi scheme puzzle: A history and
  analysis of con artists and victims}}.
\newblock \bibinfo{publisher}{Oxford University Press}.
\newblock


\bibitem[Fscs(2022)]%
        {fscs_2022}
\bibfield{author}{\bibinfo{person}{Fscs}.} \bibinfo{year}{2022}\natexlab{}.
\newblock \showarticletitle{The worrying rise in online financial scams}.
\newblock \bibinfo{journal}{\emph{Financial Services Compensation Scheme Ltd.}}
  (\bibinfo{date}{Dec} \bibinfo{year}{2022}).
\newblock
\urldef\tempurl%
\url{https://www.fscs.org.uk/news/fraud/worrying-rise-in-online-financial-scams/}
\showURL{%
\tempurl}


\bibitem[Gaganis(2009)]%
        {gaganis2009classification01}
\bibfield{author}{\bibinfo{person}{Chrysovalantis Gaganis}.}
  \bibinfo{year}{2009}\natexlab{}.
\newblock \showarticletitle{Classification techniques for the identification of
  falsified financial statements: a comparative analysis}.
\newblock \bibinfo{journal}{\emph{Intelligent Systems in Accounting, Finance \&
  Management: International Journal}} \bibinfo{volume}{16}, \bibinfo{number}{3}
  (\bibinfo{year}{2009}), \bibinfo{pages}{207--229}.
\newblock


\bibitem[Gai et~al\mbox{.}(2018)]%
        {fintec_gov12309}
\bibfield{author}{\bibinfo{person}{Keke Gai}, \bibinfo{person}{Meikang Qiu},
  {and} \bibinfo{person}{Xiaotong Sun}.} \bibinfo{year}{2018}\natexlab{}.
\newblock \showarticletitle{A survey on FinTech}.
\newblock \bibinfo{journal}{\emph{Journal of Network and Computer
  Applications}}  \bibinfo{volume}{103} (\bibinfo{year}{2018}),
  \bibinfo{pages}{262--273}.
\newblock


\bibitem[Gainous et~al\mbox{.}(2019)]%
        {gainous20199traditional}
\bibfield{author}{\bibinfo{person}{Jason Gainous}, \bibinfo{person}{Jason~P
  Abbott}, {and} \bibinfo{person}{Kevin~M Wagner}.}
  \bibinfo{year}{2019}\natexlab{}.
\newblock \showarticletitle{Traditional versus internet media in a restricted
  information environment: How trust in the medium matters}.
\newblock \bibinfo{journal}{\emph{Political Behavior}}  \bibinfo{volume}{41}
  (\bibinfo{year}{2019}), \bibinfo{pages}{401--422}.
\newblock


\bibitem[Gamble et~al\mbox{.}(2014)]%
        {gamble2014ccauses}
\bibfield{author}{\bibinfo{person}{Keith~Jacks Gamble},
  \bibinfo{person}{Patricia Boyle}, \bibinfo{person}{Lei Yu}, {and}
  \bibinfo{person}{David Bennett}.} \bibinfo{year}{2014}\natexlab{}.
\newblock \showarticletitle{The causes and consequences of financial fraud
  among older Americans}.
\newblock \bibinfo{journal}{\emph{Boston College Center for Retirement Research
  WP}}  \bibinfo{volume}{13} (\bibinfo{year}{2014}).
\newblock


\bibitem[Garg et~al\mbox{.}(2003)]%
        {gargg2003quantifying}
\bibfield{author}{\bibinfo{person}{Ashish Garg}, \bibinfo{person}{Jeffrey
  Curtis}, {and} \bibinfo{person}{Hilary Halper}.}
  \bibinfo{year}{2003}\natexlab{}.
\newblock \showarticletitle{Quantifying the financial impact of IT security
  breaches}.
\newblock \bibinfo{journal}{\emph{Information Management \& Computer Security}}
  \bibinfo{volume}{11}, \bibinfo{number}{2} (\bibinfo{year}{2003}),
  \bibinfo{pages}{74--83}.
\newblock


\bibitem[Garman and Forgue(2014)]%
        {garman2014personal}
\bibfield{author}{\bibinfo{person}{E~Thomas Garman} {and}
  \bibinfo{person}{Raymond Forgue}.} \bibinfo{year}{2014}\natexlab{}.
\newblock \bibinfo{booktitle}{\emph{Personal finance}}.
\newblock \bibinfo{publisher}{Cengage Learning}.
\newblock


\bibitem[Gastwirth(1977)]%
        {gastwirth1977probability}
\bibfield{author}{\bibinfo{person}{Joseph~L Gastwirth}.}
  \bibinfo{year}{1977}\natexlab{}.
\newblock \showarticletitle{A probability model of a pyramid scheme}.
\newblock \bibinfo{journal}{\emph{The American Statistician}}
  \bibinfo{volume}{31}, \bibinfo{number}{2} (\bibinfo{year}{1977}),
  \bibinfo{pages}{79--82}.
\newblock


\bibitem[Gastwirth and Bhattacharya(1984)]%
        {gastwirth1984two}
\bibfield{author}{\bibinfo{person}{Joseph~L Gastwirth} {and}
  \bibinfo{person}{PK Bhattacharya}.} \bibinfo{year}{1984}\natexlab{}.
\newblock \showarticletitle{Two probability models of pyramid or chain letter
  schemes demonstrating that their promotional claims are unreliable}.
\newblock \bibinfo{journal}{\emph{Operations Research}} \bibinfo{volume}{32},
  \bibinfo{number}{3} (\bibinfo{year}{1984}), \bibinfo{pages}{527--536}.
\newblock


\bibitem[Glaeser and Ujhelyi(2010)]%
        {royalsociety_2022}
\bibfield{author}{\bibinfo{person}{Edward~L Glaeser} {and}
  \bibinfo{person}{Gergely Ujhelyi}.} \bibinfo{year}{2010}\natexlab{}.
\newblock \showarticletitle{Regulating misinformation}.
\newblock \bibinfo{journal}{\emph{Journal of public Economics}}
  \bibinfo{volume}{94}, \bibinfo{number}{3-4} (\bibinfo{year}{2010}),
  \bibinfo{pages}{247--257}.
\newblock


\bibitem[Green and Choi(1997)]%
        {green1997assessing}
\bibfield{author}{\bibinfo{person}{Brian~Patrick Green} {and}
  \bibinfo{person}{Jae~Hwa Choi}.} \bibinfo{year}{1997}\natexlab{}.
\newblock \showarticletitle{Assessing the risk of management fraud through
  neural network technology}.
\newblock \bibinfo{journal}{\emph{Auditing}}  \bibinfo{volume}{16}
  (\bibinfo{year}{1997}), \bibinfo{pages}{14--28}.
\newblock


\bibitem[Griffin~III(2022)]%
        {griffin02022promises}
\bibfield{author}{\bibinfo{person}{O~Hayden Griffin~III}.}
  \bibinfo{year}{2022}\natexlab{}.
\newblock \showarticletitle{Promises, deceit and white-collar criminality
  within the Theranos scandal}.
\newblock \bibinfo{journal}{\emph{Journal of White Collar and Corporate Crime}}
  \bibinfo{volume}{3}, \bibinfo{number}{2} (\bibinfo{year}{2022}),
  \bibinfo{pages}{109--121}.
\newblock


\bibitem[Harris and Tayler(2019)]%
        {harvard_business_review_2019001283}
\bibfield{author}{\bibinfo{person}{Michael Harris} {and} \bibinfo{person}{Bill
  Tayler}.} \bibinfo{year}{2019}\natexlab{}.
\newblock \showarticletitle{Don't let metrics undermine your business: an
  obsession with the numbers can sink your strategy}.
\newblock \bibinfo{journal}{\emph{Harvard Business Review}}
  \bibinfo{volume}{97}, \bibinfo{number}{5} (\bibinfo{year}{2019}),
  \bibinfo{pages}{62--70}.
\newblock


\bibitem[Harrison and Leopold(2021)]%
        {harvard_business_review_2021aaa}
\bibfield{author}{\bibinfo{person}{Kathryn Harrison} {and}
  \bibinfo{person}{Amelia Leopold}.} \bibinfo{year}{2021}\natexlab{}.
\newblock \showarticletitle{How blockchain can help combat disinformation}.
\newblock \bibinfo{journal}{\emph{Harvard Business Review}}
  (\bibinfo{year}{2021}).
\newblock


\bibitem[Hasso et~al\mbox{.}(2022)]%
        {hasso2022participated}
\bibfield{author}{\bibinfo{person}{Tim Hasso}, \bibinfo{person}{Daniel
  M{\"u}ller}, \bibinfo{person}{Matthias Pelster}, {and} \bibinfo{person}{Sonja
  Warkulat}.} \bibinfo{year}{2022}\natexlab{}.
\newblock \showarticletitle{Who participated in the GameStop frenzy? Evidence
  from brokerage accounts}.
\newblock \bibinfo{journal}{\emph{Finance Research Letters}}
  \bibinfo{volume}{45} (\bibinfo{year}{2022}), \bibinfo{pages}{102140}.
\newblock


\bibitem[Heady(2000)]%
        {headdyy2000finance}
\bibfield{author}{\bibinfo{person}{Robert~K Heady}.}
  \bibinfo{year}{2000}\natexlab{}.
\newblock \showarticletitle{FINANCE: Beware of Credit Repair Outfits}.
\newblock \bibinfo{journal}{\emph{Network Journal}} \bibinfo{volume}{7},
  \bibinfo{number}{7} (\bibinfo{year}{2000}), \bibinfo{pages}{28}.
\newblock


\bibitem[Hebert et~al\mbox{.}(2023)]%
        {hebert_hernandez_perkins_puig_tressler_ftc_2023}
\bibfield{author}{\bibinfo{person}{Amy Hebert}, \bibinfo{person}{Alesha
  Hernandez}, \bibinfo{person}{Rhonda Perkins}, \bibinfo{person}{Alvaro Puig},
  \bibinfo{person}{Colleen Tressler}, {and} \bibinfo{person}{Ftc}.}
  \bibinfo{year}{2023}\natexlab{}.
\newblock \showarticletitle{It's Financial Literacy Month: Learn how to keep
  your money safe from scammers}.
\newblock \bibinfo{journal}{\emph{Consumer Advice}} (\bibinfo{date}{Apr}
  \bibinfo{year}{2023}).
\newblock
\urldef\tempurl%
\url{https://consumer.ftc.gov/consumer-alerts/2023/04/its-financial-literacy-month-learn-how-keep-your-money-safe-scammers}
\showURL{%
\tempurl}


\bibitem[Hidajat et~al\mbox{.}(2020)]%
        {hidajat2020people}
\bibfield{author}{\bibinfo{person}{Taofik Hidajat}, \bibinfo{person}{Ina
  Primiana}, \bibinfo{person}{Sulaeman Rahman}, {and} \bibinfo{person}{Erie
  Febrian}.} \bibinfo{year}{2020}\natexlab{}.
\newblock \showarticletitle{Why are people trapped in Ponzi and pyramid
  schemes?}
\newblock \bibinfo{journal}{\emph{Journal of Financial Crime}}
  \bibinfo{volume}{28}, \bibinfo{number}{1} (\bibinfo{year}{2020}),
  \bibinfo{pages}{187--203}.
\newblock


\bibitem[Hossain et~al\mbox{.}(2020)]%
        {hossain2020banfakenews}
\bibfield{author}{\bibinfo{person}{Md~Zobaer Hossain},
  \bibinfo{person}{Md~Ashraful Rahman}, \bibinfo{person}{Md~Saiful Islam},
  {and} \bibinfo{person}{Sudipta Kar}.} \bibinfo{year}{2020}\natexlab{}.
\newblock \showarticletitle{Banfakenews: A dataset for detecting fake news in
  bangla}.
\newblock \bibinfo{journal}{\emph{arXiv preprint arXiv:2004.08789}}
  (\bibinfo{year}{2020}).
\newblock


\bibitem[Huang et~al\mbox{.}(2017)]%
        {huang_lin_chiu_yenxc_2016}
\bibfield{author}{\bibinfo{person}{Shaio~Yan Huang}, \bibinfo{person}{Chi-Chen
  Lin}, \bibinfo{person}{An-An Chiu}, {and} \bibinfo{person}{David~C Yen}.}
  \bibinfo{year}{2017}\natexlab{}.
\newblock \showarticletitle{Fraud detection using fraud triangle risk factors}.
\newblock \bibinfo{journal}{\emph{Information Systems Frontiers}}
  \bibinfo{volume}{19} (\bibinfo{year}{2017}), \bibinfo{pages}{1343--1356}.
\newblock


\bibitem[Huberman and Regev(2001)]%
        {HUBERMAN_REGEV2004}
\bibfield{author}{\bibinfo{person}{Gur Huberman} {and} \bibinfo{person}{Tomer
  Regev}.} \bibinfo{year}{2001}\natexlab{}.
\newblock \showarticletitle{Contagious speculation and a cure for cancer: A
  nonevent that made stock prices soar}.
\newblock \bibinfo{journal}{\emph{The Journal of Finance}}
  \bibinfo{volume}{56}, \bibinfo{number}{1} (\bibinfo{year}{2001}),
  \bibinfo{pages}{387--396}.
\newblock


\bibitem[Jahanbakhsh et~al\mbox{.}(2023)]%
        {jahanbakhsh2023exploring}
\bibfield{author}{\bibinfo{person}{Farnaz Jahanbakhsh}, \bibinfo{person}{Yannis
  Katsis}, \bibinfo{person}{Dakuo Wang}, \bibinfo{person}{Lucian Popa}, {and}
  \bibinfo{person}{Michael Muller}.} \bibinfo{year}{2023}\natexlab{}.
\newblock \showarticletitle{Exploring the Use of Personalized AI for
  Identifying Misinformation on Social Media}. In
  \bibinfo{booktitle}{\emph{Proceedings of the 2023 CHI Conference on Human
  Factors in Computing Systems}}. \bibinfo{pages}{1--27}.
\newblock


\bibitem[Jarvis(2000)]%
        {jarvis200000rise}
\bibfield{author}{\bibinfo{person}{Chris Jarvis}.}
  \bibinfo{year}{2000}\natexlab{}.
\newblock \showarticletitle{The rise and fall of the pyramid schemes in
  Albania}.
\newblock \bibinfo{journal}{\emph{IMF staff papers}} \bibinfo{volume}{47},
  \bibinfo{number}{1} (\bibinfo{year}{2000}), \bibinfo{pages}{1--29}.
\newblock


\bibitem[John and Narayanan(1997)]%
        {john1997kmarket}
\bibfield{author}{\bibinfo{person}{Kose John} {and} \bibinfo{person}{Ranga
  Narayanan}.} \bibinfo{year}{1997}\natexlab{}.
\newblock \showarticletitle{Market manipulation and the role of insider trading
  regulations}.
\newblock \bibinfo{journal}{\emph{The Journal of Business}}
  \bibinfo{volume}{70}, \bibinfo{number}{2} (\bibinfo{year}{1997}),
  \bibinfo{pages}{217--247}.
\newblock


\bibitem[Jureviene and Ivanova(2013)]%
        {Jureviien2013BehaviouralFTI}
\bibfield{author}{\bibinfo{person}{Daiva Jureviene} {and}
  \bibinfo{person}{Olga~Yu. Ivanova}.} \bibinfo{year}{2013}\natexlab{}.
\newblock \showarticletitle{Behavioural Finance: Theory and Survey}.
\newblock \bibinfo{journal}{\emph{Mokslas - Lietuvos Ateitis}}
  \bibinfo{volume}{5} (\bibinfo{year}{2013}), \bibinfo{pages}{53--58}.
\newblock


\bibitem[Kahneman(2011)]%
        {kahneman2011thinking}
\bibfield{author}{\bibinfo{person}{Daniel Kahneman}.}
  \bibinfo{year}{2011}\natexlab{}.
\newblock \bibinfo{booktitle}{\emph{Thinking, fast and slow}}.
\newblock \bibinfo{publisher}{Farrar, Straus and Giroux}.
\newblock


\bibitem[Kaminski et~al\mbox{.}(2004)]%
        {kaminski2004can01}
\bibfield{author}{\bibinfo{person}{Kathleen~A Kaminski}, \bibinfo{person}{T
  Sterling~Wetzel}, {and} \bibinfo{person}{Liming Guan}.}
  \bibinfo{year}{2004}\natexlab{}.
\newblock \showarticletitle{Can financial ratios detect fraudulent financial
  reporting?}
\newblock \bibinfo{journal}{\emph{Managerial Auditing Journal}}
  \bibinfo{volume}{19}, \bibinfo{number}{1} (\bibinfo{year}{2004}),
  \bibinfo{pages}{15--28}.
\newblock


\bibitem[Kar et~al\mbox{.}(2022)]%
        {Kar2022HowDMII}
\bibfield{author}{\bibinfo{person}{Arpan~Kumar Kar},
  \bibinfo{person}{Shalini~Nath Tripathi}, \bibinfo{person}{Nishtha Malik},
  \bibinfo{person}{Shivam Gupta}, {and} \bibinfo{person}{Uthayasankar
  Sivarajah}.} \bibinfo{year}{2022}\natexlab{}.
\newblock \showarticletitle{How Does Misinformation and Capricious Opinions
  Impact the Supply Chain - A Study on the Impacts During the Pandemic}.
\newblock \bibinfo{journal}{\emph{Annals of Operations Research}}
  (\bibinfo{year}{2022}), \bibinfo{pages}{1 -- 22}.
\newblock


\bibitem[Keep and Vander~Nat(2014)]%
        {keep02014multilevel}
\bibfield{author}{\bibinfo{person}{William~W Keep} {and}
  \bibinfo{person}{Peter~J Vander~Nat}.} \bibinfo{year}{2014}\natexlab{}.
\newblock \showarticletitle{Multilevel marketing and pyramid schemes in the
  United States: An historical analysis}.
\newblock \bibinfo{journal}{\emph{Journal of Historical Research in Marketing}}
  (\bibinfo{year}{2014}).
\newblock


\bibitem[Kieffer and Mottola(2017)]%
        {finraorg_2021}
\bibfield{author}{\bibinfo{person}{Christine Kieffer} {and}
  \bibinfo{person}{Gary Mottola}.} \bibinfo{year}{2017}\natexlab{}.
\newblock \showarticletitle{Understanding and combating investment fraud}.
\newblock \bibinfo{journal}{\emph{Financial decision making and retirement
  security in an aging world}}  \bibinfo{volume}{185} (\bibinfo{year}{2017}).
\newblock


\bibitem[Kim et~al\mbox{.}(2019)]%
        {kim_moravec_dennis_2019ass}
\bibfield{author}{\bibinfo{person}{Antino Kim}, \bibinfo{person}{Patricia~L
  Moravec}, {and} \bibinfo{person}{Alan~R Dennis}.}
  \bibinfo{year}{2019}\natexlab{}.
\newblock \showarticletitle{Combating fake news on social media with source
  ratings: The effects of user and expert reputation ratings}.
\newblock \bibinfo{journal}{\emph{Journal of Management Information Systems}}
  \bibinfo{volume}{36}, \bibinfo{number}{3} (\bibinfo{year}{2019}),
  \bibinfo{pages}{931--968}.
\newblock


\bibitem[Kim and Mei(1994)]%
        {Kim_Mei_1999}
\bibfield{author}{\bibinfo{person}{Harold~Y Kim} {and}
  \bibinfo{person}{Jianping Mei}.} \bibinfo{year}{1994}\natexlab{}.
\newblock \showarticletitle{Political risk and stock returns: The case of Hong
  Kong}.
\newblock  (\bibinfo{year}{1994}).
\newblock


\bibitem[Kirilenko et~al\mbox{.}(2019)]%
        {kirilenko2019regulating}
\bibfield{author}{\bibinfo{person}{Andrei~A Kirilenko},
  \bibinfo{person}{Margaret~P Kyle}, \bibinfo{person}{Mehrzad Samadi}, {and}
  \bibinfo{person}{Tayfun Tuzun}.} \bibinfo{year}{2019}\natexlab{}.
\newblock \showarticletitle{Regulating fintech: Lessons from early policy
  experiments}.
\newblock \bibinfo{journal}{\emph{The Review of Financial Studies}}
  \bibinfo{volume}{32}, \bibinfo{number}{5} (\bibinfo{year}{2019}),
  \bibinfo{pages}{1868--1912}.
\newblock
\urldef\tempurl%
\url{https://doi.org/10.1093/rfs/hhz058}
\showDOI{\tempurl}


\bibitem[Kirkos et~al\mbox{.}(2007)]%
        {kirkos2007data01}
\bibfield{author}{\bibinfo{person}{Efstathios Kirkos},
  \bibinfo{person}{Charalambos Spathis}, {and} \bibinfo{person}{Yannis
  Manolopoulos}.} \bibinfo{year}{2007}\natexlab{}.
\newblock \showarticletitle{Data mining techniques for the detection of
  fraudulent financial statements}.
\newblock \bibinfo{journal}{\emph{Expert systems with applications}}
  \bibinfo{volume}{32}, \bibinfo{number}{4} (\bibinfo{year}{2007}),
  \bibinfo{pages}{995--1003}.
\newblock


\bibitem[Knight and Knight(1992)]%
        {knightth1992criminal}
\bibfield{author}{\bibinfo{person}{Ray~A Knight} {and} \bibinfo{person}{Lee~G
  Knight}.} \bibinfo{year}{1992}\natexlab{}.
\newblock \showarticletitle{Criminal Tax Fraud: An Analytical Review}.
\newblock \bibinfo{journal}{\emph{Mo. L. REv.}}  \bibinfo{volume}{57}
  (\bibinfo{year}{1992}), \bibinfo{pages}{175}.
\newblock


\bibitem[Kogan et~al\mbox{.}(2020)]%
        {kogan2020fake}
\bibfield{author}{\bibinfo{person}{Shimon Kogan}, \bibinfo{person}{Tobias~J
  Moskowitz}, {and} \bibinfo{person}{Marina Niessner}.}
  \bibinfo{year}{2020}\natexlab{}.
\newblock \bibinfo{booktitle}{\emph{Fake news in financial markets}}.
\newblock \bibinfo{publisher}{SSRN}.
\newblock


\bibitem[Kogan et~al\mbox{.}(2022)]%
        {10.1093/rof/rfac058}
\bibfield{author}{\bibinfo{person}{Shimon Kogan}, \bibinfo{person}{Tobias~J
  Moskowitz}, {and} \bibinfo{person}{Marina Niessner}.}
  \bibinfo{year}{2022}\natexlab{}.
\newblock \showarticletitle{{Social Media and Financial News Manipulation*}}.
\newblock \bibinfo{journal}{\emph{Review of Finance}} (\bibinfo{date}{10}
  \bibinfo{year}{2022}).
\newblock
\showISSN{1572-3097}
\urldef\tempurl%
\url{https://doi.org/10.1093/rof/rfac058}
\showDOI{\tempurl}
\showeprint{https://academic.oup.com/rof/advance-article-pdf/doi/10.1093/rof/rfac058/46695048/rfac058.pdf}
\newblock
\shownote{rfac058}.


\bibitem[Lappas et~al\mbox{.}(2016)]%
        {lappas_sabnis_valkanas_2016as}
\bibfield{author}{\bibinfo{person}{Theodoros Lappas}, \bibinfo{person}{Gaurav
  Sabnis}, {and} \bibinfo{person}{Georgios Valkanas}.}
  \bibinfo{year}{2016}\natexlab{}.
\newblock \showarticletitle{The impact of fake reviews on online visibility: A
  vulnerability assessment of the hotel industry}.
\newblock \bibinfo{journal}{\emph{Information Systems Research}}
  \bibinfo{volume}{27}, \bibinfo{number}{4} (\bibinfo{year}{2016}),
  \bibinfo{pages}{940--961}.
\newblock


\bibitem[Lawrence(2021)]%
        {lawrence2021mmisleading}
\bibfield{author}{\bibinfo{person}{Adewale Lawrence}.}
  \bibinfo{year}{2021}\natexlab{}.
\newblock \showarticletitle{Misleading Advertisement and its regulation in the
  EU Medicine Promotion legal framework}.
\newblock \bibinfo{journal}{\emph{ScienceOpen Preprints}}
  (\bibinfo{year}{2021}).
\newblock


\bibitem[Lennox et~al\mbox{.}(2013)]%
        {lennoxxxon2013tax}
\bibfield{author}{\bibinfo{person}{Clive Lennox}, \bibinfo{person}{Petro
  Lisowsky}, {and} \bibinfo{person}{Jeffrey Pittman}.}
  \bibinfo{year}{2013}\natexlab{}.
\newblock \showarticletitle{Tax aggressiveness and accounting fraud}.
\newblock \bibinfo{journal}{\emph{Journal of accounting research}}
  \bibinfo{volume}{51}, \bibinfo{number}{4} (\bibinfo{year}{2013}),
  \bibinfo{pages}{739--778}.
\newblock


\bibitem[Leonard and Loonin(2019)]%
        {leonarddf2019credit}
\bibfield{author}{\bibinfo{person}{Attorneys~Robin Leonard} {and}
  \bibinfo{person}{Deanne Loonin}.} \bibinfo{year}{2019}\natexlab{}.
\newblock \bibinfo{booktitle}{\emph{Credit Repair}}.
\newblock


\bibitem[Lewandowsky et~al\mbox{.}(2012)]%
        {lewandowsky2012misinformation}
\bibfield{author}{\bibinfo{person}{Stephan Lewandowsky},
  \bibinfo{person}{Ullrich~K Ecker}, \bibinfo{person}{Colleen~M Seifert},
  \bibinfo{person}{Norbert Schwarz}, {and} \bibinfo{person}{John Cook}.}
  \bibinfo{year}{2012}\natexlab{}.
\newblock \showarticletitle{Misinformation and its correction: Continued
  influence and successful debiasing}.
\newblock \bibinfo{journal}{\emph{Psychological science in the public
  interest}} \bibinfo{volume}{13}, \bibinfo{number}{3} (\bibinfo{year}{2012}),
  \bibinfo{pages}{106--131}.
\newblock


\bibitem[Li et~al\mbox{.}(2023)]%
        {li2023doeeoes}
\bibfield{author}{\bibinfo{person}{Dun Li}, \bibinfo{person}{Dezhi Han},
  \bibinfo{person}{Zibin Zheng}, \bibinfo{person}{Tien-Hsiung Weng},
  \bibinfo{person}{Kuan-Ching Li}, \bibinfo{person}{Ming Li}, {and}
  \bibinfo{person}{Shaokang Cai}.} \bibinfo{year}{2023}\natexlab{}.
\newblock \showarticletitle{Does Short-and-Distort Scheme Really Exist? A
  Bitcoin Futures Audit Scheme through BIRCH \& BPNN Approach}.
\newblock \bibinfo{journal}{\emph{Computational Economics}}
  (\bibinfo{year}{2023}), \bibinfo{pages}{1--23}.
\newblock


\bibitem[Li et~al\mbox{.}(2020)]%
        {li2020multii}
\bibfield{author}{\bibinfo{person}{Qian Li}, \bibinfo{person}{Qingyuan Hu},
  \bibinfo{person}{Youshui Lu}, \bibinfo{person}{Yue Yang}, {and}
  \bibinfo{person}{Jingxian Cheng}.} \bibinfo{year}{2020}\natexlab{}.
\newblock \showarticletitle{Multi-level word features based on CNN for fake
  news detection in cultural communication}.
\newblock \bibinfo{journal}{\emph{Personal and Ubiquitous Computing}}
  \bibinfo{volume}{24} (\bibinfo{year}{2020}), \bibinfo{pages}{259--272}.
\newblock


\bibitem[Li(2010)]%
        {lii2010case}
\bibfield{author}{\bibinfo{person}{Yuhao Li}.} \bibinfo{year}{2010}\natexlab{}.
\newblock \showarticletitle{The case analysis of the scandal of Enron}.
\newblock \bibinfo{journal}{\emph{International Journal of business and
  management}} \bibinfo{volume}{5}, \bibinfo{number}{10}
  (\bibinfo{year}{2010}), \bibinfo{pages}{37}.
\newblock


\bibitem[Lin et~al\mbox{.}(2003)]%
        {lin2003fuzzy}
\bibfield{author}{\bibinfo{person}{Jerry~W Lin}, \bibinfo{person}{Mark~I
  Hwang}, {and} \bibinfo{person}{Jack~D Becker}.}
  \bibinfo{year}{2003}\natexlab{}.
\newblock \showarticletitle{A fuzzy neural network for assessing the risk of
  fraudulent financial reporting}.
\newblock \bibinfo{journal}{\emph{Managerial Auditing Journal}}
  (\bibinfo{year}{2003}).
\newblock


\bibitem[Liu and Moss(2022)]%
        {liu2022role02}
\bibfield{author}{\bibinfo{person}{Betty Liu} {and} \bibinfo{person}{Austin
  Moss}.} \bibinfo{year}{2022}\natexlab{}.
\newblock \showarticletitle{The Role of Accounting Information in an Era of
  Fake News}.
\newblock \bibinfo{journal}{\emph{Available at SSRN 4399543}}
  (\bibinfo{year}{2022}).
\newblock


\bibitem[Liu et~al\mbox{.}(2020)]%
        {liu2020financial}
\bibfield{author}{\bibinfo{person}{Cong Liu}, \bibinfo{person}{Jian Huang},
  {and} \bibinfo{person}{Li Zhao}.} \bibinfo{year}{2020}\natexlab{}.
\newblock \showarticletitle{Financial education, literacy and fraud: Evidence
  from a randomized controlled trial in China}.
\newblock \bibinfo{journal}{\emph{Journal of Banking \& Finance}}
  \bibinfo{volume}{110} (\bibinfo{year}{2020}), \bibinfo{pages}{105710}.
\newblock
\urldef\tempurl%
\url{https://doi.org/10.1016/j.jbankfin.2019.105710}
\showDOI{\tempurl}


\bibitem[Liu et~al\mbox{.}(2023)]%
        {liu2023tracking118}
\bibfield{author}{\bibinfo{person}{Rong Liu}, \bibinfo{person}{Jujun Huang},
  {and} \bibinfo{person}{Zhongju Zhang}.} \bibinfo{year}{2023}\natexlab{}.
\newblock \showarticletitle{Tracking disclosure change trajectories for
  financial fraud detection}.
\newblock \bibinfo{journal}{\emph{Production and Operations Management}}
  \bibinfo{volume}{32}, \bibinfo{number}{2} (\bibinfo{year}{2023}),
  \bibinfo{pages}{584--602}.
\newblock


\bibitem[Lopez-Gonzalez and Griffiths(2018)]%
        {lopezz2018betting}
\bibfield{author}{\bibinfo{person}{Hibai Lopez-Gonzalez} {and}
  \bibinfo{person}{Mark~D Griffiths}.} \bibinfo{year}{2018}\natexlab{}.
\newblock \showarticletitle{Betting, forex trading, and fantasy gaming
  sponsorships—a responsible marketing inquiry into the
  ‘gamblification’of English football}.
\newblock \bibinfo{journal}{\emph{International Journal of Mental Health and
  Addiction}}  \bibinfo{volume}{16} (\bibinfo{year}{2018}),
  \bibinfo{pages}{404--419}.
\newblock


\bibitem[Lu et~al\mbox{.}(2022)]%
        {lu2022eheffects}
\bibfield{author}{\bibinfo{person}{Zhuoran Lu}, \bibinfo{person}{Patrick Li},
  \bibinfo{person}{Weilong Wang}, {and} \bibinfo{person}{Ming Yin}.}
  \bibinfo{year}{2022}\natexlab{}.
\newblock \showarticletitle{The Effects of AI-based Credibility Indicators on
  the Detection and Spread of Misinformation under Social Influence}.
\newblock \bibinfo{journal}{\emph{Proceedings of the ACM on Human-Computer
  Interaction}} \bibinfo{volume}{6}, \bibinfo{number}{CSCW2}
  (\bibinfo{year}{2022}), \bibinfo{pages}{1--27}.
\newblock


\bibitem[Lucangeli(2021)]%
        {lucangeli2021gamestop}
\bibfield{author}{\bibinfo{person}{Giorgio Lucangeli}.}
  \bibinfo{year}{2021}\natexlab{}.
\newblock \showarticletitle{The GameStop case: how a reddit group shocked the
  stock market}.
\newblock  (\bibinfo{year}{2021}).
\newblock


\bibitem[Malz(2021)]%
        {university89937337}
\bibfield{author}{\bibinfo{person}{Allan~M Malz}.}
  \bibinfo{year}{2021}\natexlab{}.
\newblock \showarticletitle{The GameStop Episode: What Happened and What Does
  It Mean?}
\newblock \bibinfo{journal}{\emph{Journal of Applied Corporate Finance}}
  \bibinfo{volume}{33}, \bibinfo{number}{4} (\bibinfo{year}{2021}),
  \bibinfo{pages}{87--97}.
\newblock


\bibitem[Markham(2015)]%
        {markham2015financial}
\bibfield{author}{\bibinfo{person}{Jerry~W Markham}.}
  \bibinfo{year}{2015}\natexlab{}.
\newblock \bibinfo{booktitle}{\emph{A financial history of modern US corporate
  scandals: From Enron to reform}}.
\newblock \bibinfo{publisher}{Routledge}.
\newblock


\bibitem[Marta~Lazo and Farias~Batlle(2019)]%
        {marta2019information09}
\bibfield{author}{\bibinfo{person}{Carmen Marta~Lazo} {and}
  \bibinfo{person}{Pedro Farias~Batlle}.} \bibinfo{year}{2019}\natexlab{}.
\newblock \showarticletitle{Information quality and trust: From traditional
  media to cybermedia}.
\newblock \bibinfo{journal}{\emph{Communication: Innovation \& Quality}}
  (\bibinfo{year}{2019}), \bibinfo{pages}{185--206}.
\newblock


\bibitem[Mazar et~al\mbox{.}(2008)]%
        {doi:10.1509/jmkr.45.6.633}
\bibfield{author}{\bibinfo{person}{Nina Mazar}, \bibinfo{person}{On Amir},
  {and} \bibinfo{person}{Dan Ariely}.} \bibinfo{year}{2008}\natexlab{}.
\newblock \showarticletitle{The Dishonesty of Honest People: A Theory of
  Self-Concept Maintenance}.
\newblock \bibinfo{journal}{\emph{Journal of Marketing Research}}
  \bibinfo{volume}{45}, \bibinfo{number}{6} (\bibinfo{year}{2008}),
  \bibinfo{pages}{633--644}.
\newblock
\urldef\tempurl%
\url{https://doi.org/10.1509/jmkr.45.6.633}
\showDOI{\tempurl}
\showeprint{https://doi.org/10.1509/jmkr.45.6.633}


\bibitem[McBarnet(1991)]%
        {mcbarnet1991whiter}
\bibfield{author}{\bibinfo{person}{Doreen McBarnet}.}
  \bibinfo{year}{1991}\natexlab{}.
\newblock \showarticletitle{Whiter than white collar crime: tax, fraud
  insurance and the management of stigma}.
\newblock \bibinfo{journal}{\emph{British Journal of Sociology}}
  (\bibinfo{year}{1991}), \bibinfo{pages}{323--344}.
\newblock


\bibitem[Mendes et~al\mbox{.}(2012)]%
        {mendes2012fforex}
\bibfield{author}{\bibinfo{person}{Lu{\'\i}s Mendes}, \bibinfo{person}{Pedro
  Godinho}, {and} \bibinfo{person}{Joana Dias}.}
  \bibinfo{year}{2012}\natexlab{}.
\newblock \showarticletitle{A Forex trading system based on a genetic
  algorithm}.
\newblock \bibinfo{journal}{\emph{Journal of Heuristics}}  \bibinfo{volume}{18}
  (\bibinfo{year}{2012}), \bibinfo{pages}{627--656}.
\newblock


\bibitem[Metzger and Flanagin(2007)]%
        {metzger2007making}
\bibfield{author}{\bibinfo{person}{Miriam~J Metzger} {and}
  \bibinfo{person}{Andrew~J Flanagin}.} \bibinfo{year}{2007}\natexlab{}.
\newblock \showarticletitle{Making sense of credibility on the web: Models for
  evaluating online information and recommendations for future research}.
\newblock \bibinfo{journal}{\emph{Journal of the American Society for
  Information Science and Technology}} \bibinfo{volume}{58},
  \bibinfo{number}{13} (\bibinfo{year}{2007}), \bibinfo{pages}{2078--2091}.
\newblock


\bibitem[Mindell(1961)]%
        {how_news_mar110982}
\bibfield{author}{\bibinfo{person}{Joseph Mindell}.}
  \bibinfo{year}{1961}\natexlab{}.
\newblock \showarticletitle{How news affects market trends}.
\newblock \bibinfo{journal}{\emph{Financial Analysts Journal}}
  \bibinfo{volume}{17}, \bibinfo{number}{1} (\bibinfo{year}{1961}),
  \bibinfo{pages}{31--34}.
\newblock


\bibitem[Mitchell and Mulherin(1994)]%
        {Mitchell_Mulherin_impact}
\bibfield{author}{\bibinfo{person}{Mark~L Mitchell} {and}
  \bibinfo{person}{J~Harold Mulherin}.} \bibinfo{year}{1994}\natexlab{}.
\newblock \showarticletitle{The impact of public information on the stock
  market}.
\newblock \bibinfo{journal}{\emph{The Journal of Finance}}
  \bibinfo{volume}{49}, \bibinfo{number}{3} (\bibinfo{year}{1994}),
  \bibinfo{pages}{923--950}.
\newblock


\bibitem[Mitts(2020)]%
        {mitts2020short}
\bibfield{author}{\bibinfo{person}{Joshua Mitts}.}
  \bibinfo{year}{2020}\natexlab{}.
\newblock \showarticletitle{Short and distort}.
\newblock \bibinfo{journal}{\emph{The Journal of Legal Studies}}
  \bibinfo{volume}{49}, \bibinfo{number}{2} (\bibinfo{year}{2020}),
  \bibinfo{pages}{287--334}.
\newblock


\bibitem[Mohankumar et~al\mbox{.}(2023)]%
        {mohankumar2023financial01}
\bibfield{author}{\bibinfo{person}{Padmapriya Mohankumar},
  \bibinfo{person}{Ashraf Kamal}, \bibinfo{person}{Vishal~Kumar Singh}, {and}
  \bibinfo{person}{Amrish Satish}.} \bibinfo{year}{2023}\natexlab{}.
\newblock \showarticletitle{Financial Fake News Detection via Context-Aware
  Embedding and Sequential Representation using Cross-Joint Networks}. In
  \bibinfo{booktitle}{\emph{2023 15th International Conference on COMmunication
  Systems \& NETworkS (COMSNETS)}}. IEEE, \bibinfo{pages}{780--784}.
\newblock


\bibitem[Mohd~Padil et~al\mbox{.}(2022)]%
        {MohdPadil2022}
\bibfield{author}{\bibinfo{person}{Hazlina Mohd~Padil},
  \bibinfo{person}{Eley~Suzana Kasim}, \bibinfo{person}{Salwa Muda},
  \bibinfo{person}{Norhidayah Ismail}, {and} \bibinfo{person}{Norlaila
  Md~Zin}.} \bibinfo{year}{2022}\natexlab{}.
\newblock \showarticletitle{Financial literacy and awareness of investment
  scams among University students}.
\newblock \bibinfo{journal}{\emph{Journal of Financial Crime}}
  \bibinfo{volume}{29}, \bibinfo{number}{1} (\bibinfo{date}{01 Jan}
  \bibinfo{year}{2022}), \bibinfo{pages}{355--367}.
\newblock
\showISSN{1359-0790}
\urldef\tempurl%
\url{https://doi.org/10.1108/JFC-01-2021-0012}
\showDOI{\tempurl}


\bibitem[Moore et~al\mbox{.}(2012)]%
        {moore20121postmodern}
\bibfield{author}{\bibinfo{person}{Tyler Moore}, \bibinfo{person}{Jie Han},
  {and} \bibinfo{person}{Richard Clayton}.} \bibinfo{year}{2012}\natexlab{}.
\newblock \showarticletitle{The postmodern Ponzi scheme: Empirical analysis of
  high-yield investment programs}. In \bibinfo{booktitle}{\emph{Financial
  Cryptography and Data Security: 16th International Conference, FC 2012,
  Kralendijk, Bonaire, Februray 27-March 2, 2012, Revised Selected Papers 16}}.
  Springer, \bibinfo{pages}{41--56}.
\newblock


\bibitem[Muhammed~T and Mathew(2022)]%
        {Muhammed_T2022}
\bibfield{author}{\bibinfo{person}{Sadiq Muhammed~T} {and}
  \bibinfo{person}{Saji~K Mathew}.} \bibinfo{year}{2022}\natexlab{}.
\newblock \showarticletitle{The disaster of misinformation: a review of
  research in social media}.
\newblock \bibinfo{journal}{\emph{Int J Data Sci Anal}} \bibinfo{volume}{13},
  \bibinfo{number}{4} (\bibinfo{date}{Feb.} \bibinfo{year}{2022}),
  \bibinfo{pages}{271--285}.
\newblock


\bibitem[Muncy(2004)]%
        {muncy2004ethicall}
\bibfield{author}{\bibinfo{person}{James~A Muncy}.}
  \bibinfo{year}{2004}\natexlab{}.
\newblock \showarticletitle{Ethical issues in multilevel marketing: Is it a
  legitimate business or just another pyramid scheme?}
\newblock \bibinfo{journal}{\emph{Marketing Education Review}}
  \bibinfo{volume}{14}, \bibinfo{number}{3} (\bibinfo{year}{2004}),
  \bibinfo{pages}{47--53}.
\newblock


\bibitem[Najee-Ullah et~al\mbox{.}(2021)]%
        {najee2021towards}
\bibfield{author}{\bibinfo{person}{Ahmad Najee-Ullah}, \bibinfo{person}{Luis
  Landeros}, \bibinfo{person}{Yaroslav Balytskyi}, {and}
  \bibinfo{person}{Sang-Yoon Chang}.} \bibinfo{year}{2021}\natexlab{}.
\newblock \showarticletitle{Towards detection of AI-generated texts and
  misinformation}. In \bibinfo{booktitle}{\emph{International Workshop on
  Socio-Technical Aspects in Security}}. Springer, \bibinfo{pages}{194--205}.
\newblock


\bibitem[Nan et~al\mbox{.}(2021)]%
        {nan2021mdfend}
\bibfield{author}{\bibinfo{person}{Qiong Nan}, \bibinfo{person}{Juan Cao},
  \bibinfo{person}{Yongchun Zhu}, \bibinfo{person}{Yanyan Wang}, {and}
  \bibinfo{person}{Jintao Li}.} \bibinfo{year}{2021}\natexlab{}.
\newblock \showarticletitle{MDFEND: Multi-domain fake news detection}. In
  \bibinfo{booktitle}{\emph{Proceedings of the 30th ACM International
  Conference on Information \& Knowledge Management}}.
  \bibinfo{pages}{3343--3347}.
\newblock


\bibitem[Nasdaq({[n.\,d.]})]%
        {publisher_nasdaq}
\bibfield{author}{\bibinfo{person}{Publisher Nasdaq}.}
  \bibinfo{year}{[n.\,d.]}\natexlab{}.
\newblock \showarticletitle{How does Social Media Influence Financial Markets?}
\newblock \bibinfo{journal}{\emph{Nasdaq}} (\bibinfo{year}{[n.\,d.]}).
\newblock
\urldef\tempurl%
\url{https://www.nasdaq.com/articles/how-does-social-media-influence-financial-markets-2019-10-14}
\showURL{%
\tempurl}


\bibitem[Nat and Keep(2002)]%
        {nat2002marketingg}
\bibfield{author}{\bibinfo{person}{Peter J~Vander Nat} {and}
  \bibinfo{person}{William~W Keep}.} \bibinfo{year}{2002}\natexlab{}.
\newblock \showarticletitle{Marketing fraud: An approach for differentiating
  multilevel marketing from pyramid schemes}.
\newblock \bibinfo{journal}{\emph{Journal of Public Policy \& Marketing}}
  \bibinfo{volume}{21}, \bibinfo{number}{1} (\bibinfo{year}{2002}),
  \bibinfo{pages}{139--151}.
\newblock


\bibitem[Neely and Weller(2013)]%
        {neelyy2013lessons}
\bibfield{author}{\bibinfo{person}{Christopher~J Neely} {and}
  \bibinfo{person}{Paul~A Weller}.} \bibinfo{year}{2013}\natexlab{}.
\newblock \showarticletitle{Lessons from the evolution of foreign exchange
  trading strategies}.
\newblock \bibinfo{journal}{\emph{Journal of Banking \& Finance}}
  \bibinfo{volume}{37}, \bibinfo{number}{10} (\bibinfo{year}{2013}),
  \bibinfo{pages}{3783--3798}.
\newblock


\bibitem[Nehf(1991)]%
        {nehfl1991legislative}
\bibfield{author}{\bibinfo{person}{James~P Nehf}.}
  \bibinfo{year}{1991}\natexlab{}.
\newblock \showarticletitle{Legislative Framework for Reducing Fraud in the
  Credit Repair Industry, A}.
\newblock \bibinfo{journal}{\emph{NCL Rev.}}  \bibinfo{volume}{70}
  (\bibinfo{year}{1991}), \bibinfo{pages}{781}.
\newblock


\bibitem[Ng et~al\mbox{.}(2021)]%
        {doi:10.1080/07421222.2021.1990612}
\bibfield{author}{\bibinfo{person}{Ka~Chung Ng}, \bibinfo{person}{Jie Tang},
  {and} \bibinfo{person}{Dongwon Lee}.} \bibinfo{year}{2021}\natexlab{}.
\newblock \showarticletitle{The Effect of Platform Intervention Policies on
  Fake News Dissemination and Survival: An Empirical Examination}.
\newblock \bibinfo{journal}{\emph{Journal of Management Information Systems}}
  \bibinfo{volume}{38}, \bibinfo{number}{4} (\bibinfo{year}{2021}),
  \bibinfo{pages}{898--930}.
\newblock
\urldef\tempurl%
\url{https://doi.org/10.1080/07421222.2021.1990612}
\showDOI{\tempurl}
\showeprint{https://doi.org/10.1080/07421222.2021.1990612}


\bibitem[Nguyen et~al\mbox{.}(2022)]%
        {nguyen2022mindsponge00}
\bibfield{author}{\bibinfo{person}{Minh-Hoang Nguyen}, \bibinfo{person}{Quy~Van
  Khuc}, \bibinfo{person}{Viet-Phuong La}, \bibinfo{person}{Tam-Tri Le},
  \bibinfo{person}{Quang-Loc Nguyen}, \bibinfo{person}{Ruining Jin},
  \bibinfo{person}{Phuong-Tri Nguyen}, {and} \bibinfo{person}{Quan-Hoang
  Vuong}.} \bibinfo{year}{2022}\natexlab{}.
\newblock \showarticletitle{Mindsponge-Based Reasoning of Households’
  Financial Resilience during the COVID-19 Crisis}.
\newblock \bibinfo{journal}{\emph{Journal of Risk and Financial Management}}
  \bibinfo{volume}{15}, \bibinfo{number}{11} (\bibinfo{year}{2022}),
  \bibinfo{pages}{542}.
\newblock


\bibitem[Nyhan and Reifler(2010)]%
        {nyhan2010corrections}
\bibfield{author}{\bibinfo{person}{Brendan Nyhan} {and} \bibinfo{person}{Jason
  Reifler}.} \bibinfo{year}{2010}\natexlab{}.
\newblock \showarticletitle{Corrections of misinformation through information
  and source cues}.
\newblock \bibinfo{journal}{\emph{Journal of politics}} \bibinfo{volume}{72},
  \bibinfo{number}{3} (\bibinfo{year}{2010}), \bibinfo{pages}{685--692}.
\newblock


\bibitem[OECD(2020)]%
        {oecdcovidd199123}
\bibfield{author}{\bibinfo{person}{OECD}.} \bibinfo{year}{2020}\natexlab{}.
\newblock \bibinfo{title}{Covid-19 and International Trade: Issues and
  Actions}.
\newblock
\newblock


\bibitem[Pennycook and Rand(2019)]%
        {pennycook2019falls}
\bibfield{author}{\bibinfo{person}{Gordon Pennycook} {and}
  \bibinfo{person}{David~G Rand}.} \bibinfo{year}{2019}\natexlab{}.
\newblock \showarticletitle{Who falls for fake news? The roles of bullshit
  receptivity, overclaiming, familiarity, and analytic thinking}.
\newblock \bibinfo{journal}{\emph{Journal of Personality}}
  \bibinfo{volume}{88}, \bibinfo{number}{2} (\bibinfo{year}{2019}),
  \bibinfo{pages}{185--200}.
\newblock
\urldef\tempurl%
\url{https://doi.org/10.1111/jopy.12394}
\showDOI{\tempurl}


\bibitem[Peretti(2008)]%
        {perretti2008data}
\bibfield{author}{\bibinfo{person}{Kimberly~Kiefer Peretti}.}
  \bibinfo{year}{2008}\natexlab{}.
\newblock \showarticletitle{Data breaches: what the underground world of
  carding reveals}.
\newblock \bibinfo{journal}{\emph{Santa Clara Computer \& High Tech. LJ}}
  \bibinfo{volume}{25} (\bibinfo{year}{2008}), \bibinfo{pages}{375}.
\newblock


\bibitem[Perols(2011)]%
        {perols2011ffinancial}
\bibfield{author}{\bibinfo{person}{Johan Perols}.}
  \bibinfo{year}{2011}\natexlab{}.
\newblock \showarticletitle{Financial statement fraud detection: An analysis of
  statistical and machine learning algorithms}.
\newblock \bibinfo{journal}{\emph{Auditing: A Journal of Practice \& Theory}}
  \bibinfo{volume}{30}, \bibinfo{number}{2} (\bibinfo{year}{2011}),
  \bibinfo{pages}{19--50}.
\newblock


\bibitem[Persons et~al\mbox{.}(1995)]%
        {persons1995using}
\bibfield{author}{\bibinfo{person}{Obeua~S Persons} {et~al\mbox{.}}}
  \bibinfo{year}{1995}\natexlab{}.
\newblock \showarticletitle{Using financial statement data to identify factors
  associated with fraudulent financial reporting}.
\newblock \bibinfo{journal}{\emph{Journal of Applied Business Research (JABR)}}
  \bibinfo{volume}{11}, \bibinfo{number}{3} (\bibinfo{year}{1995}),
  \bibinfo{pages}{38--46}.
\newblock


\bibitem[Petrick and Scherer(2003)]%
        {petrick2003enron}
\bibfield{author}{\bibinfo{person}{Joseph~A Petrick} {and}
  \bibinfo{person}{Robert~F Scherer}.} \bibinfo{year}{2003}\natexlab{}.
\newblock \showarticletitle{The Enron scandal and the neglect of management
  integrity capacity}.
\newblock \bibinfo{journal}{\emph{American Journal of Business}}
  \bibinfo{volume}{18}, \bibinfo{number}{1} (\bibinfo{year}{2003}),
  \bibinfo{pages}{37--50}.
\newblock


\bibitem[Polak(2012)]%
        {Polak2012THEMEA}
\bibfield{author}{\bibinfo{person}{Mateusz Polak}.}
  \bibinfo{year}{2012}\natexlab{}.
\newblock \showarticletitle{The misinformation effect in financial markets: An
  emerging issue in behavioural fianance}.
\newblock \bibinfo{journal}{\emph{e-Finanse: Financial Internet Quarterly}}
  \bibinfo{volume}{8}, \bibinfo{number}{3} (\bibinfo{year}{2012}),
  \bibinfo{pages}{55--61}.
\newblock


\bibitem[Putni{\c{n}}{\v{s}}(2012)]%
        {putnicnvs20122mmarket}
\bibfield{author}{\bibinfo{person}{T{\=a}lis~J Putni{\c{n}}{\v{s}}}.}
  \bibinfo{year}{2012}\natexlab{}.
\newblock \showarticletitle{Market manipulation: A survey}.
\newblock \bibinfo{journal}{\emph{Journal of economic surveys}}
  \bibinfo{volume}{26}, \bibinfo{number}{5} (\bibinfo{year}{2012}),
  \bibinfo{pages}{952--967}.
\newblock


\bibitem[Quisenberry(2017)]%
        {quisenberry2017pponzi}
\bibfield{author}{\bibinfo{person}{William~L Quisenberry}.}
  \bibinfo{year}{2017}\natexlab{}.
\newblock \showarticletitle{Ponzi of all Ponzis: critical analysis of the
  Bernie Madoff scheme}.
\newblock \bibinfo{journal}{\emph{International Journal of Econometrics and
  Financial Management}} \bibinfo{volume}{5}, \bibinfo{number}{1}
  (\bibinfo{year}{2017}), \bibinfo{pages}{1--6}.
\newblock


\bibitem[Rahmatika et~al\mbox{.}(2019)]%
        {rahmatika2019detection}
\bibfield{author}{\bibinfo{person}{Dien~Noviany Rahmatika},
  \bibinfo{person}{Maulida~Dwi Kartikasari}, \bibinfo{person}{Dewi Indriasih},
  \bibinfo{person}{Inayah~Adi Sari}, {and} \bibinfo{person}{Armya Mulia}.}
  \bibinfo{year}{2019}\natexlab{}.
\newblock \showarticletitle{Detection of Fraudulent Financial Statement; Can
  Perspective of Fraud Diamond Theory be applied to Property, Real Estate, and
  Building Construction Companies in Indonesia?}
\newblock \bibinfo{journal}{\emph{European Journal of Business and Management
  Research}} \bibinfo{volume}{4}, \bibinfo{number}{6} (\bibinfo{year}{2019}).
\newblock


\bibitem[Ramsey(1927)]%
        {ramsey1927contribution}
\bibfield{author}{\bibinfo{person}{Frank~P Ramsey}.}
  \bibinfo{year}{1927}\natexlab{}.
\newblock \showarticletitle{A Contribution to the Theory of Taxation}.
\newblock \bibinfo{journal}{\emph{The economic journal}} \bibinfo{volume}{37},
  \bibinfo{number}{145} (\bibinfo{year}{1927}), \bibinfo{pages}{47--61}.
\newblock


\bibitem[Rathinaraj and Chendroyaperumal(2010)]%
        {rathinarajjj2010financial}
\bibfield{author}{\bibinfo{person}{Daniel Rathinaraj} {and}
  \bibinfo{person}{Chendrayan Chendroyaperumal}.}
  \bibinfo{year}{2010}\natexlab{}.
\newblock \showarticletitle{Financial fraud, cyber scams and India--A small
  survey of popular recent cases}.
\newblock \bibinfo{journal}{\emph{Cyber Scams and India--A Small Survey of
  Popular Recent Cases (May 12, 2010)}} (\bibinfo{year}{2010}).
\newblock


\bibitem[Rauchs and Hileman(2017)]%
        {RePEcjbsaltfin201704_gcbs}
\bibfield{author}{\bibinfo{person}{Michel Rauchs} {and}
  \bibinfo{person}{Garrick Hileman}.} \bibinfo{year}{2017}\natexlab{}.
\newblock \bibinfo{booktitle}{\emph{{Global Cryptocurrency Benchmarking
  Study}}}.
\newblock Number 201704-gcbs in \bibinfo{series}{Cambridge Centre for
  Alternative Finance Reports}. \bibinfo{publisher}{Cambridge Centre for
  Alternative Finance, Cambridge Judge Business School, University of
  Cambridge}.
\newblock
\urldef\tempurl%
\url{https://ideas.repec.org/b/jbs/altfin/201704-gcbs.html}
\showURL{%
\tempurl}


\bibitem[Renault(2014)]%
        {renault2014pump}
\bibfield{author}{\bibinfo{person}{Thomas Renault}.}
  \bibinfo{year}{2014}\natexlab{}.
\newblock \showarticletitle{Pump-and-dump or news? Stock market manipulation on
  social media}.
\newblock  (\bibinfo{year}{2014}).
\newblock


\bibitem[Reurink(2016)]%
        {article1}
\bibfield{author}{\bibinfo{person}{Arjan Reurink}.}
  \bibinfo{year}{2016}\natexlab{}.
\newblock \showarticletitle{Financial Fraud: A Literature Review}.
\newblock \bibinfo{journal}{\emph{MPIfG Discussion Paper Series}}
  \bibinfo{volume}{16} (\bibinfo{date}{05} \bibinfo{year}{2016}).
\newblock
\urldef\tempurl%
\url{https://doi.org/10.1111/joes.12294}
\showDOI{\tempurl}


\bibitem[Reurink(2019)]%
        {reurink2019financial}
\bibfield{author}{\bibinfo{person}{Arjan Reurink}.}
  \bibinfo{year}{2019}\natexlab{}.
\newblock \showarticletitle{Financial fraud: a literature review}.
\newblock \bibinfo{journal}{\emph{Contemporary Topics in Finance: A Collection
  of Literature Surveys}} (\bibinfo{year}{2019}), \bibinfo{pages}{79--115}.
\newblock


\bibitem[Reznik(2012)]%
        {reznik2012identity}
\bibfield{author}{\bibinfo{person}{Maksim Reznik}.}
  \bibinfo{year}{2012}\natexlab{}.
\newblock \showarticletitle{Identity theft on social networking sites:
  Developing issues of internet impersonation}.
\newblock \bibinfo{journal}{\emph{Touro L. Rev.}}  \bibinfo{volume}{29}
  (\bibinfo{year}{2012}), \bibinfo{pages}{455}.
\newblock


\bibitem[Rogal(2020)]%
        {rogal2020ssecrets}
\bibfield{author}{\bibinfo{person}{Lauren Rogal}.}
  \bibinfo{year}{2020}\natexlab{}.
\newblock \showarticletitle{Secrets, Lies, and Lessons from the Theranos
  Scandal}.
\newblock \bibinfo{journal}{\emph{Hastings LJ}}  \bibinfo{volume}{72}
  (\bibinfo{year}{2020}), \bibinfo{pages}{1663}.
\newblock


\bibitem[Romanosky et~al\mbox{.}(2010)]%
        {romanoskyy2010data}
\bibfield{author}{\bibinfo{person}{Sasha Romanosky},
  \bibinfo{person}{Alessandro Acquisti}, {and} \bibinfo{person}{Richard
  Sharp}.} \bibinfo{year}{2010}\natexlab{}.
\newblock \showarticletitle{Data breaches and identity theft: When is mandatory
  disclosure optimal?} TPRC.
\newblock


\bibitem[Rosmani et~al\mbox{.}(2020)]%
        {Rosmani2020}
\bibfield{author}{\bibinfo{person}{Arifah~Fasha Rosmani},
  \bibinfo{person}{Ariffin~Abdul Mutalib}, {and} \bibinfo{person}{Siti~Mahfuzah
  Sarif}.} \bibinfo{year}{2020}\natexlab{}.
\newblock \showarticletitle{The evolution of information dissemination,
  communication media and technology in Malaysia}. In
  \bibinfo{booktitle}{\emph{Journal of Physics: Conference Series}},
  Vol.~\bibinfo{volume}{1529}. IOP Publishing, \bibinfo{pages}{022044}.
\newblock


\bibitem[Rubin(2010)]%
        {rubin2010financiall}
\bibfield{author}{\bibinfo{person}{Lee Rubin}.}
  \bibinfo{year}{2010}\natexlab{}.
\newblock \showarticletitle{Financial services and sponsored links}.
\newblock \bibinfo{journal}{\emph{Journal of Direct, Data and Digital Marketing
  Practice}} \bibinfo{volume}{11}, \bibinfo{number}{3} (\bibinfo{year}{2010}),
  \bibinfo{pages}{241--244}.
\newblock


\bibitem[Rubin(2022)]%
        {rubin2022misinformationnk}
\bibfield{author}{\bibinfo{person}{Victoria~L Rubin}.}
  \bibinfo{year}{2022}\natexlab{}.
\newblock \bibinfo{booktitle}{\emph{Misinformation and Disinformation:
  Detecting Fakes with the Eye and AI}}.
\newblock \bibinfo{publisher}{Springer Nature}.
\newblock


\bibitem[Salanie(2011)]%
        {salanie2011economics}
\bibfield{author}{\bibinfo{person}{Bernard Salanie}.}
  \bibinfo{year}{2011}\natexlab{}.
\newblock \bibinfo{booktitle}{\emph{The economics of taxation}}.
\newblock \bibinfo{publisher}{MIT press}.
\newblock


\bibitem[Sannikov et~al\mbox{.}(2016)]%
        {sannikov2016dynamic}
\bibfield{author}{\bibinfo{person}{Yuliy Sannikov}, \bibinfo{person}{Andrzej
  Skrzypacz}, {et~al\mbox{.}}} \bibinfo{year}{2016}\natexlab{}.
\newblock \showarticletitle{Dynamic trading: price inertia and front-running}.
\newblock \bibinfo{journal}{\emph{Preprint}} (\bibinfo{year}{2016}).
\newblock


\bibitem[Sarna(2010)]%
        {sarna2010hhistory}
\bibfield{author}{\bibinfo{person}{David~EY Sarna}.}
  \bibinfo{year}{2010}\natexlab{}.
\newblock \bibinfo{booktitle}{\emph{History of greed: Financial fraud from
  tulip mania to Bernie Madoff}}.
\newblock \bibinfo{publisher}{John Wiley \& Sons}.
\newblock


\bibitem[Schleppegrell(2013)]%
        {schleppegrell2013systemic}
\bibfield{author}{\bibinfo{person}{Mary~J Schleppegrell}.}
  \bibinfo{year}{2013}\natexlab{}.
\newblock \showarticletitle{Systemic functional linguistics}.
\newblock In \bibinfo{booktitle}{\emph{The Routledge handbook of discourse
  analysis}}. \bibinfo{publisher}{Routledge}, \bibinfo{pages}{21--34}.
\newblock


\bibitem[Scopino(2014)]%
        {scopino2014questionable}
\bibfield{author}{\bibinfo{person}{Gregory Scopino}.}
  \bibinfo{year}{2014}\natexlab{}.
\newblock \showarticletitle{The (questionable) legality of high-speed pinging
  and front running in the futures market}.
\newblock \bibinfo{journal}{\emph{Conn. L. Rev.}}  \bibinfo{volume}{47}
  (\bibinfo{year}{2014}), \bibinfo{pages}{607}.
\newblock


\bibitem[Securities et~al\mbox{.}(2017)]%
        {secemblem_2017}
\bibfield{author}{\bibinfo{person}{US Securities}, \bibinfo{person}{Exchange
  Commission}, {et~al\mbox{.}}} \bibinfo{year}{2017}\natexlab{}.
\newblock \showarticletitle{Investor Bulletin: Initial Coin Offerings}.
\newblock \bibinfo{journal}{\emph{SEC. gov website, July}}
  \bibinfo{volume}{25} (\bibinfo{year}{2017}).
\newblock


\bibitem[Shao et~al\mbox{.}(2018)]%
        {shao_ciampaglia_varol_yang_flammini_menczer_2018}
\bibfield{author}{\bibinfo{person}{Chengcheng Shao},
  \bibinfo{person}{Giovanni~Luca Ciampaglia}, \bibinfo{person}{Onur Varol},
  \bibinfo{person}{Kai-Cheng Yang}, \bibinfo{person}{Alessandro Flammini},
  {and} \bibinfo{person}{Filippo Menczer}.} \bibinfo{year}{2018}\natexlab{}.
\newblock \showarticletitle{The spread of low-credibility content by social
  bots}.
\newblock \bibinfo{journal}{\emph{Nature communications}} \bibinfo{volume}{9},
  \bibinfo{number}{1} (\bibinfo{year}{2018}), \bibinfo{pages}{1--9}.
\newblock


\bibitem[Shin et~al\mbox{.}(2018)]%
        {shin_jian_driscoll_bar_2018df}
\bibfield{author}{\bibinfo{person}{Jieun Shin}, \bibinfo{person}{Lian Jian},
  \bibinfo{person}{Kevin Driscoll}, {and} \bibinfo{person}{Fran{\c{c}}ois
  Bar}.} \bibinfo{year}{2018}\natexlab{}.
\newblock \showarticletitle{The diffusion of misinformation on social media:
  Temporal pattern, message, and source}.
\newblock \bibinfo{journal}{\emph{Computers in Human Behavior}}
  \bibinfo{volume}{83} (\bibinfo{year}{2018}), \bibinfo{pages}{278--287}.
\newblock


\bibitem[Shoup(2017)]%
        {shoup2017public}
\bibfield{author}{\bibinfo{person}{Carl Shoup}.}
  \bibinfo{year}{2017}\natexlab{}.
\newblock \bibinfo{booktitle}{\emph{Public finance}}.
\newblock \bibinfo{publisher}{Routledge}.
\newblock


\bibitem[Shu et~al\mbox{.}(2017)]%
        {shu2017fake}
\bibfield{author}{\bibinfo{person}{Kai Shu}, \bibinfo{person}{Rohan Mahajan},
  \bibinfo{person}{Suhang Wang}, {and} \bibinfo{person}{Dongwon Luo}.}
  \bibinfo{year}{2017}\natexlab{}.
\newblock \showarticletitle{Fake news detection on social media: A data mining
  perspective}.
\newblock \bibinfo{journal}{\emph{ACM SIGKDD Explorations Newsletter}}
  \bibinfo{volume}{19}, \bibinfo{number}{1} (\bibinfo{year}{2017}),
  \bibinfo{pages}{22--36}.
\newblock


\bibitem[Simon(1971)]%
        {simon1971designing}
\bibfield{author}{\bibinfo{person}{Herbert~A Simon}.}
  \bibinfo{year}{1971}\natexlab{}.
\newblock \bibinfo{booktitle}{\emph{Designing organizations for an
  information-rich world}}.
\newblock \bibinfo{publisher}{Carnegie-Mellon Univ Pittsburgh PA Graduate
  School of Industrial Administration}.
\newblock


\bibitem[Spathis et~al\mbox{.}(2002)]%
        {spathis2002detecting012}
\bibfield{author}{\bibinfo{person}{Ch Spathis}, \bibinfo{person}{Michael
  Doumpos}, {and} \bibinfo{person}{Constantine Zopounidis}.}
  \bibinfo{year}{2002}\natexlab{}.
\newblock \showarticletitle{Detecting falsified financial statements: a
  comparative study using multicriteria analysis and multivariate statistical
  techniques}.
\newblock \bibinfo{journal}{\emph{European Accounting Review}}
  \bibinfo{volume}{11}, \bibinfo{number}{3} (\bibinfo{year}{2002}),
  \bibinfo{pages}{509--535}.
\newblock


\bibitem[Spathis(2002)]%
        {spathis2002detecting}
\bibfield{author}{\bibinfo{person}{Charalambos~T Spathis}.}
  \bibinfo{year}{2002}\natexlab{}.
\newblock \showarticletitle{Detecting false financial statements using
  published data: some evidence from Greece}.
\newblock \bibinfo{journal}{\emph{Managerial Auditing Journal}}
  (\bibinfo{year}{2002}).
\newblock


\bibitem[Stadler(2011)]%
        {staddller2011predatory}
\bibfield{author}{\bibinfo{person}{William~A Stadler}.}
  \bibinfo{year}{2011}\natexlab{}.
\newblock \showarticletitle{Predatory lending: is the credit CARD act enough?}
\newblock \bibinfo{journal}{\emph{Journal of Financial Crime}}
  \bibinfo{volume}{19}, \bibinfo{number}{1} (\bibinfo{year}{2011}),
  \bibinfo{pages}{99--111}.
\newblock


\bibitem[Str{\"o}mb{\"a}ck et~al\mbox{.}(2020)]%
        {stromback2020news1}
\bibfield{author}{\bibinfo{person}{Jesper Str{\"o}mb{\"a}ck},
  \bibinfo{person}{Yariv Tsfati}, \bibinfo{person}{Hajo Boomgaarden},
  \bibinfo{person}{Alyt Damstra}, \bibinfo{person}{Elina Lindgren},
  \bibinfo{person}{Rens Vliegenthart}, {and} \bibinfo{person}{Torun Lindholm}.}
  \bibinfo{year}{2020}\natexlab{}.
\newblock \showarticletitle{News media trust and its impact on media use:
  Toward a framework for future research}.
\newblock \bibinfo{journal}{\emph{Annals of the International Communication
  Association}} \bibinfo{volume}{44}, \bibinfo{number}{2}
  (\bibinfo{year}{2020}), \bibinfo{pages}{139--156}.
\newblock


\bibitem[Summers and Sweeney(1998)]%
        {summers1998fraudulently}
\bibfield{author}{\bibinfo{person}{Scott~L Summers} {and}
  \bibinfo{person}{John~T Sweeney}.} \bibinfo{year}{1998}\natexlab{}.
\newblock \showarticletitle{Fraudulently misstated financial statements and
  insider trading: An empirical analysis}.
\newblock \bibinfo{journal}{\emph{Accounting Review}} (\bibinfo{year}{1998}),
  \bibinfo{pages}{131--146}.
\newblock


\bibitem[Tardelli et~al\mbox{.}(2022)]%
        {tardelli_avvenutii_tesconii_cresci_2022}
\bibfield{author}{\bibinfo{person}{Serena Tardelli}, \bibinfo{person}{Marco
  Avvenuti}, \bibinfo{person}{Maurizio Tesconi}, {and} \bibinfo{person}{Stefano
  Cresci}.} \bibinfo{year}{2022}\natexlab{}.
\newblock \showarticletitle{Detecting inorganic financial campaigns on
  Twitter}.
\newblock \bibinfo{journal}{\emph{Information Systems}}  \bibinfo{volume}{103}
  (\bibinfo{year}{2022}), \bibinfo{pages}{101769}.
\newblock


\bibitem[Tay et~al\mbox{.}(2022)]%
        {tay2022ccomparison}
\bibfield{author}{\bibinfo{person}{Li~Qian Tay}, \bibinfo{person}{Mark~J
  Hurlstone}, \bibinfo{person}{Tim Kurz}, {and} \bibinfo{person}{Ullrich~KH
  Ecker}.} \bibinfo{year}{2022}\natexlab{}.
\newblock \showarticletitle{A comparison of prebunking and debunking
  interventions for implied versus explicit misinformation}.
\newblock \bibinfo{journal}{\emph{British Journal of Psychology}}
  \bibinfo{volume}{113}, \bibinfo{number}{3} (\bibinfo{year}{2022}),
  \bibinfo{pages}{591--607}.
\newblock


\bibitem[Tetlock(2010)]%
        {Tetlock_2008}
\bibfield{author}{\bibinfo{person}{Paul~C Tetlock}.}
  \bibinfo{year}{2010}\natexlab{}.
\newblock \showarticletitle{Does public financial news resolve asymmetric
  information?}
\newblock \bibinfo{journal}{\emph{The Review of Financial Studies}}
  \bibinfo{volume}{23}, \bibinfo{number}{9} (\bibinfo{year}{2010}),
  \bibinfo{pages}{3520--3557}.
\newblock


\bibitem[Torgler(2008)]%
        {torglerr2008we}
\bibfield{author}{\bibinfo{person}{Benno Torgler}.}
  \bibinfo{year}{2008}\natexlab{}.
\newblock \showarticletitle{What Do We Know About Tax Fraud?: an Overview of
  Recent Developments}.
\newblock \bibinfo{journal}{\emph{Social Research: An International Quarterly}}
  \bibinfo{volume}{75}, \bibinfo{number}{4} (\bibinfo{year}{2008}),
  \bibinfo{pages}{1239--1270}.
\newblock


\bibitem[Trozze et~al\mbox{.}(2022)]%
        {Trozze2022}
\bibfield{author}{\bibinfo{person}{Arianna Trozze}, \bibinfo{person}{Josh
  Kamps}, \bibinfo{person}{Eray~Arda Akartuna}, \bibinfo{person}{Florian~J
  Hetzel}, \bibinfo{person}{Bennett Kleinberg}, \bibinfo{person}{Toby Davies},
  {and} \bibinfo{person}{Shane~D Johnson}.} \bibinfo{year}{2022}\natexlab{}.
\newblock \showarticletitle{Cryptocurrencies and future financial crime}.
\newblock \bibinfo{journal}{\emph{Crime Science}}  \bibinfo{volume}{11}
  (\bibinfo{year}{2022}), \bibinfo{pages}{1--35}.
\newblock


\bibitem[Turcotte et~al\mbox{.}(2015)]%
        {turcotte20151news}
\bibfield{author}{\bibinfo{person}{Jason Turcotte}, \bibinfo{person}{Chance
  York}, \bibinfo{person}{Jacob Irving}, \bibinfo{person}{Rosanne~M Scholl},
  {and} \bibinfo{person}{Raymond~J Pingree}.} \bibinfo{year}{2015}\natexlab{}.
\newblock \showarticletitle{News recommendations from social media opinion
  leaders: Effects on media trust and information seeking}.
\newblock \bibinfo{journal}{\emph{Journal of computer-mediated communication}}
  \bibinfo{volume}{20}, \bibinfo{number}{5} (\bibinfo{year}{2015}),
  \bibinfo{pages}{520--535}.
\newblock


\bibitem[Van~der Linden et~al\mbox{.}(2015)]%
        {van2015debunking}
\bibfield{author}{\bibinfo{person}{Sander Van~der Linden},
  \bibinfo{person}{Anthony~A Leiserowitz}, {and} \bibinfo{person}{Edward~W
  Maibach}.} \bibinfo{year}{2015}\natexlab{}.
\newblock \showarticletitle{Debunking: A meta-analysis of the psychological
  efficacy of messages countering misinformation}.
\newblock \bibinfo{journal}{\emph{Psychological science}} \bibinfo{volume}{26},
  \bibinfo{number}{10} (\bibinfo{year}{2015}), \bibinfo{pages}{1531--1546}.
\newblock


\bibitem[Vedova and Technology(2021)]%
        {vedova_technology_2021}
\bibfield{author}{\bibinfo{person}{Holly Vedova} {and} \bibinfo{person}{The FTC
  Office~of Technology}.} \bibinfo{year}{2021}\natexlab{}.
\newblock \showarticletitle{FTC data shows huge spike in cryptocurrency
  investment scams}.
\newblock \bibinfo{journal}{\emph{Federal Trade Commission}}
  (\bibinfo{date}{May} \bibinfo{year}{2021}).
\newblock
\urldef\tempurl%
\url{https://www.ftc.gov/news-events/news/press-releases/2021/05/ftc-data-shows-huge-spike-cryptocurrency-investment-scams}
\showURL{%
\tempurl}


\bibitem[Vedova and Technology(2023)]%
        {vedova_technology_2023}
\bibfield{author}{\bibinfo{person}{Holly Vedova} {and} \bibinfo{person}{The FTC
  Office~of Technology}.} \bibinfo{year}{2023}\natexlab{}.
\newblock \showarticletitle{Coronavirus advice for consumers}.
\newblock \bibinfo{journal}{\emph{Federal Trade Commission}}
  (\bibinfo{date}{Mar} \bibinfo{year}{2023}).
\newblock
\urldef\tempurl%
\url{https://www.ftc.gov/news-events/features/coronavirus/scams-consumer-advice}
\showURL{%
\tempurl}


\bibitem[Venkataramakrishnan(2023)]%
        {venkataramakrishnan_2023}
\bibfield{author}{\bibinfo{person}{Siddharth Venkataramakrishnan}.}
  \bibinfo{year}{2023}\natexlab{}.
\newblock \showarticletitle{FCA to gain powers to crack down on 'buy now, pay
  later firms}.
\newblock \bibinfo{journal}{\emph{Subscribe to read | Financial Times}}
  (\bibinfo{date}{Feb} \bibinfo{year}{2023}).
\newblock
\urldef\tempurl%
\url{https://www.ft.com/content/87399d17-8d93-4a6a-a056-b062ed07ce79}
\showURL{%
\tempurl}


\bibitem[Verma et~al\mbox{.}(2021)]%
        {verma2021wwelfake}
\bibfield{author}{\bibinfo{person}{Pawan~Kumar Verma}, \bibinfo{person}{Prateek
  Agrawal}, \bibinfo{person}{Ivone Amorim}, {and} \bibinfo{person}{Radu
  Prodan}.} \bibinfo{year}{2021}\natexlab{}.
\newblock \showarticletitle{WELFake: Word embedding over linguistic features
  for fake news detection}.
\newblock \bibinfo{journal}{\emph{IEEE Transactions on Computational Social
  Systems}} \bibinfo{volume}{8}, \bibinfo{number}{4} (\bibinfo{year}{2021}),
  \bibinfo{pages}{881--893}.
\newblock


\bibitem[Vishwanath(2015)]%
        {vishwanath_2014_social}
\bibfield{author}{\bibinfo{person}{Arun Vishwanath}.}
  \bibinfo{year}{2015}\natexlab{}.
\newblock \showarticletitle{Diffusion of deception in social media: Social
  contagion effects and its antecedents}.
\newblock \bibinfo{journal}{\emph{Information Systems Frontiers}}
  \bibinfo{volume}{17} (\bibinfo{year}{2015}), \bibinfo{pages}{1353--1367}.
\newblock


\bibitem[Vosoughi et~al\mbox{.}(2018)]%
        {vosoughi_roy_aral_2018adf}
\bibfield{author}{\bibinfo{person}{Soroush Vosoughi}, \bibinfo{person}{Deb
  Roy}, {and} \bibinfo{person}{Sinan Aral}.} \bibinfo{year}{2018}\natexlab{}.
\newblock \showarticletitle{The spread of true and false news online}.
\newblock \bibinfo{journal}{\emph{science}} \bibinfo{volume}{359},
  \bibinfo{number}{6380} (\bibinfo{year}{2018}), \bibinfo{pages}{1146--1151}.
\newblock


\bibitem[Vyas et~al\mbox{.}(2021)]%
        {vyas2021fake}
\bibfield{author}{\bibinfo{person}{Piyush Vyas}, \bibinfo{person}{Jun Liu},
  {and} \bibinfo{person}{Omar El-Gayar}.} \bibinfo{year}{2021}\natexlab{}.
\newblock \showarticletitle{Fake news detection on the web: An LSTM-based
  approach}.
\newblock  (\bibinfo{year}{2021}).
\newblock


\bibitem[Wang et~al\mbox{.}(2020)]%
        {wang2020grover}
\bibfield{author}{\bibinfo{person}{Rowan Wang}, \bibinfo{person}{Jieyu Zhang},
  {and} \bibinfo{person}{Quoc~V Le}.} \bibinfo{year}{2020}\natexlab{}.
\newblock \showarticletitle{Grover: A state-of-the-art defense against neural
  fake news}. In \bibinfo{booktitle}{\emph{Proceedings of the 2019 Conference
  on Empirical Methods in Natural Language Processing and the 9th International
  Joint Conference on Natural Language Processing (EMNLP-IJCNLP)}}. Association
  for Computational Linguistics, \bibinfo{pages}{54--64}.
\newblock


\bibitem[Weiner et~al\mbox{.}(2019)]%
        {weiner2019sec}
\bibfield{author}{\bibinfo{person}{Perrie~Michael Weiner},
  \bibinfo{person}{Edward~D Totino}, {and} \bibinfo{person}{Aaron Goodman}.}
  \bibinfo{year}{2019}\natexlab{}.
\newblock \showarticletitle{SEC issues warning to analysts profiting from
  “short and distort” schemes, opens the door for civil claims}.
\newblock \bibinfo{journal}{\emph{Journal of Investment Compliance}}
  \bibinfo{volume}{20}, \bibinfo{number}{2} (\bibinfo{year}{2019}),
  \bibinfo{pages}{34--38}.
\newblock


\bibitem[Weiner et~al\mbox{.}(2017)]%
        {weiner2017growing}
\bibfield{author}{\bibinfo{person}{Perrie~M Weiner}, \bibinfo{person}{Robert~D
  Weber}, {and} \bibinfo{person}{Kirby Hsu}.} \bibinfo{year}{2017}\natexlab{}.
\newblock \showarticletitle{The growing menace of ‘short and
  distort’campaigns}.
\newblock \bibinfo{journal}{\emph{Thomson Reuters Exp. Anal}}
  (\bibinfo{year}{2017}).
\newblock


\bibitem[Wicaksari et~al\mbox{.}(2023)]%
        {wicaksari2023diamond}
\bibfield{author}{\bibinfo{person}{Erisa~Aprilia Wicaksari},
  \bibinfo{person}{Syam Widia}, {and} \bibinfo{person}{Vini~Wiratno Putri}.}
  \bibinfo{year}{2023}\natexlab{}.
\newblock \showarticletitle{The Diamond Fraud Theory for Property and Real
  Estate to Detect Financial Report Fraud}.
\newblock \bibinfo{journal}{\emph{Management Analysis Journal}}
  \bibinfo{volume}{12}, \bibinfo{number}{2} (\bibinfo{year}{2023}),
  \bibinfo{pages}{144--156}.
\newblock


\bibitem[Widiyati et~al\mbox{.}(2021)]%
        {widiyati2021role}
\bibfield{author}{\bibinfo{person}{Dian Widiyati}, \bibinfo{person}{Riyan~Harbi
  Valdiansyah}, \bibinfo{person}{M Meidijati}, {and} \bibinfo{person}{H
  Hendra}.} \bibinfo{year}{2021}\natexlab{}.
\newblock \showarticletitle{The Role of Public Accountants in Fraud Prevention
  and Detection in the Taxation Sector during Covid-19}.
\newblock \bibinfo{journal}{\emph{Golden Ratio of Auditing Research}}
  \bibinfo{volume}{1}, \bibinfo{number}{2} (\bibinfo{year}{2021}),
  \bibinfo{pages}{70--82}.
\newblock


\bibitem[Williams({[n.\,d.]})]%
        {williams6383930}
\bibfield{author}{\bibinfo{person}{Geoff Williams}.}
  \bibinfo{year}{[n.\,d.]}\natexlab{}.
\newblock \showarticletitle{Financial misinformation and social media: 5
  questions to ask before spending money}.
\newblock \bibinfo{journal}{\emph{MoneyGeek.com}} (\bibinfo{year}{[n.\,d.]}).
\newblock
\urldef\tempurl%
\url{https://www.moneygeek.com/financial-planning/advice/navigating-financial-misinformation/}
\showURL{%
\tempurl}


\bibitem[Williams et~al\mbox{.}(2019)]%
        {williams20191preferences}
\bibfield{author}{\bibinfo{person}{Susan~Lee Williams}, \bibinfo{person}{Kate
  Ames}, {and} \bibinfo{person}{Celeste Lawson}.}
  \bibinfo{year}{2019}\natexlab{}.
\newblock \showarticletitle{Preferences and trust in traditional and
  non-traditional sources of health information--a study of middle to older
  aged Australian adults}.
\newblock \bibinfo{journal}{\emph{Journal of Communication in Healthcare}}
  \bibinfo{volume}{12}, \bibinfo{number}{2} (\bibinfo{year}{2019}),
  \bibinfo{pages}{134--142}.
\newblock


\bibitem[Wu et~al\mbox{.}(2023)]%
        {wu2023bloomberggpt}
\bibfield{author}{\bibinfo{person}{Shijie Wu}, \bibinfo{person}{Ozan Irsoy},
  \bibinfo{person}{Steven Lu}, \bibinfo{person}{Vadim Dabravolski},
  \bibinfo{person}{Mark Dredze}, \bibinfo{person}{Sebastian Gehrmann},
  \bibinfo{person}{Prabhanjan Kambadur}, \bibinfo{person}{David Rosenberg},
  {and} \bibinfo{person}{Gideon Mann}.} \bibinfo{year}{2023}\natexlab{}.
\newblock \showarticletitle{Bloomberggpt: A large language model for finance}.
\newblock \bibinfo{journal}{\emph{arXiv preprint arXiv:2303.17564}}
  (\bibinfo{year}{2023}).
\newblock


\bibitem[Xiao and Benbasat(2011)]%
        {xiao_benbasat_2011asdd}
\bibfield{author}{\bibinfo{person}{Bo Xiao} {and} \bibinfo{person}{Izak
  Benbasat}.} \bibinfo{year}{2011}\natexlab{}.
\newblock \showarticletitle{Product-related deception in e-commerce: A
  theoretical perspective}.
\newblock \bibinfo{journal}{\emph{Mis Quarterly}} (\bibinfo{year}{2011}),
  \bibinfo{pages}{169--195}.
\newblock


\bibitem[Zhang et~al\mbox{.}(2022)]%
        {zhang2022theory009}
\bibfield{author}{\bibinfo{person}{Xiaohui Zhang}, \bibinfo{person}{Qianzhou
  Du}, {and} \bibinfo{person}{Zhongju Zhang}.} \bibinfo{year}{2022}\natexlab{}.
\newblock \showarticletitle{A theory-driven machine learning system for
  financial disinformation detection}.
\newblock \bibinfo{journal}{\emph{Production and Operations Management}}
  \bibinfo{volume}{31}, \bibinfo{number}{8} (\bibinfo{year}{2022}),
  \bibinfo{pages}{3160--3179}.
\newblock


\bibitem[Zhi et~al\mbox{.}(2021)]%
        {zhi2021financial01}
\bibfield{author}{\bibinfo{person}{Xiaofan Zhi}, \bibinfo{person}{Li Xue},
  \bibinfo{person}{Wengang Zhi}, \bibinfo{person}{Ziye Li}, \bibinfo{person}{Bo
  Zhao}, \bibinfo{person}{Yanzhen Wang}, {and} \bibinfo{person}{Zhen Shen}.}
  \bibinfo{year}{2021}\natexlab{}.
\newblock \showarticletitle{Financial fake news detection with multi fact
  CNN-LSTM model}. In \bibinfo{booktitle}{\emph{2021 IEEE 4th International
  Conference on Electronics Technology (ICET)}}. IEEE,
  \bibinfo{pages}{1338--1341}.
\newblock


\bibitem[Zhou et~al\mbox{.}(2023)]%
        {zhou2023synthetic}
\bibfield{author}{\bibinfo{person}{Jiawei Zhou}, \bibinfo{person}{Yixuan
  Zhang}, \bibinfo{person}{Qianni Luo}, \bibinfo{person}{Andrea~G Parker},
  {and} \bibinfo{person}{Munmun De~Choudhury}.}
  \bibinfo{year}{2023}\natexlab{}.
\newblock \showarticletitle{Synthetic lies: Understanding ai-generated
  misinformation and evaluating algorithmic and human solutions}. In
  \bibinfo{booktitle}{\emph{Proceedings of the 2023 CHI Conference on Human
  Factors in Computing Systems}}. \bibinfo{pages}{1--20}.
\newblock


\bibitem[Zhou and Zhang(2008)]%
        {zhouu_zhangg_2008}
\bibfield{author}{\bibinfo{person}{Lina Zhou} {and} \bibinfo{person}{Dongsong
  Zhang}.} \bibinfo{year}{2008}\natexlab{}.
\newblock \showarticletitle{Following linguistic footprints: Automatic
  deception detection in online communication}.
\newblock \bibinfo{journal}{\emph{Commun. ACM}} \bibinfo{volume}{51},
  \bibinfo{number}{9} (\bibinfo{year}{2008}), \bibinfo{pages}{119--122}.
\newblock


\end{thebibliography}
\end{document}